\documentclass[acmlarge,screen]{acmart}

\AtBeginDocument{%
  \providecommand\BibTeX{{%
    \normalfont B\kern-0.5em{\scshape i\kern-0.25em b}\kern-0.8em\TeX}}}

\setcopyright{rightsretained}
\acmJournal{IMWUT}
\acmYear{2023} \acmVolume{7} \acmNumber{4} \acmArticle{189} \acmMonth{12} \acmPrice{}\acmDOI{10.1145/3631441}

\usepackage{enumitem}
\usepackage{makecell}
\usepackage{multirow}
\usepackage{caption}
\usepackage{subcaption}
\begin{document}

\title{"Reading Between the Heat": Co-Teaching Body Thermal Signatures for Non-intrusive Stress Detection}


\author{\href{https://orcid.org/0000-0002-5261-5440}{Yi Xiao}}
\email{yxiao54@syr.edu}
\affiliation{%
  \institution{Syracuse University}
  \city{Syracuse}
  \state{New York}
  \country{USA}
  }

\author{\href{https://orcid.org/0000-0002-7016-6220}{Harshit Sharma}}
\email{hsharm04@syr.edu}
\affiliation{%
  \institution{Syracuse University}
  \city{Syracuse}
  \state{New York}
  \country{USA}
  }

\author{\href{https://orcid.org/0000-0003-4349-252X}{Zhongyang Zhang}}
\email{zhz138@ucsd.edu}
\affiliation{%
  \institution{University of California San Diego}
  \city{San Diego}
  \state{California}
  \country{USA}
  }
\author{\href{https://orcid.org/0000-0002-8852-732X}{Dessa Bergen-Cico}}
\email{dkbergen@syr.edu}
\affiliation{%
  \institution{Syracuse University}
  \city{Syracuse}
  \state{New York}
  \country{USA}
  }
  
\author{\href{https://orcid.org/0000-0003-1981-6395}{Tauhidur Rahman}}
\email{trahman@ucsd.edu}
\affiliation{%
  \institution{University of California San Diego}
  \city{San Diego}
  \state{California}
  \country{USA}
  }
  
\author{\href{https://orcid.org/0000-0002-0807-8967}{Asif Salekin}}
\thanks{Asif Salekin is the corresponding author}
\email{asalekin@syr.edu}
\affiliation{%
  \institution{Syracuse University}
  \city{Syracuse}
  \state{New York}
  \country{USA}
  }

\renewcommand{\shortauthors}{Xiao et al.}

\begin{abstract}

Stress impacts our physical and mental health as well as our social life. A passive and contactless indoor stress monitoring system can unlock numerous important applications such as workplace productivity assessment, smart homes, and personalized mental health monitoring. While the thermal signatures from a user's body captured by a thermal camera can provide important information about the ``fight-flight'' response of the sympathetic and parasympathetic nervous system, relying solely on thermal imaging for training a stress prediction model often lead to overfitting and consequently a suboptimal performance. This paper addresses this challenge by introducing ThermaStrain, a novel co-teaching framework that achieves high-stress prediction performance by transferring knowledge from the wearable modality to the contactless thermal modality. During training, ThermaStrain incorporates a wearable electrodermal activity (EDA) sensor to generate stress-indicative representations from thermal videos, emulating stress-indicative representations from a wearable EDA sensor. During testing, only thermal sensing is used, and stress-indicative patterns from thermal data and emulated EDA representations are extracted to improve stress assessment. The study collected a comprehensive dataset with thermal video and EDA data under various stress conditions and distances. ThermaStrain achieves an F1 score of $0.8293$ in binary stress classification, outperforming the thermal-only baseline approach by over 9\%. Extensive evaluations highlight ThermaStrain's effectiveness in recognizing stress-indicative attributes, its adaptability across distances and stress scenarios, real-time executability on edge platforms, its applicability to multi-individual sensing, ability to function on limited visibility and unfamiliar conditions, and the advantages of its co-teaching approach. These evaluations validate ThermaStrain's fidelity and its potential for enhancing stress assessment.


\end{abstract}

\ccsdesc[500]{Human-centered computing~Ubiquitous and mobile computing}
\ccsdesc[500]{Computing methodologies~Biometrics}
\ccsdesc[500]{Applied computing~Health care information systems}
\ccsdesc[300]{Computing methodologies~Learning latent representations}

\keywords{ Thermal sensing, Stress detection, Affective Computing, Health Sensing, Machine Learning, Multimodality, Co-teaching, Contactless sensing.}

\maketitle

\section{Introduction}\label{Intro}
Stress is an intense emotional phenomenon that can be triggered by external stressors or stimuli \cite{khosrowabadi2018stress,gross2014emotion,liu2017many}.  It elicits spontaneous physiological responses governed by our autonomic nervous system's ``fight or flight'' response \cite{buijs2000integration}. These responses may include changes in skin temperature \cite{engert2014exploring,sonkusare2019detecting}, skin conductance \cite{mauri2010psychophysiological}, and other indicators. Chronic stress poses significant risks to both physical and mental health, emphasizing the importance of monitoring and managing stress \cite{picard2016automating,reiche2004stress}.
\par 
Smart wearables have been explored for stress sensing \cite{ollander2016comparison,menghini2019stressing,boucsein2012electrodermal}. However, such wearable-based solutions typically require close proximity to the user's body or skin for capturing different physiological parameters (e.g., electrodermal activities - EDA, skin temperature), which can be burdensome and invasive for the users. Furthermore, their sensing scope is limited to the wearer, restricting their applicability in indoor environments with multiple occupants. Passive and contactless indoor stress monitoring, on the other hand, offers the potential to unlock numerous applications that are challenging to achieve through EDA or other wearable-based solutions. For instance, passive and contactless stress monitoring for elderly dementia patients \cite{fook2007automated, konig2015validation, salekin2017dave} or employees in smart workplaces \cite{leone2020multi,alberdi2018using,mantello2023emotional,lee2019clara} can provide valuable insights. These approaches can facilitate feedback on stress, including bio-feedback \cite{yu2018biofeedback}, interventions for stress management, enhancements in well-being, and customization of user experiences based on stress-related data \cite{lee2020toward}. Further discussion on specific application scenarios for contactless passive stress sensing is provided in Section \ref{application-scenarios}.
\par 
To address the challenge, several RGB camera-based stress sensing systems \cite{gavrilescu2019predicting,zhang2020video,giannakakis} have been developed; however, their efficacy depends on lighting conditions and is fraught with privacy concerns \cite{griffiths2018privacy,nielsen2014taking}. Similarly, remote  photoplethysmography (PPG) based on RGB cameras fails to work well under varying lighting conditions \cite{cho2019physiological}. Balancing accuracy, privacy, and adaptability to environmental variations remains a significant challenge for developing stress sensing systems.
\par 
Infrared thermography, which utilizes thermal cameras for stress sensing, can offer a viable solution. Thermal cameras can capture changes in skin temperature \cite{james2014reliability}, heart rate through facial skin blood flow \cite{kim2018remote}, which are indicative of physiological stress responses \cite{abdelrahman2017cognitive,puri}. Unlike RGB cameras, thermal cameras are robust to different light conditions \cite{al2021affective}. Prior works have shown promising human sensing results using thermal imaging in poor lighting and even at night \cite{eum2013human,kim2019pedestrian,nielsen2014taking}. Additionally, thermal videos/imaging are typically considered more privacy-preserving compared to RGB imaging, preventing the inadvertent exposure of environmental/contextual information like personal items, addresses, displayed documents, and content within photo frames, among others \cite{chiu2023privacy,nielsen2014taking,griffiths2018privacy}.
 These attributes enhance the appeal of thermal cameras for stress sensing. 
\par 
Several studies \cite{al2021affective,cross2013thermal,perez2013thermal} have explored the use of thermal sensing to assess stress. However, the efficacy of stress detection achieved solely through thermal sensing is lower compared to leveraging other single modalities such as EEG \cite{thammasan2017familiarity}, ECG \cite{keshan2015machine}, and PPG \cite{heo2021stress}. Recognizing this limitation, recent studies \cite{cho2019instant,ghosh2022classification,zhang2022real} are focusing on multi-modality approaches to increase the efficacy of thermal stress sensing. These approaches typically require combining thermal data with other physiological signals from different modalities, such as EEG, ECG, or PPG, which increases computational cost, user burden and thus limits scalability. 
\par 
This paper explores \emph{"whether a system that utilizes stress-indicative physiological signals (from wearable sensors) during model development but relies solely on thermal sensing during evaluation or deployment can outperform uni-modal thermal camera-based stress sensing approaches."} - Uni-modal approaches use only thermal information in all model development and evaluation phases, discussed in Section \ref{related-uni-modal}. 
\par 
To address the above-mentioned question, we introduce \emph{ThermaStrain}, a first-of-its-kind end-to-end co-teaching framework that enhances the efficacy of infrared thermography-based stress sensing. By incorporating electrodermal activity (EDA) sensing (collected from wearable) in the model training phase, which is a reliable method for measuring human stress response in real-time \cite{alberdi2016towards,jukiewicz2021electrodermal}, \emph{ThermaStrain} improves the accuracy of stress detection. During training, the model utilizes EDA sensing to generate a stress-indicative latent representation from thermal videos, emulating the stress-indicative signal patterns obtained from a real wearable EDA sensor. During test/evaluation time, when only thermal sensing is available, the stress-indicative information extracted from the thermal videos using the emulated EDA representation is used for stress detection. By integrating EDA sensing and learning to extract EDA-guided stress-indicative information from thermal videos, \emph{ThermaStrain} offers an accurate and non-intrusive solution. It is a pioneering approach that addresses the research question while maintaining the simplicity and effectiveness of thermal sensing in stress assessment.
\par 
The main contributions of this work are:
\begin{itemize}
    \item The paper presents \emph{ThermaStrain}, a novel co-teaching-based solution that surpasses existing uni-modal and co-teaching baselines in thermal stress sensing (Sections \ref {EVal-comp-uni-multi} and \ref{eval-co-teachin}). What sets \emph{ThermaStrain} apart is its ability to achieve superior performance (Section \ref{eval-real-time}) using only thermal sensing during evaluation/testing, thus maintaining the non-intrusive appeal of thermal sensing. \emph{ThermaStrain} solution opens up new possibilities for enhancing ubiquitous computing applications where non-intrusive stress assessment plays a crucial role in promoting well-being and optimizing experiences, such as workplace productivity assessment, smart home systems, and personalized and passive mental health monitoring, including depression \cite{hammen2005stress}. 

    \item To our knowledge, no existing public/available dataset contains variable distance, full or partial body thermal camera data, and physiological parameters in different stress conditions. To overcome this limitation, we collected a comprehensive dataset consisting of infrared thermography sensing (thermal video data) and electrodermal activity (EDA) physiological parameter sensing data. The dataset was gathered from 32 individuals who performed four distinct stress-inducing tasks, each associated with different stressors. Importantly, data was collected from varying distances of 5-11 feet. This dataset's unique characteristics, including the diverse set of stressors and variable distances, allow the paper to develop and evaluate models that are capable of generalizing to different distances (Section \ref{distance-eval})) and stress situations (Section \ref{Eval:generalizability}). De-identified data will be made public. By addressing the gap in available datasets, this work enables more robust and applicable research in thermal stress sensing.

    \item The paper presents thorough evaluations and discussions (Sections \ref{Discussion-sec} and \ref{Discussion-deployment}) that delve into the benefits of co-teaching (i.e., \emph{ThermaStrain}'s approach) in developing an improved stress-sensing solution. Section \ref{Discussion-sec}'s evaluations encompassed various aspects, including understanding how co-teaching facilitates better solution development and the effectiveness of \emph{ThermaStrain} in extracting stress-indicative information from thermal frames. Furthermore, Section \ref{Discussion-deployment} delves deeply into the potential applications and challenges of deploying \emph{ThermaStrain} in real-world scenarios. This encompasses scenarios with multiple individuals, limited visibility, and conditions that are unseen during training, such as camera angles, distances, postures, stress conditions, backgrounds, and ethical concerns. The evaluations establish the fidelity of \emph{ThermaStrain} to the co-teaching paradigm and validate its ability to enhance stress sensing.
    
\end{itemize}

\section{Motivation and Vision Model of Co-teaching-based Thermal Stress Sensing}\label{Motivation-of-the-paper-co-teaching}
This section discusses the vision model of thermal sensing and the motivation or justification of the presented co-teaching solution.
\paragraph{Vision Model of the Infrared Thermography Sensing} While tracking the thermal signatures of a target object, the thermal energy reaching the thermal camera sensors is formulated by Kylili et al. \cite{kylili2014infrared} as:
\begin{equation}
    I= I_{EM}+I_{REF}+I_{ATM}
\end{equation}
Here $I_{EM}$ is the energy emitted by the object, $I_{REF}$ is the energy reflected by the surrounding and intercepted by the object, and $I_{ATM}$ is a term that accounts for atmospheric influence due to attenuation of thermal radiation. Here, the camera determines the $I_{REF}$ and $I_{ATM}$ during calibration. This allows the sensor to get information about the amount of thermal energy emitted by the target object, which is human body in this paper's scope.

\begin{figure}[h]
  \centering
  \includegraphics[width=0.8\textwidth]{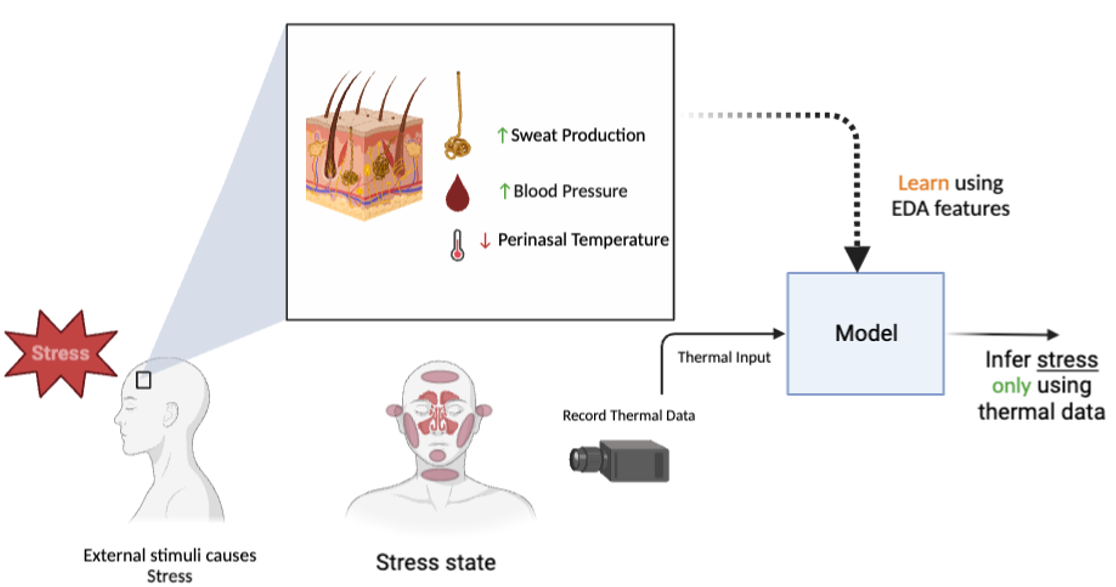}
 \vskip -3ex
 \caption{Vision Model of Co-teaching based Thermal Sensing}

\label{vision}
\vskip -4ex
\end{figure}

\paragraph{Thermal-based Stress Sensing} While facing a stressful situation, our autonomic nervous system (ANS) changes the blood perfusion to the skin surface, which changes the skin temperature \cite{engert2014exploring,sonkusare2019detecting}. Prior works \cite{pavlidis2012fast,sonkusare2019detecting} have demonstrated that human body thermal signatures, i.e., skin temperature, particularly those from the face and neck regions \cite{sonkusare2019detecting}, can provide an insight into human physiology under stressful conditions, that thermal camera captures through $I_{EM}$.

\paragraph{EDA-based Stress Sensing} Prior studies in psychophysiology have shown that electrodermal activity (EDA) or skin conductance is a gold standard to measure human stress response in real-time \cite{alberdi2016towards,jukiewicz2021electrodermal}. While experiencing a stressful situation, there is an increase in the skin conductance levels, which is caused by the activation of the eccrine sweat glands \cite{jukiewicz2021electrodermal,krzywicki2014non}.

\paragraph{Limitation: Latency in Thermal Sensing}\label{motivation-paper-thermal-limit} Thermal sensing can help to model the understanding of the stress response but cannot outperform the EDA-based assessment since skin conductance provides a rapid response profile (having a delay of 1-3 seconds from the stimulus onset) \cite{braithwaite2013guide}. In contrast, thermal responses have a relatively higher latency of $4-5$ seconds from the stimulus onset \cite{merla2007thermal,sonkusare2019detecting}.

\paragraph{\textbf{Co-teaching Goal}} The Co-teaching goal is to learn the rapid patterns emerging through skin conductance, i.e., activation of eccrine sweat glands through the thermal energy measurement from the human body $I_{EM}$. The eccrine sweat glands are composed of a single tubular structure, and the volume of liquid in the tubular part of the eccrine glands increases when activated \cite{hodge2018anatomy}. Studies \cite{neema2021infrared,quesada2015effect} have shown that an increase in sweating, i.e., water on the skin surface, results in the perception of lower temperature by the infrared thermography sensing (i.e., thermal camera information) than the thermal contact sensor (or actual skin temperature) \cite{neema2021infrared,quesada2015effect}. \emph{Meaning there exists a latent thermal signature of skin conductance.} The co-teaching goal of the \emph{ThermaStrain} approach is to teach thermal sensing modality to extract such latent patterns with the guidance of EDA modality, resulting in better stress assessment performance. Our presented end-to-end co-teaching approach effectively teaches such patterns, resulting in higher stress assessment performance from thermal sensing alone during testing or evaluation.

\section{Related works}\label{related-works}
Thermal imaging has shown promising results for physiology-based affective state and stress detection techniques in recent years \cite{al2021affective,rusli2020implementation,cho2019instant,zhang2022real,perez2013thermal,cross2013thermal}. \emph{ThermaStrain} solution builds upon two components of prior work (1) Thermal imaging for understanding electrodermal activity (EDA) responses. (2) Stress detection using thermal responses. These are discussed below:
 
 \subsection{Thermal Imaging for Measuring EDA Response }
Recent works \cite{pavlidis2012fast,cho2019physiological,krzywicki2014non} showed that thermal imaging of skin areas with a high density of sweat glands like the palm or the perinasal regions \cite{cho2019physiological} could be used to monitor the EDA response or the activation of the sweat glands. The activation of the sweat glands leads to a change in the skin temperature \cite{al2021affective,cho2019physiological}, which is captured using the thermal camera. 
A recent study \cite{pavlidis2012fast} found that there were high correlations between the galvanic skin response (GSR) extracted from the electrodermal activity (EDA) and the thermal signals extracted from the finger and perinasal region (correlation coefficient r=0.94 and r=0.96 respectively). Another work \cite{krzywicki2014non} studied the active pores on the skin surface using high-resolution thermal imaging and found a high correlation (correlation coefficient r=0.7) between the pore activation index measured using the thermal images \cite{krzywicki2014non} and skin conductance response measured from the finger. These works show the potential of thermal imaging for studying human physiological processes. Our work leverages the correlations between thermal responses and skin conductance, i.e., EDA, which these prior works have established.

\subsection{Thermal Imaging for Stress Detection}\label{Rel-worl-termal}
Studies have utilized different physiological modalities like PPG \cite{heo2021stress,cho2019instant}, ECG \cite{keshan2015machine,zhang2022real}, RGB image or video data \cite{zhang2022real,walambe2021employing}, EDA \cite{setz2009discriminating,zhu2022feasibility}, and thermal imaging \cite{cho2019instant,zhang2022real,perez2013thermal,cross2013thermal} to detect human stress. 
Recently, researchers have successfully used a combination of thermal imaging and different physiological sensors to detect individuals’ affective state \cite{al2021affective,rusli2020implementation}, cognitive load \cite{abdelrahman2017cognitive}, stress \cite{cho2019instant,zhang2022real,perez2013thermal,cross2013thermal} and even deception \cite{pavlidis2002seeing}. 
\par 
This section focuses on the prior literature on human stress detection, with a particular focus on thermal imaging based studies. These works can be divided into three strands based on their methodology.

\subsubsection{Uni-Modal Approaches to Stress Detection:}\label{related-uni-modal}
Uni-modal approaches in the literature use a single physiological modality like EEG \cite{thammasan2017familiarity}, ECG \cite{keshan2015machine}, PPG \cite{heo2021stress} to detect human stress. In uni-modal thermal stress sensing scope, works like Cross et al. \cite{cross2013thermal} used only thermal imaging to track regions of interest in the facial area to detect human stress using an LDA classifier, achieving 89.3\% accuracy. Another work \cite{perez2013thermal} used a thermal imaging-based approach to extract thermal maps corresponding to the facial, neck, and shoulder region for detecting positive and negative affective states with 90\% accuracy by using statistical descriptors like average, minimum, maximum, and standard deviation and the difference between the minimum and the maximum temperature for the face, neck, and shoulder region in the thermal image frames. However, these approaches require the thermal cameras to be up close to an individual’s face, hence have limited practical use in non-intrusive stress monitoring.

\subsubsection{Multi-Modal Approaches to Stress Detection:} 
Multi-modal approaches use two or more modalities for the human stress detection task. These techniques have also been widely studied in the literature. 

Cho et al. \cite{cho2019instant} proposed a human stress measurement system using smartphone camera–based PPG and thermal video, achieving an average classification accuracy of 78.3\% which outperformed the single modality baselines. 
Walambe et al. \cite{walambe2021employing} explored early fusion and late fusion techniques to predict stress from posture, physiological, and video data. Their evaluation showed that early fusion outperformed late fusion by 5\%. Can et al. \cite{can2019continuous} tried different schemes for modality fusion and found that using multiple modalities improved the performance of their stress detection systems in all scenarios. Ghosh et al. \cite{ghosh2022classification} converted data into Gramian Angular Field (GAF) before fusing multi-modality data, which can represent temporal correlations between each timestamp. They achieved significantly better performance than using raw data. Zhang et al. \cite{zhang2022real} fused ECG, voice, and RGB facial video for acute stress detection. Their ablation study showed that the overall performance was improved using ResNet 50 and Inflated 3D-CNN. 

Finally, multi-modal machine learning approaches leverage information from multiple modalities and often achieve higher accuracy than uni-modal approaches. However, not all modalities may be available in real-life scenarios and can be costly and comparatively more invasive, limiting their practical use \cite{zheng2021deep}.

\subsubsection{Co-Teaching Approaches}
To address the limitation of multi-modality, many studies have attempted to reconstruct missing modalities from existing ones to address the issue of missing modalities at inference time. These approaches fall under the domain of Co-Teaching. E.g., Zheng et al. \cite{zheng2021deep} trained a prototype network to learn meta-sensory representations by modeling knowledge retention mechanisms. Rajan et al. \cite{rajan2021robust} proposed a modality translator to translate the weak modality of strong modality, so that weak modality alone can achieve better performance during evaluation. Fortin et al. \cite{fortin2019multimodal} proposed a multi-task learning framework that prepares multiple classifiers depending on the availability of modalities. Wang et al. \cite{wang2022adversarial} designed a Generative Adversarial Network (GAN) to reconstruct the missing modality. Li et al. \cite{li2022valhalla} trained a Visual Hallucination Transformer that maps text to images and showed that visualizing scenes from the text can improve machine translation systems. This paper considers the multi-task learning \cite{fortin2019multimodal} and `Hallucination Transformer' \cite{li2022valhalla} as baselines.
\par 
None of the above-discussed studies are on thermal imaging or stress. The closest state-of-the-art work to co-teaching on thermal imaging is StressNet \cite{kumar2021stressnet}, which obtained ECG attributes from thermal input and utilized the extracted ECG-relevant thermal embedding to predict stress. The study utilizes only closed facial thermal frames, extracts ECG-relevant embedding using a ResNet, and captures temporal dynamics through an LSTM backbone. Even though it is not exactly co-teaching, due to its similarity to the concept, it is considered one of the baselines of this paper.

\section{Description of the Dataset and our Data Collection Procedure }
\label{Data-collection}
To address the lack of an available dataset containing variable distance, full or partial body thermal camera data, and physiological parameters in different stress conditions, the paper collected data in an indoor setting. Participants were engaged in various non-stress and stress-inducing tasks. The tasks were carefully designed in collaboration with a behavioral psychologist and approved by the X University Institutional Review Board (IRB) to ensure ethical compliance.
\par
\textbf{Participants:} The participants in the dataset were $32$ undergrad and graduate students enrolled at X University. They comprised $12$ male and $20$ female participants ($22-32$ years of age). All data were collected in a single laboratory visit, and participants signed informed consent before initiating the study. This limited age group may not be generalizable to older adults or children.
\subsection{Sensing Modalities} During the experiment, we collected thermal videos and electrodermal activity (EDA) physiological parameters. The following devices were utilized for the data collection:
\par 
\textbf{Thermal Imaging:} The Seek Thermal CompactPRO thermal camera\footnote{\url{https://www.thermal.com/compact-series.html}} \cite{kirimtat2020flir} was used to capture thermal video data (i.e., sequence of thermal frames). Thermal frames were captured with a 240$\times$320 pixel resolution, a $32$-degree field of view, and at $5$ frames per second (fps). We use libseek\_thermal \cite{libseek} API for data collection. Though the thermal camera can capture $10$ fps, our observation showed that at $5$ fps, frame rates are the most stable.
\par 
\textbf{The Empatica E4 Wristband:} The Empatica E4 Wristband\footnote{\url{https://www.empatica.com/research/e4/}} is a wearable device designed to monitor physiological signals and gather data about an individual's physical and emotional well-being. E4 wristband encompasses an EDA sensor and a PPG (Photoplethysmography) sensor, where EDA data is collected at a sampling rate of 4 fps, while the PPG data is captured at 64 fps.

\subsection{Data-Collection Experimental Procedure}
\label{procedure}
\begin{figure}
     \centering
     \begin{subfigure}[b]{0.45\textwidth}
         \centering
         \includegraphics[width=\textwidth]{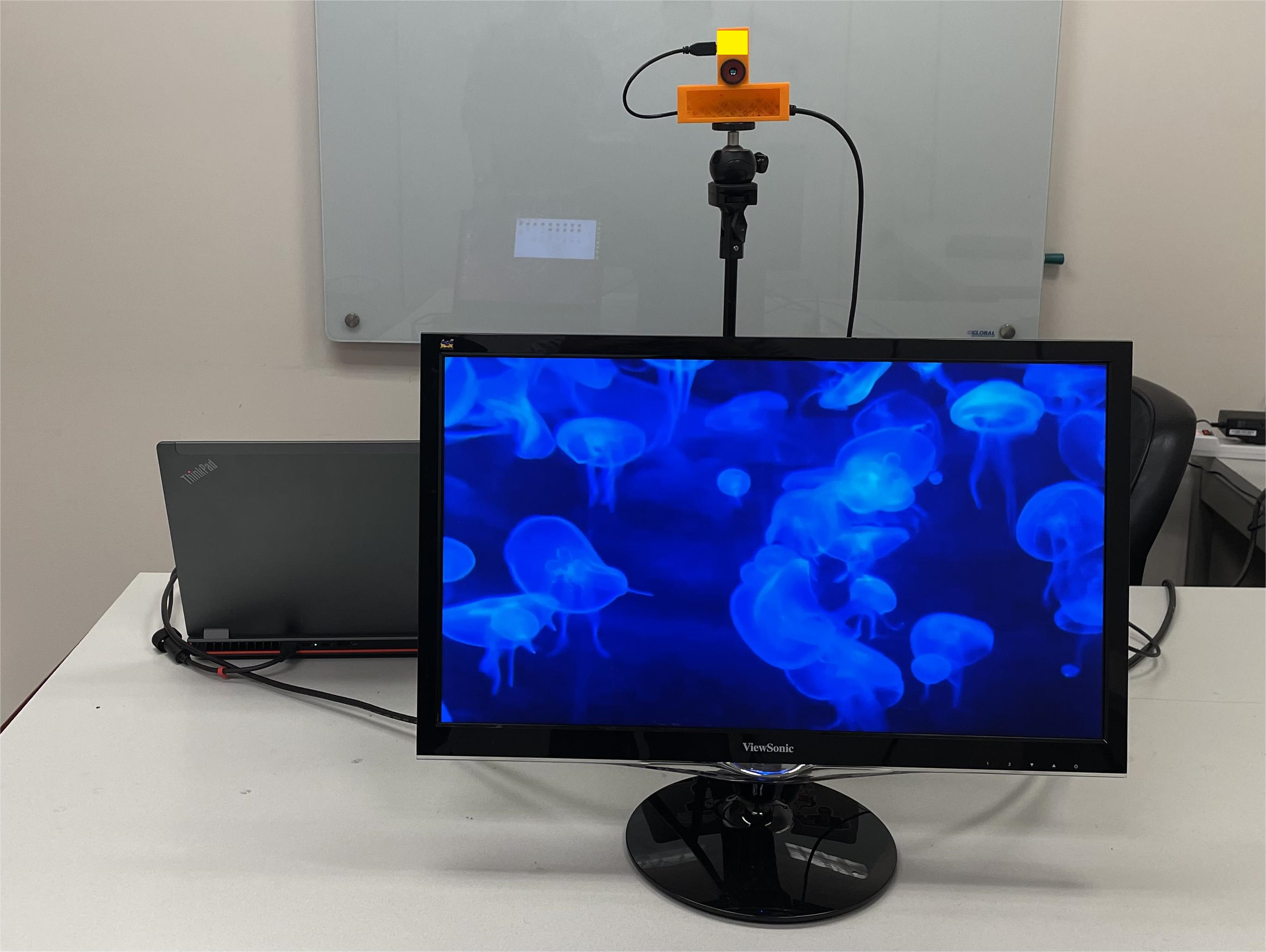}
         \caption{Frontend}
         \label{fig:frontend}
     \end{subfigure}
     \hfill
     \begin{subfigure}[b]{0.45\textwidth}
         \centering
         \includegraphics[width=\textwidth]{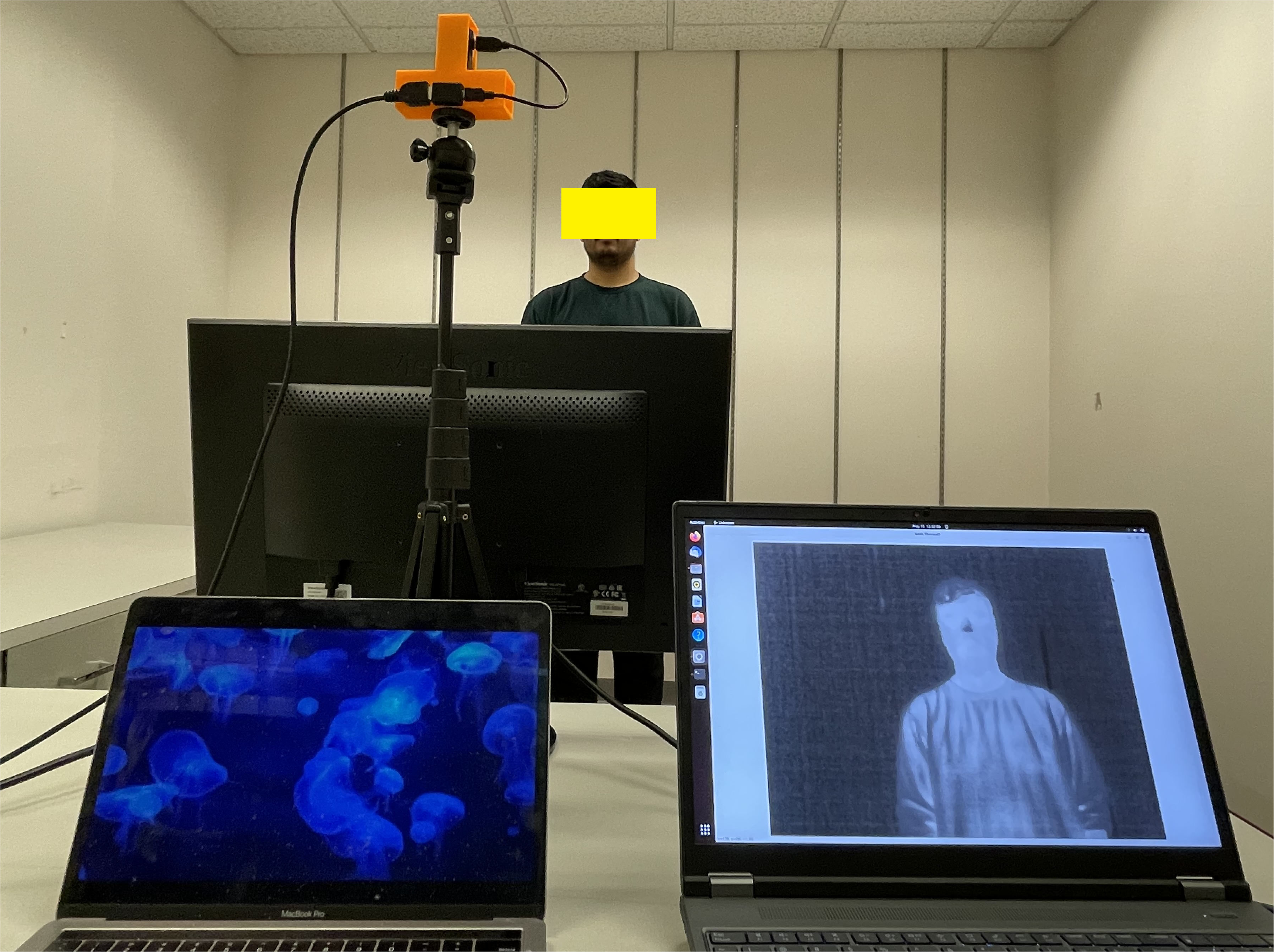}
         \caption{Backend}
         \label{fig:backend}
     \end{subfigure}
\caption{Data collection setup}
\label{fig:data-collection-setup}
\end{figure}

This section discusses the experimental procedure we followed during the data collection session. Upon arrival, the participants were given time to read and sign the consent form. 
Next, participants were asked to stand in front of a computer screen and a Seek Thermal CompactPRO camera. Additionally, they wore an Empatica E4 wristband on their left hand, which recorded their EDA responses during the data collection. Figure \ref{fig:data-collection-setup} illustrates the data collection setup in the laboratory.

\textit{Distances:} Three distance lines were marked from the thermal camera at $5$, $7$, and $9$ feet. Each participant was randomly asked to stand and encouraged to limit their movement within $1-2$ feet behind (i.e., farther from the sensor) one of these lines. In total, $10$ participants stood at the $5$ feet line, $12$ participants at the $7$ feet line, and $10$ participants at the $9$ feet line.
\par
\textit{Ambient Temperature:}  Data collection was performed in indoor rooms of the X University over two years, across all seasons. However, indoor temperatures were regulated. During the heating season (September 15 - May 15), the AC was set to 68 degrees Fahrenheit; during the cooling season (May 16 - September 14), it was set to 76 degrees Fahrenheit.

\par


\textit{Study Protocol: }
Thermal and EDA physiological response data were collected for non-stress and stressful conditions. Notably, this study's data collection protocol did not adhere to a 1-to-1 non-stress vs. stress challenge design. Instead, it followed a protocol similar to the Trier Social Stress Test (TSST) \cite{goodman2017meta,iqbal2022stress}, where one or more non-stress-inducing baselines are used for comparison with stress challenges. For instance, Iqbal et al. 2022's \cite{iqbal2022stress} protocol had a single non-stress-inducing baseline task followed by three stress-inducing tasks in a fixed sequence. Similarly, in this paper’s protocol, participants performed two non-stress tasks first, followed by four stress-inducing tasks known to elicit physiological responses \cite{brouwer2014new,goodman2017meta,iqbal2022stress}. While two non-stress tasks enable establishing baselines from various conditions and participants’ activities, each stress-inducing task presented unique stimuli to the participants, eliciting stress responses across various conditions and activities, as discussed below. The collected data from this protocol enables the development of a non-stress vs. stress detection approach applicable across various conditions.
\par
Moreover, human physiological responses, including skin temperature changes, are not momentary concerning the onset of a stressor; rather, they may persist for several minutes \cite{herborn2015skin}. Following the literature \cite{brouwer2014new,goodman2017meta,iqbal2022stress}, to avoid any bias from residual stress effects, non-stress-inducing tasks are conducted initially, and later, a fixed sequence of stress-inducing tasks are performed sequentially. Additionally, as identified in \cite{goodman2017meta}, no interfering activities, such as questionnaires, occurred at least 15 mins before introducing the four stress-inducing tasks.
\par
\begin{figure}[h]
  \centering
  \includegraphics[width=0.6\textwidth]{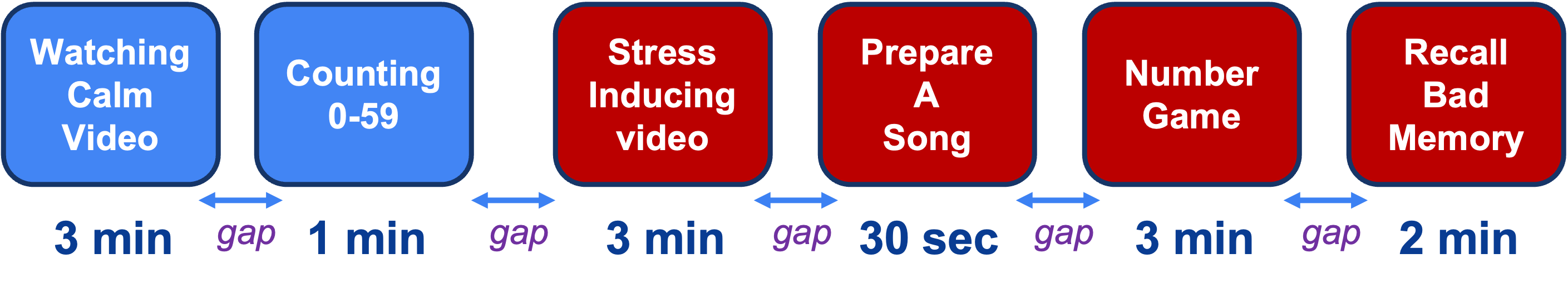}
 \vskip -3ex
 \caption{Study Protocol}

\label{study-protocol}
\vskip -4ex
\end{figure}

 
The study protocol is presented in Figure \ref{study-protocol}. On average, the data collection session was 15 minutes incorporating the gaps between the tasks. Participants self-reported their subjective stress levels on a scale from 0 (no stress) to 5 (extreme stress). Participants self-reported their subjective stress levels every 30 seconds during the watching calm video and stress-inducing video tasks. However, for tasks involving counting task, preparing a song, playing a number game, and recalling a negative memory, self-reporting was only performed at the end of each task. This approach was chosen to prevent potential disruption during task execution, which could affect the stressor's effectiveness.
The details of the (1) Non-Stress Inducing and (2) Stress-inducing tasks are below.

\paragraph{Non-Stress Inducing Tasks:}
\label{non-stress-stress}
The participants performed two tasks in the non-stressful condition data collection phase.
\begin{enumerate}
    \item \textbf{Watching a Calm video:} Participants watched a $3$ min video with jellyfish floating in the ocean. Recent stress and behavioral research studies \cite{sharma2022psychophysiological,tumanova2019autonomic} have used similar videos to establish a non-stress-inducing measurement baseline. The self-reporting mean score was $0.73$ during this task.

    \item \textbf{Counting task:} We asked the participants to slowly count from $0$ to $59$. This task was designed to reflect a non-stressful speaking condition. This task took, on average, 1 min for each participant. At the end of this task, the self-reporting mean score was $0.83$.

\end{enumerate}
\paragraph{Stress Inducing Tasks:}
\label{stress-task}
The participants performed four stress-inducing tasks in this stressful condition data collection phase. The tasks were designed based on prior studies \cite{brouwer2014new,goodman2017meta,schaefer2010assessing} in psychology and behavioral science. 

\begin{enumerate}
    \item \textbf {Passive stress induce video}: Participants watched four stress-inducing video clips (for a total 3min) from the emotional stimuli database \cite{schaefer2010assessing}, which contains movie clips with labels: stressful, scary, fearful, and disgust, and are annotated and ranked by 50 film experts and 364 volunteers. During this task, the self-reporting mean score was $2.99$.

    \item \textbf{Sing-a-Song Stress Test (SSST)}: During the SSST \cite{brouwer2014new} task, participants were asked to prepare a song in 30 seconds without any prior notification in the presence of the task by the experiment coordinators. Subsequently, they sing a song for up to 30 seconds. It's important to note that this study used only the 30-second preparation phase data, excluding any data during the actual singing. This is due to research indicating that the SSST task, like song preparation, induces stress through social evaluation and uncertainty in the confederate's reaction to the participant's performance \cite{dickerson2004acute}. At the end of this task, the self-reporting mean score was $1.42$.
    
    \item \textbf{Trier Stress Task (TST)}. This task follows the TST task \cite{kirschbaum1993trier}, where the participants were given a surprise arithmetic task to count backward from a large number by `17'. For example, if the starting number is 1000, the participant should say: 983, 966, 949,$...$, etc. Every time the participants made a mistake or took longer, they were asked to start from the beginning. Studies \cite{goodman2017meta,helminen2019meta} have reported that TST is the gold standard protocol that leads to a reliable high-stress response. The task took, on average, about 3 mins for each participant. At the end of this task, the self-reporting mean score was $3.33$.

    \item \textbf{Recalling a Bad memory:} According to literature \cite{connolly2018negative}, when we reminisce about negative events, our bodies respond as if we are experiencing those events again, activating the fight or flight response and releasing stress hormones such as cortisol and adrenaline. This can lead to high-stress response \cite{kross2012asking}. The task took, on average, about 2 mins for each participant. At the end of this task, the self-reporting mean score was $1.8$.

\end{enumerate}

\subsection{Validating Stress-Response Due to the Stress-Inducing Tasks}\label{Stress=validation-preprocessing}
To verify induced stress through the stress-inducing tasks, we assessed Heart Rate Variability (HRV) using Empatica E4 wristband data. Literature \cite{kim2018stress,sarsenbayeva2019measuring, von2017resolving, campanella2023method} link HRV changes to stress, often tied to reduced parasympathetic activity, seen as decreased High Frequency (HF) and increased Low Frequency (LF). We compute LF/HF ratio by dividing LF power by HF power, which rises under stress \cite{kim2018stress}. Following \cite{campanella2023method}; we extract photoplethysmogram (PPG) features from Empatica E4, with $3$-minute windows and $1$-second steps, aligning with recommended HRV analysis window sizes \cite{makowski2021neurokit2}.
\par 
First, we separate the non-stress-inducing (i.e., 3-min windows belonging to the first two tasks in Figure \ref{study-protocol}) and stress-inducing task windows (i.e., 3-min windows belonging to the last four tasks in Figure \ref{study-protocol}), then filter all using a Chebyshev II order-$4$ filter ($20$ dB stopband attenuation, $0.5-5$ Hz passband). PPG signals become heartbeat intervals, removing outliers beyond $500-1200$ ms (heart rates $50-120$ bpm) and filling gaps with linear interpolation. Finally, we extract LF/HF HRV values from processed features.

\begin{figure}[h]
  \centering
  \includegraphics[width=0.4\textwidth]{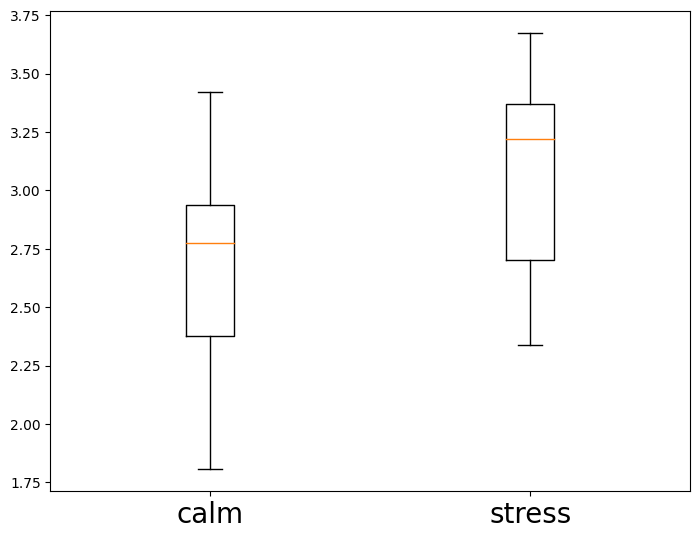}
 \vskip -3ex
 \caption{LF/HF ratio during non-stress vs. stress-inducing tasks}

\label{stress-validation}
\vskip -2ex
\end{figure}

Figure \ref{stress-validation} shows the LF/HF ratio during the non-stress and stress-inducing tasks. We applied a one-way ANOVA, which yielded a statistically significant effect of stress on heart rate variability, i.e., LF/HF ratio (p-value=0.0077). Consistent with findings by \cite{kim2018stress}, an elevated LF/HF ratio is noticeable during stress-inducing tasks, indicating heightened participant stress levels resulting from the introduced challenges through these four tasks.

\subsection{Data Preprocessing}\label{Data-preprocessing}
Notably, throughout our assessment of the \emph{ThermaStrain}, we refrained from excluding any data segments based on self-reported information (as described in Section \ref{procedure}) or stress validation through HRV (discussed in Section \ref{Stress=validation-preprocessing}). This decision was made because the physiological stress response may persist for several minutes concerning the onset of a stressor \cite{herborn2015skin}, and in each of the stress-inducing tasks, both self-reports and HRV analysis indicated heightened stress levels, confirming the efficacy of the stressors. Given that our experimental protocol entails stress-inducing tasks lasting between 30 seconds and 3 minutes each, assuming a participant is stressed, only a portion of that duration may introduce bias.
\par 
In this section, we discuss our data preprocessing steps for the raw thermal and EDA data.
\par
\textbf{Normalizing the EDA data:} Research shows that individuals from different populations may exhibit different levels of skin conductance due to various factors such as genetics, skin thickness, and environmental factors. Therefore, following previous research \cite{braithwaite2013guide}, z-score normalization was applied to each participant to control for individual differences in EDA level.
\par
\textbf{Extraction of stress event detection windows:} A window size of 5 seconds with 2 seconds of overlap was used for real-time stress detection. We determined this window size empirically (through grid search), aiming to balance high-stress assessment efficacy with the real-time usability of the sensing system. Larger window sizes yielded similar effectiveness, while smaller sizes compromised stress-assessment performance. Notably, this aligns with the discussion in Section \ref{Motivation-of-the-paper-co-teaching}. Given that thermal response latency is under 5 seconds, and the latency gap between EDA and thermal is approximately 2 seconds, a 5-second detection window can capture the physiological stress response at the onset of the stressor and facilitate knowledge transfer from EDA to Thermal during training.
\par 
During data collection, each thermal and EDA data point was marked with an absolute global time to facilitate synchronization between modalities. We synchronized the thermal and EDA data by selecting 5 seconds of thermal data and retrieving simultaneous EDA data according to the absolute global timestamp.  After pre-processing, 5 seconds of thermal data had a dimension of  [25 × 1 × 240 × 320] (an individual thermal frame was of the shape [1×240×320]), and the EDA data had a dimension of [5 × 4].
\par
\textbf{Human body detection:} The human body constitutes a fraction of the thermal frames. Since only the human body thermal information is pertinent to the stress, we applied a human body segmentation (i.e., body-region identification) algorithm named DetectorRS \cite{detectors} on thermal frames. We pre-trained the DetectorRS model on Microsoft COCO dataset \cite{lin2014microsoft} and evaluated its performance on our manually labeled (with pixel-wise body area and bounding box labels) thermal dataset. We used IOU as the evaluation metrics \cite{rosebrock2016intersection} that represent the ratio of overlap vs. union of the predicted and ground truth image segmentation regions. The DetectorRS body segmentation model achieves an average of 85.02\% IOU score, which is reasonably high. After identifying the body region, the thermal frame pixel values outside the human body segmentation mask are zeroed out. Finally, a window-wise z-score normalization is applied to the thermal images. In all of the evaluations, background masked-out thermal frames are utilized during training and testing. 

\par 
Such background masking enables \emph{ThermaStrain} to be generalizable and readily deployable in unknown scenarios. E.g., as discussed in Section \ref{discussion-multi-person}, such masking allows for identifying high stress in scenarios when multiple individuals are present in front of the thermal camera.

\section{Proposed Co-Teaching Approach}\label{Approach}
\subsection{Problem Statement} 
Given a sequence of thermal frames, i.e., thermal video $t\in T$, where $t= (t_{1}, . . . , t_{k})$ and synchronized EDA values, $e \in E$, where $e = (e_{1}, . . . , e_{l})$, our goal is to train a model that can predict $y \in Y$, where $y=$\textit{(`stress' or `non-stress')}, from only thermal video $t$ without requiring the EDA values at inference time. Since thermal video and EDA sampling rates are not necessarily the same, for a fixed stress detection window, $k \ne l$.

\subsection{Approach Overview}\label{overview}
This section presents a novel co-teaching approach named \emph{ThermaStrain}. Since EDA is a strong indicator of stress \cite{setz2009discriminating}, the \emph{ThermaStrain} approach simultaneously learns separate stress-indicative embeddings from EDA and thermal video; however, enforcing them to be similar for the same objective, inferring stress vs. non-stress class. \textit{Such enforcement enables the extraction of knowledge from thermal video similar to stress-indicative EDA physiological parameters alongside other thermal attributes indicative of stress, resulting in a higher stress inference performance.}

\par 
The \emph{ThermaStrain} model, shown in Figure \ref{fig:framework}, comprises three neural network modules. A thermal encoder $F_{T}$ is used to map an input thermal video $t$ to thermal embedding $z_{t}$; an EDA encoder $F_{E}$ is utilized to map the synchronized EDA sequence $e$ to EDA embedding $z_{e}$; And a classifier module $F_{C}$ that predicts the corresponding inference $y$ taking $z_{t}$ and $z_{e}$ separately. 
\par 
\subsubsection{During training} Both the thermal videos and corresponding synchronized EDA values are available during training. Meaning, the training dataset $D_{train}={(t,e,y)}$, where $t\in T$, $e\in E$, and $y\in Y$. The stress vs. non-stress output is inferred through two streams. The thermal embedding $z_{t}$ and EDA embedding $z_{e}$ are generated separately. Notably, the generated $z_t$ and $z_e$ embeddings have the same dimension $d$. The classifier module $F_{C}$ infers $y^t$ and $y^e$ by taking the $z_{t}$ and $z_{e}$ separately as follows:
\begin{equation}
\begin{split}
y^t = p(y|t,F_{T}, F_{C})=p(y|z_{t},F_{C})p(z_{t}|t,F_{T})\\
y^e = p(y|e,F_{E}, F_{C})=p(y|z_{e},F_{C})p(z_{e}|e,F_{E})
\label{eq:coteach1}
\end{split}
\end{equation}

\textit{Task losses}: One of the training objectives is to teach the model to infer $y^t$ and $y^e$ as close as the target output $y$. To attain the objective, the proposed approach introduces two task losses $l_t$ and $l_e$, which are cross-entropy losses between the target output $y$ and the generated inferences $y^t$ and $y^e$ through the two streams, thermal, and EDA. \textit{Minimizing $l_t$ and $l_e$ ensures better embeddings $z_{t}$ and $z_{e}$ extraction encompassing the respective mortality's stress-indicative markers.}

\begin{equation}
\begin{split}
l_{e}(y^e)=-log p (y|e,F_{E})\\
l_{t}(y^t)=-log p (y|t,F_{T})
\label{eq:task-loss1}
\end{split}
\end{equation}

\textit{Similarity loss $l_{s}$ }: Since the \emph{ThermaStrain} approach also aims to \textit{learn the extraction of knowledge from thermal video similar to the knowledge from EDA information that is highly predictive of stress, it is important to enforce the embeddings $z_{t}$ and $z_{e}$ to be similar.} Hence, \emph{ThermaStrain} introduces a similarity loss $l_{s}$ to maximize the joint likelihood of $z_{t}$ and $z_{e}$ during training. Here we use the mean square error to measure the similarity of embeddings.
\begin{equation}
l_{s}(z_{t},z_{e})=\sum (z_{t}^{2}-z_{e}^{2})
\label{eq:loss_coteach}
\end{equation}

\textit{Consistency loss $l_{c}$ }:
Another training objective is to encourage consistency between inferences $y^t$ and $y^e$. Considering that with the utilization of similarity loss $l_{s}$ during training, the thermal and EDA embeddings $z_{t}$ and $z_{e}$ would be similar; however, not the same, classifier module $F_C$ may generate mismatched $y^t$ and $y^e$. This may result in inferior performance during testing when the EDA sequence $e$ would not be available. Hence, to enforce consistency between $y^t$ and $y^e$, We define a consistency loss $l_{c}$.
\begin{equation}
l_{c}(y^{t},y^{e})=y^{t} log\frac{y^{t}}{y^{e}}= p(y|t,F_{T},F_{C})log\frac{p(y|t,F_{T},F_{C})}{ p(y|t,F_{E}, F_{C})}
\label{eq:loss_consistency1}
\end{equation}
Here, Equation \ref{eq:loss_consistency1} is the Kullback-Leibler divergence between the two conditional distributions $y^t$ and $y^e$.

\textit{Overall Training Loss $L$}: The overall optimization objective, i.e., overall training loss $L$ of the \emph{ThermaStrain} approach, is finally defined as a weighted sum of the two task losses, similarity loss, and consistency loss:

\begin{equation}
L=l_{t}+l_{e}+\alpha  l_{s}+\beta l_{c}
\label{eq:finalloss}
\end{equation}

Where $\alpha$ and $\beta$ are hyperparameters that control the weight of co-teaching and consistency objectives during training.

\subsubsection{During Test/Evaluation} 
\begin{figure}[]
\centering
\includegraphics[width=\linewidth]{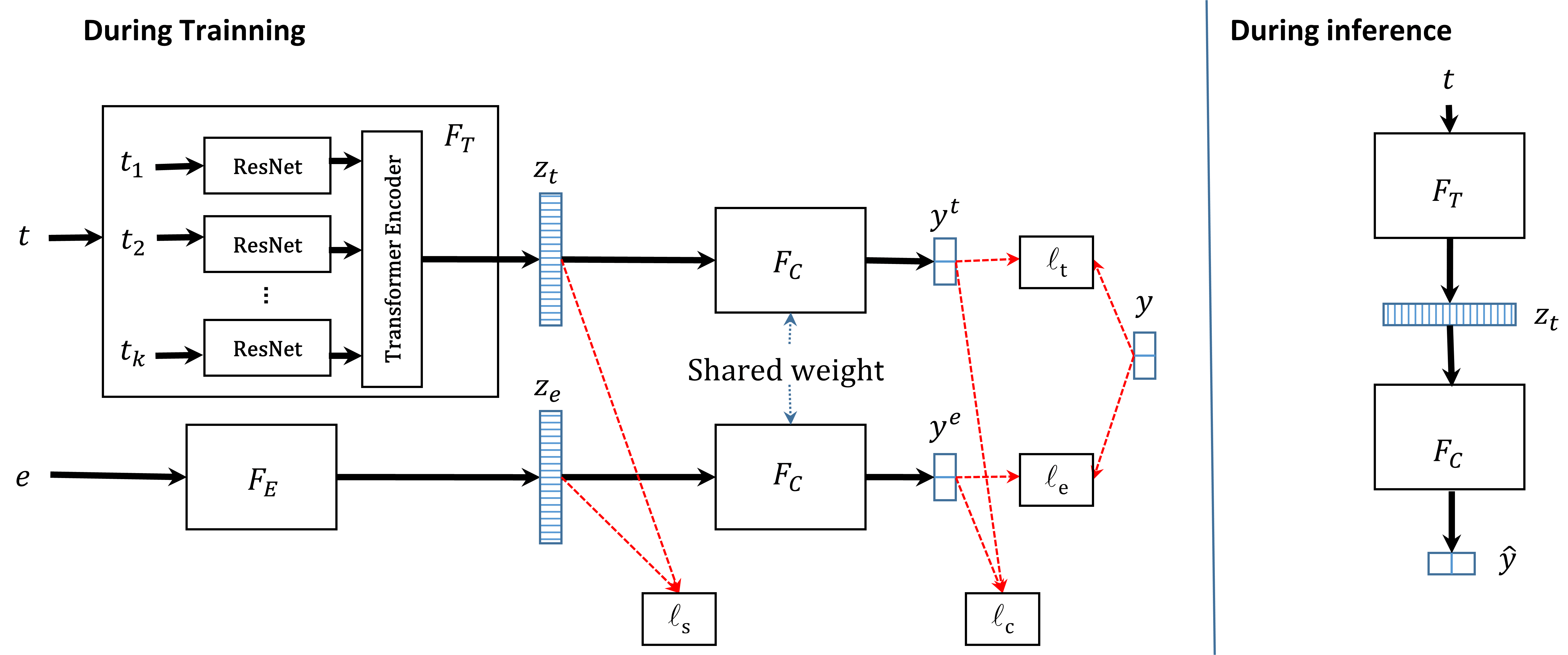}
\vskip -1ex
\caption{The framework of \emph{ThermaStrain} approach.}
\label{fig:framework}
\end{figure}

Only the thermal videos are available during testing or evaluation. Meaning, the test dataset $D_{test}={(t,y)}$, where $t\in T$, and $y\in Y$. As shown in Figure \ref{fig:framework}, during testing, the stress vs. non-stress output is inferred through the thermal stream, generating inference $\hat{y}$ taking $t$ as input using the equation below:

\begin{equation}
\hat{y} = p(y|t,F_{T}, F_{C})=p(y|z_{t},F_{C})p(z_{t}|t,F_{T})
\label{eq:test}
\end{equation}



\subsection{Discussion of the Modules}
The modules of the \emph{ThermaStrain} approach are discussed below. 

\subsubsection{Thermal Encoder $F_T$}
It takes a thermal video in the form of a sequence of thermal frames, $t= (t_{1}, . . . , t_{k})$ as input, and generates an aggregated embedding $z_{t}$ comprising stress indicative thermal markers/information from each frame, and the temporal thermal attributes depicted through the frame sequence. The module comprises a ResNet followed by a Transformer network. 

As shown in Figure \ref{fig:framework}, ResNet takes each frame $t_i$ to generate a framewise embedding $z^i_t$, representing the stress indicative thermal information from the respective frame. Later, the Transformer takes all the framewise embeddings ($z^1_t$,…,$z^T_t$) to aggregate the temporal information and generate the thermal video embedding  $z_{t}$.


\subsubsection{EDA Encoder $F_E$}
The input EDA values $e$ within a detection window have a $[5 \times 4]$ dimension representation. First, we extract the 6 features from EDA values, including mean, min, max, median, variability, and standard deviation. Then these six features was fed into two linear layers followed by a ReLU activation function to generate the EDA embeddings $z_{e}$.

\subsubsection{Classifier $F_C$}
In the \emph{ThermaStrain} model, the $F_C$ module is represented by a simple network comprising linear layers with a ReLU activation function. Notably, the generated thermal and EDA embeddings $z_{t}$ and $z_{e}$ have the same dimension $d$. During training, the $F_C$ takes both the embeddings separately to predict $y^t$ and $y^e$, respectively. Finally during test/evaluation, $F_C$ predicts the stress vs. non-stress inference $\hat{y}$ during the test or evaluation. 

\section{Evaluations}\label{evaluation}
This section discusses the efficiency and applicability of \emph{ThermaStrain} by investigating some key questions. The presented network parameter configurations were optimized by performing a grid search of the possible parameter values. Presented evaluation results are end-to-end, incorporating the inaccuracy due to pre-processing errors. Finally, evaluations are presented with metrics: sensitivity, specificity, accuracy (\%), and F1 scores.

\textit{Evaluation Dataset-split}: For each of the evaluations, we followed \textbf{\textit{the person-disjoint hold-out method}} \cite{cawley2010over}. Our collected data includes 32 sessions (each with a different participant). In this study, we performed a 5-fold evaluation where-in each fold, we split the dataset into validation (5 sessions) and training set (rest of the sessions). Training and validation sets were disjoint concerning participants and sessions. Our dataset is imbalanced, having more stress samples than non-stress. Therefore, in each fold, we performed under-sampling on the stress samples to make the dataset balanced.

\subsection{Implementation of the Presented Approach} 
As discussed in Section \ref{overview}, presented approach has three modules, discussed below:

\begin{itemize}
\item \textit{The Thermal Encoder}: Comprises a ResNet and a Transformer Encoder. The ResNet starts with a $7\times7$ convolutional layer, followed by three residual blocks. Each residual block includes two convolutional layers, followed by batch normalization and ReLU activation. At the end of the ResNet, an adaptive pooling layer pools the feature map into a size of $2\times2$. Finally, we set the channel of the last convolutional layer as $64$, resulting in an output embedding of shape $2\times2\times64$, which is $256$ dimensions. The embedding of all frames is then fed into a transformer encoder to aggregate information over time. After the transformer, we perform mean pooling on the sequential dimension and use the resulting embeddings for downstream tasks.

\item \textit{The EDA Encoder}: Considering the input EDA data has a simple feature dimension of $6$. We design a simple EDA encoder that comprises two linear layers with ReLU activation and dropout. 

\item \textit{The Classifier}: The classifier module comprises two linear layers with a ReLU activation function and dropout.

\end{itemize}

\subsection{Optimization} 
The weights $\alpha$ and $\beta$ in Equation \ref{eq:finalloss}, and the learning rate are hyper-parameters that were identified through the python toolkit Optuna\cite{optuna_2019}. It uses a Bayesian Optimization algorithm called Tree-Structured Parzen estimator to identify the optimum set of values.

\subsection{Comparison of Co-Teaching with Uni- and Multi-Modal Approaches}\label{EVal-comp-uni-multi}
This section compares the \emph{ThermaStrain} approach with uni-modal and multi-modal approaches leveraging the modalities in hand, thermal video, and EDA. 
To ensure a fair comparison among the approaches being compared, we set the complexity of each component to be the same, including the number of neurons in linear layers, the number of layers in the classifier, and the number of kernels in the CNN layers. We then use Optuna, a hyperparameter optimization framework, to comprehensively evaluate hyperparameters such as the learning rate to determine the optimal settings. Finally, we present the best results obtained for each model.
\par 

\emph{Implementations:} The thermal baseline takes only a 5-second thermal video as input and predicts stress. It uses ResNet as a feature extractor, a transformer to aggregate information over time, and a multi-layer Perceptron classifier to make the classification. 

The EDA baseline takes 5 seconds of EDA data and predicts stress. It uses a multi-layer perceptron followed by a transformer to extract features and a multi-layer perceptron classifier to make the classification. 

The multimodal baseline shares a similar structure to our \emph{ThermaStrain} implementation, with a thermal feature extractor, EDA encoder, and classifier. The only difference is that, instead of enforcing the $z_{T}$ and $z_{E}$ embeddings to be similar, classifier taking each of them separately, the classifier takes the concatenated embedding of $z_{T}$ and $z_{E}$ to make inferences.


\emph{Evaluation Result Discussion:} 
Table \ref{table:baselines} presents the results of our stress vs. non-stress binary classification evaluation. The \emph{ThermaStrain} approach achieved an accuracy of $83.17$\% and an F1-score of $0.8293$. In comparison, the uni-modal thermal baseline model only achieved $76.2$\% accuracy and $0.7592$ F1 scores. The \emph{ThermaStrain} model outperforms the thermal baseline model by over $9$\%. 

The EDA baseline and multi-modality model also achieved $0.8568$ and $0.8897$ F1 scores, respectively. These scores are higher than our thermal baseline and \emph{ThermaStrain} model.

The evaluation coherent with EDA is a strong indicator of stress. The co-teaching approach successfully extracts EDA-relevant embedding, i.e., skin conductance-relevant information from thermal video, resulting in a significant performance improvement over the uni-modal thermal baseline. However, such extraction is lossy; hence co-teaching still cannot outperform multi-modality or EDA-based stress assessment approaches. Notably, compared to these approaches, EDA is not required by the \emph{ThermaStrain} approach at the inference time, enabling contactless deployment, i.e., less obtrusive sensing.

\begin{table}[h]
\caption{Evaluation of \textit{Co-teaching} compared with Uni- and Multi-Modal Approaches}
\resizebox{0.6\textwidth}{!}{

\begin{tabular}{|c|l|l|l|l|}
\hline
Model     & \multicolumn{1}{c|}{Sensitivity} & \multicolumn{1}{c|}{Specificity} & \multicolumn{1}{c|}{Accuracy} & \multicolumn{1}{c|}{F1 score} \\ \hline
        \textit{Thermal baseline}    & 0.8397                           & 0.6663                            & 76.2\%                        & 0.7592

        \\ \hline
\textit{EDA baseline} & 0.8559                           & 0.8575                          & 85.71\%                        & 0.8568     
\\ \hline
\textit{\emph{ThermaStrain}} & 0.8911                          & 0.7598                           & 83.17\%                        & 0.8293 
		
\\ \hline
\textit{Multi-modality baseline} & 0.9202                           & 0.8541                           & \textbf{89.13\%}                        & \textbf{0.8897}	  

\\ \hline
\end{tabular}

}

\label{table:baselines}
\end{table}

\subsection{Generalizability Evaluation}\label{Eval:generalizability}
A generalizable stress-sensing solution needs to be robust and perform similarly in previously unseen stress conditions, meaning in the stressful situations that were not present in the training dataset. 

To evaluate the generalizability of the \emph{ThermaStrain} approach in unseen stress conditions, we performed a stress-task disjoint evaluation over the 5-folds (discussed in Section \ref{evaluation}). As mentioned in Section \ref{procedure}, each participant performed four distinct stress-inducing tasks in each data collection session, simulating four stress conditions. In this evaluation, during training, only the `Passive stress induce video’ and `TST’ task data were used as stress samples, while during evaluation, only the `SSST’ and `recalling bad memories’ task data were used. 

Like the previous section, \emph{ThermaStrain}’s performance is compared with EDA and thermal video-based uni-modal and multi-modal approaches; the results are shown in table \ref{table:unknown_task}. 

The uni-modal thermal baseline's performance decreased by 3\% in accuracy and F1 scores compared to the baseline evaluation in table \ref{table:baselines}. In contrast, \emph{ThermaStrain}
model's performance decreased by 2\%, but it is still significantly better than the thermal baseline.

The EDA and multi-modality baselines achieved even higher accuracy. These evaluations indicate that during the `SSST’ and `recalling bad memories’ tasks, EDA is an even stronger indicator of stress than thermal modality.

It is important to note that, according to our study protocol discussed in Section \ref{procedure}, the stress responses for each task are not completely disjoint. This is due to the potential partial influence of residual physiological stress responses from stress-inducing tasks on one another. Nonetheless, the evaluation in this section demonstrates the relatively greater generalizability of the \emph{ThermaStrain} approach compared to the unimodal thermal baseline, thereby showcasing its improved utility.


\begin{table}[h]
\caption{Evaluate on unknown (during training) stress tasks}
\resizebox{0.6\textwidth}{!}{
\begin{tabular}{|c|l|l|l|l|}
\hline
Model     & \multicolumn{1}{c|}{Sensitivity} & \multicolumn{1}{c|}{Specificity} & \multicolumn{1}{c|}{Accuracy} & \multicolumn{1}{c|}{F1 score} \\ \hline
        \textit{Thermal baseline}    & 0.8373                           & 0.6666                          & 73.24\%                        & 0.7289

        \\ \hline
\textit{EDA baseline} & 0.9107                           & 0.8815                        & \textbf{89.34\%}                        & \textbf{0.8939}     
\\ \hline
\textit{\emph{ThermaStrain}} & 0.9028                          & 0.7595                           & 81.36\%                        & 0.8107

\\ \hline
\textit{Multi-modality baseline} & 0.9586                           & 0.8517                           & \textbf{89.34\%}                        & 0.8919

\\ \hline
\end{tabular}

}
\label{table:unknown_task}

\end{table}

\subsection{Comparsion with Other Co-Teaching Baselines}\label{eval-co-teachin}
This section compares \emph{ThermaStrain} approach with the existing co-teaching baselines. Following the state-of-the-art literature \cite{fortin2019multimodal,kumar2021stressnet,li2022valhalla}, we implemented three co-teaching approaches as baselines. Since none of them have leveraged thermal and EDA modalities, we followed their model structure and design but made necessary changes to fit our dataset.

The results are shown in table \ref{table:co-teaching}. As discussed in Section \ref{related-works}, the multi-task learning approach \cite{fortin2019multimodal} has multiple classifiers that fit different missing modality scenarios. In StressNet \cite{kumar2021stressnet}, thermal data was used to predict the EDA modality and then used the predicted EDA to predict stress. In the vision hallucination model \cite{li2022valhalla}, there is a hallucination network that mimics the EDA embedding. During inference, pseudo-embedding replaces the EDA embedding and concatenates with the thermal-independent embedding. 
\par 
As shown in table \ref{table:co-teaching}, all models achieved lower performance than \emph{ThermaStrain}. The reason is that the multi-task learning approach doesn't have similarity loss and consistent loss that force each modality to learn joint patterns. The StressNet only takes the reconstructed EDA modality to predict stress, which loses some independent information about the thermal modality. \textit{The inferior performance of StressNet further emphasizes the impact of presented co-teaching approach rather than just simulating physiological parameters from thermal sensing and using the simulated physiological features to assess stress.}
Finally, the vision hallucination model is too complex, leading to overfitting in our limited dataset.

\begin{table}[h]
\caption{Comparsion with other co-teaching approaches}
\resizebox{0.6\textwidth}{!}{

\begin{tabular}{|c|l|l|l|l|}
\hline
Model     & \multicolumn{1}{c|}{Sensitivity} & \multicolumn{1}{c|}{Specificity} & \multicolumn{1}{c|}{Accuracy} & \multicolumn{1}{c|}{F1 score} \\ \hline

\textit{\emph{ThermaStrain} }  & 0.8911                          & 0.7598                           & \textbf{83.17\%}                       & \textbf{0.8293}   
\\ \hline

\textit{multi-task learning}  & 0.8145                          & 0.7595                          & 79.05\%                        & 0.7891   

\\ \hline

\textit{StressNet}  & 0.8443                          & 0.6930                          & 77.48\%                        & 0.7739   

\\ \hline
\textit{Visual hallucination}  & 0.8268                          & 0.7317                          & 78.5\%                        & 0.7826   

\\ \hline
\end{tabular}

}
\label{table:co-teaching}
\end{table}
\subsection{Ablation Study}

Table \ref{table:ablation} presents the results of the ablation study, where various components of \emph{ThermaStrain} are modified while keeping the rest of the network constant. We discuss the evaluation results and observations below:

\textbf{\textit{Using Central Moment Discrepancy (CMD) as $l_{s}$}}: Both Mean Square Error (MSE) and CMD loss are popular distance metrics that measure the discrepancy between the distribution of two representations. We evaluate them thoroughly. As shown in table \ref{table:ablation}, the MSE achieves better performance. Therefore, we choose MSE as our $l_{s}$.

\textbf{\textit{Not using $l_{c}$}}: 
As there is a similarity loss $l_{s}$ that forces the two embeddings $z_{T}$ and $z_{E}$ to be similar to each other, it may be questioned whether we need the consistency loss $l_{c}$. Hence, we evaluated not using the $l_{c}$ in the \emph{ThermaStrain} implementation. The results show a drastic performance drop when we remove the consistency loss, demonstrating the importance of $l_{c}$ in encouraging consistency between inferences $y_{t}$ and $y_{e}$, which leads to better performance while EDA modality, consequently, $y_{e}$ is unavailable during evaluation.

\textbf{\textit{Use Vision Transformer to replace ResNet}}: The Vision Transformer \cite{dosovitskiy2020image} is a transformer-based image feature extractor and outperforms ResNet-based models in RGB image-based literature. We attempt to use the Vision Transformer as our feature extractor for each frame. However, accuracy and F1 scores decrease by approximately 5\%. Considering the limited size of our dataset, the complex transformer-based feature extractor may lead to overfitting.

\begin{table}[h]
\caption{Ablation study of our proposed model}
\resizebox{0.6\textwidth}{!}{

\begin{tabular}{|c|l|l|l|l|}
\hline
Model     & \multicolumn{1}{c|}{Sensitivity} & \multicolumn{1}{c|}{Specificity} & \multicolumn{1}{c|}{Accuracy} & \multicolumn{1}{c|}{F1 score} \\ \hline
\textit{ThermaStrain}  & 0.8911                          & 0.7598                           & \textbf{83.17\%}                        & \textbf{0.8293}   
\\ \hline

\textit{ $l_{s}$ use Central Moment Discrepancy}  & 0.8208                          & 0.7555                          & 79.17\%                        & 0.7909

\\ \hline
\textit{Not use $l_{c}$}  & 0.7871                          & 0.7608                          & 77.64\%                        & 0.7753   

\\ \hline

\textit{Use Vision Transformer to replace ResNet}  & 0.7646                          & 0.7581                          & 76.13\%                        & 0.7603  

\\ \hline
\end{tabular}

}
\label{table:ablation}
\end{table}
\subsection{Distance Evaluation}\label{distance-eval}
This section breaks down the performance of the \emph{ThermaStrain} based on the distance between the participant and the thermal camera. It is observed that the model performs better for closer distances. Specifically, the model achieves the highest performance in the 5-7 ft range with an accuracy of 91.36\% and an F1 score of 0.9126. However, the performance drops for larger distances, especially in the 9-11 ft range, where the accuracy is 72.58\%, and the F1 score is 0.6902. The performance deterioration for larger distances can be attributed to each pixel's reduced quality of thermal energy perception and the reduction in the quantity of thermal pixels covering the important body regions at the longer ranges.

\begin{table}[h]
\caption{Distance evaluation}
\resizebox{0.6\textwidth}{!}{

\begin{tabular}{|c|l|l|l|l|l|}
\hline
Model    & Number of sessions & \multicolumn{1}{c|}{Sensitivity} & \multicolumn{1}{c|}{Specificity} & \multicolumn{1}{c|}{Accuracy} & \multicolumn{1}{c|}{F1 score} \\ \hline

\textit{5-7 ft} &7 & 0.9304                          & 0.8904                          & \textbf{91.36}\%                        & \textbf{0.9126}

\\ \hline

\textit{7-9 ft} &8 & 0.8241                          & 0.9016                          & 87.81\%                        & 0.8634 
\\ \hline
\textit{9-11 ft} &7 & 0.8826                          & 0.5187                        & 72.58\%                       & 0.6902 

\\ \hline

\end{tabular}

}

\label{table:distance}
\end{table}

\subsection{Benchmarking for Real Time Execution}\label{eval-real-time}
To evaluate our approach’s real-time executability, we performed a run-time evaluation on an Nvidia Jetson Nano. The program reads 5-second data at a time, analyzes and infers the class results, and waits for the next 5-second data. The binary classifiers take 0.324s to process one 5-second video window on Jetson Nano. The average CPU usage is 19.38\%, and the average GPU usage is 9.64\%. The average RAM usage is 1.85 GB. According to this evaluation, our presented approach is capable of real-time execution on a Jetson Nano module. Note that the times reported are when only the stress assessment program is running. Running additional programs will affect/change these times.

\section{Discussion on \textit{ThermaStrain}'s Efficacy}\label{Discussion-sec}
This section further investigates the \emph{ThermaStrain}'s capability in identifying effective thermal information extraction (Section \ref{shapinterpretation}) and how co-teaching enables better stress sensing solution development through loss landscape analysis (Section \ref{discussion:Loss-land}). 

\subsection{Inference Interpretation Discussion: Visualizing \emph{ThermaStrain}'s Effective Information Extraction}\label{shapinterpretation}
Many studies have highlighted that stress-induced changes in temperature are primarily concentrated in the forehead, eyehole, and cheekbone regions \cite{perpetuini2021regions,nguyen2018towards}. Therefore, these regions are more critical for improving the accuracy of stress detection models. To evaluate the developed models' capability in capturing information from the critical body regions, we utilized the KernalSHAP model-agnostic interpretation framework \cite{lundberg2017unified}. The SHAP values \cite{shapley1953stochastic} indicate the contribution of each input attribute in driving the model inference closer or farther away from the true/correct inference. We divided the human body region into an $8\times8$ grid to compute the Shapley value for each grid.
\begin{figure}[]
\centering
\includegraphics[width= 0.8\linewidth]{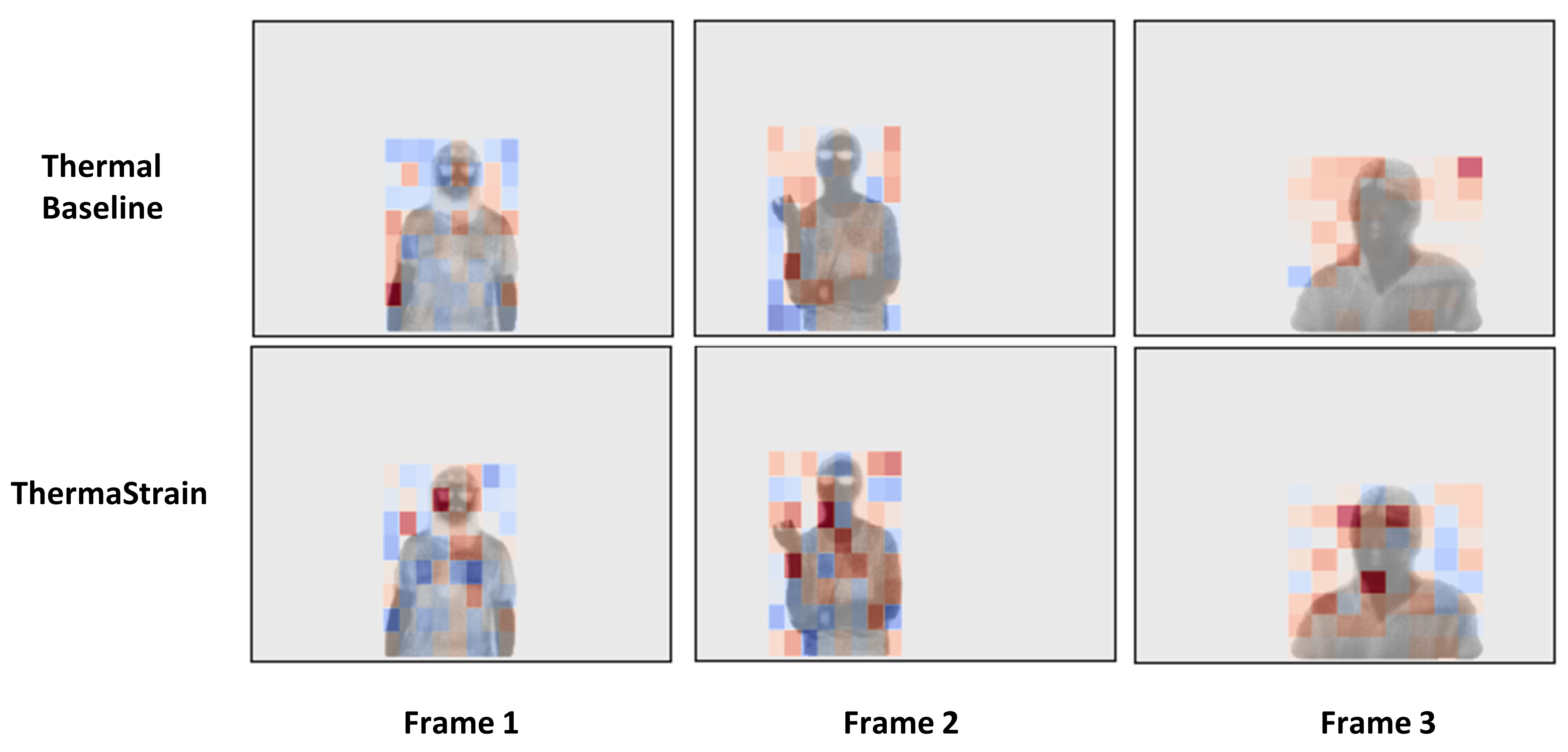}
\vskip -1ex
\caption{The SHAP interpretation to Visualize Models' Information Extraction Efficacy.}
\label{fig:SHAP}
\end{figure}

Figure \ref{fig:SHAP} shows generated explanations of the baseline and \emph{ThermaStrain} classifiers' inference for three thermal frames belonging to three different individuals. The baseline model's Shapley value appears more normally distributed, indicating that it had to select information from the entire frame. In contrast, \emph{ThermaStrain} effectively learns the critical regions to focus on, illustrated by having darker colors in the face, neck, and hand regions that are established as crucial stress-indicative body areas according to literature \cite{perpetuini2021regions,nguyen2018towards}.

This visualization demonstrates that \textit{the EDA modality effectively guides the \emph{ThermaStrain} model to extract better thermal embeddings.} This results in perceiving high-stress-indicative and physiologically relevant information by focusing on crucial visible body regions.

\subsection{Loss Landscape Visualization: Understanding How Co-teaching Facilitates Effective Model}\label{discussion:Loss-land}

Our evaluation in Section \ref{EVal-comp-uni-multi} shows that the presented Co-teaching approach outperforms the uni-modal thermal sensing baseline approach. This section investigates how the co-teaching approach enables better performance. 

Li et al. \cite{li2018visualizing} showed that visualizing the loss landscape for neural network models provides a richer understanding of how the different approaches’ design choices influence the optimization of the loss function. We used the \textit{loss-landscapes} library \cite{marcellodebernardi} to generate the 3D loss landscape plots of \emph{ThermaStrain} and the uni-modal thermal baseline model, as shown in Figure \ref{fig:3dlosslandscape}. A detailed discussion on the plot generation is in Appendix \ref{loss-landscape}.

\begin{figure}[h]
  \centering
  \includegraphics[width=0.7\textwidth]{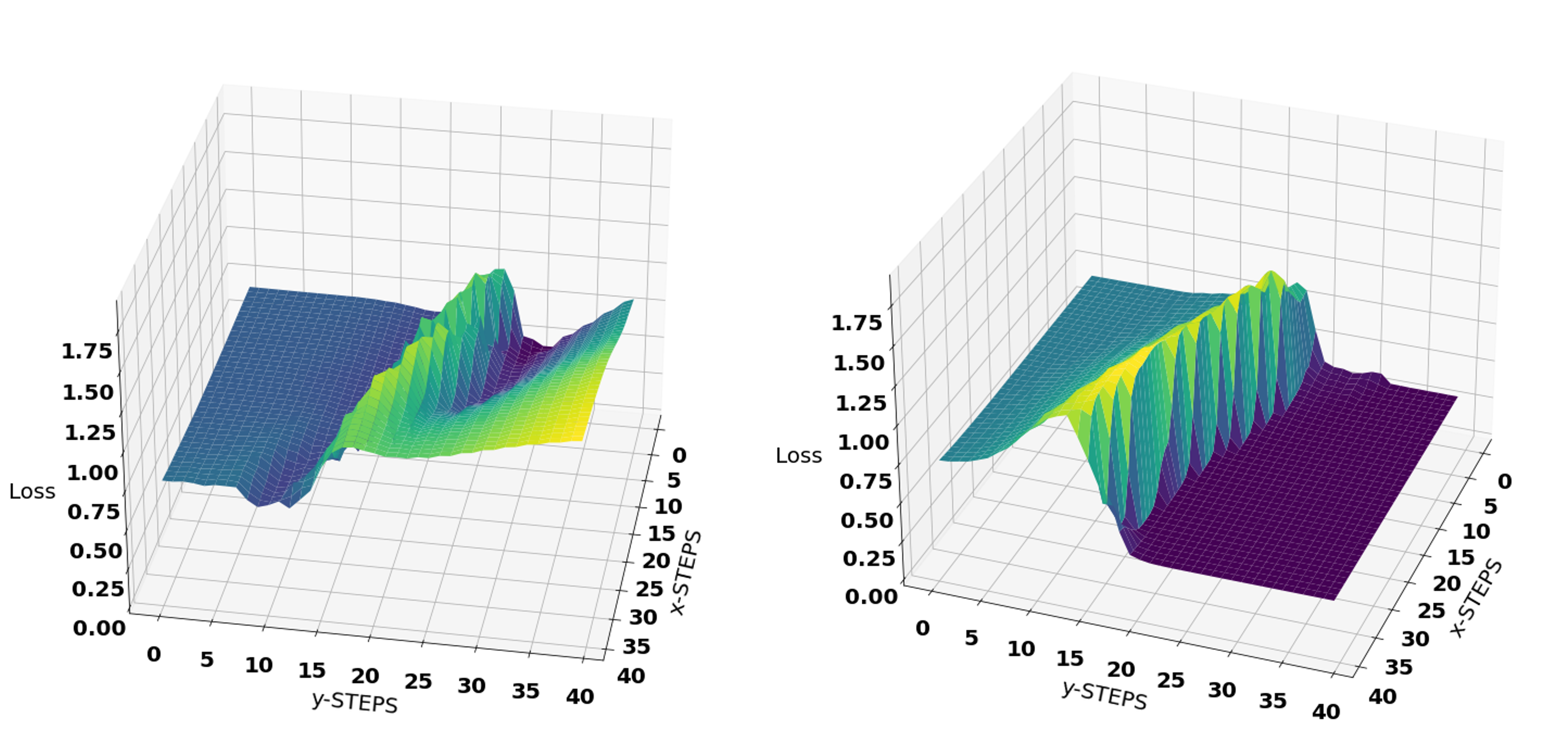}
 \caption{3D Loss Landscape Plot Comparison (Baseline is in the left vs the \emph{ThermaStrain} is in the right)}
\label{fig:3dlosslandscape}
\vskip -2ex
\end{figure}

Several prior works \cite{keskar2016large,chaudhari2019entropy,li2018visualizing,hochreiter1997flat} investigating the loss landscapes to understand the ability of neural networks to optimize better (i.e., obtaining better performance) emphasized that the `flatness’ of the loss landscape is a property of interest. Hochreiter et al. \cite{hochreiter1997flat} define the `flatness' of the loss landscape as the region around the minima where the loss remains low. Literature \cite{li2018visualizing,kawaguchi2017generalization,hochreiter1997flat} suggests that the model with flatter loss surface optimizes better, i.e., effectively identifies the minima in the loss space, hence achieves better performance.

As shown in Figure \ref{fig:3dlosslandscape}, the loss landscape is significantly `flatter’ in the blue regions (near minima) for the \emph{ThermaStrain} than the baseline approach. Meaning \textit{co-teaching enables easier propagation through the loss landscape and identification of minima, resulting in a more effective stress sensing performance} by the \emph{ThermaStrain} model.



\section{Discussion on Applications and Deployment of the \emph{ThermaStrain}}\label{Discussion-deployment}
This section discusses the scenarios where \emph{ThermaStrain} will be more effective than wearable-based stress sensing solutions (Discussed in Section \ref{application-scenarios}). Additionally, it expounds upon the deployment challenges associated with deploying \emph{ThermaStrain} in real-world settings (Discussed in Section \ref{discussion-multi-person}). This discussion encompasses scenarios involving multiple individuals, constrained visualization, and deployment conditions, camera angles, distances, backgrounds, participant's postures, etc., that were not encountered during the training phase. Finally, the Section \ref{Ethical-discussion} discusses the ethical and practical considerations for real-world deployment of \emph{ThermaStrain}.


\subsection{Application Scenarios of \emph{ThermaStrain}}\label{application-scenarios}

As shown in Section \ref{EVal-comp-uni-multi}, while \emph{ThermaStrain} outperforms the thermal-video-based state-of-the-art solutions, its efficacy is relatively lower than the EDA-based stress sensing solutions. Nevertheless, contactless thermal video-based stress detection presents distinct applications and deployment possibilities that are challenging to achieve through EDA or other wearable-based solutions. Two such example scenarios are discussed below. 

\subsubsection{Application in Smart Health} 
While wearable sensors find extensive application in healthcare monitoring scenarios \cite{bergmann2012wearable}, contactless sensors present distinct advantages over wearables in specific situations.
\par 
For instance, the non-intrusive characteristics of contactless sensors render them especially suitable for assessing stress in vulnerable populations, such as elderly individuals who may have impaired memory function, as observed in cases of Dementia \cite{fook2007automated, konig2015validation, salekin2017dave}. Wearable sensors necessitate patients to wear or carry battery-powered devices, which can lead to discomfort and inconvenience due to frequent recharging of batteries \cite{wright2022reducing}. In the case of Dementia, patients might forget to wear or charge these devices. Additionally, wearable sensors can pose risks to the safety of elderly individuals, e.g., a recent incident involved an elderly woman strangled by her fall detection pendant \footnote{\url{https://www.huffpost.com/entry/medical-necklace-strangles-woman_n_56d75817e4b0871f60edbb47}}. 
\par 
Consequently, contactless sensing solutions present an effective continuous stress assessment alternative. While RGB video-based solutions are already making their way into commercial use for monitoring elderly health \cite{Safelyyou}, privacy concerns limit their widespread adaptation \cite{guardian2021}. 
Thermal imaging offers relatively higher privacy protection compared to RGB-based alternatives. For instance, unlike RGB imaging, \textit{ThermaStrain} ensures contextual or environmental information protection by not capturing non-human body content \cite{chiu2023privacy,griffiths2018privacy}. Hence, \textit{ThermaStrain} holds the potential as a highly suitable option for such vulnerable populations. This approach requires no active participation from patients, such as device recharging or wearing, and enhances safety due to its passive contactless operation.
\par 

\subsubsection{Application in Smart Work-Place} Stress monitoring in smart workplaces, whether in manufacturing contexts \cite{leone2020multi} or smart offices \cite{ alberdi2018using}, is crucial for safeguarding employee well-being, optimizing productivity, enhancing workplace safety, reducing staff turnover, and fostering a positive work environment.
\par 
Presently, prevalent techniques often involve the measurement of EDA through disc electrodes \cite{boucsein2012electrodermal} or through the utilization of wearable devices like the Empatica E4 or the Apple Watch \cite{apple--support}. Positioning disc electrodes at the most sensitive bodily sites, such as the feet or fingers \cite{van2012emotional}, can impose significant inconvenience on users or may even be impractical, such as in office settings where hands are engaged in typing or other activities. Additionally, since workplaces involve multiple individuals, using wearables can incur substantial costs. Furthermore, some individuals find it uncomfortable to wear such wearable devices for extended durations consistently \cite{jeong2017smartwatch}. 
\par 
Hence, contactless but relatively privacy preserving \textit{ThermaStrain} can be a suitable alternative capable of simultaneously assessing stress in multiple individuals at a relatively affordable cost. Notably, stress assessments aimed at quantifying employee well-being within smart workplaces \cite{mantello2023emotional,lee2019clara} often occur in an aggregated manner rather than being conducted in real-time, such as on a per-minute basis. This characteristic ensures that the efficacy of the use case remains unaffected, even in scenarios where employees might be momentarily obscured due to occlusion. Significantly, thermal camera-based solutions are already being incorporated into workplaces for tasks like employee health screening \cite{CDW} and security measures \cite{Protech}. This trend paves the path for smoother integration of thermal-video-based stress assessment solutions like \textit{ThermaStrain} into smart workplaces.

\subsection{Deployment Procedure of \emph{ThermaStrain} in Real-World Scenarios}\label{discussion-multi-person}
Three challenges require attention for the successful practical implementation of \emph{ThermaStrain}.
\begin{enumerate}
    \item Achieving effective body segmentation and accurately identifying segments corresponding to the target users undergoing stress assessment in scenarios involving multiple individuals is crucial (Discusses in Section \ref{real-deploment-segmentation}).

    \item Due to real-world occlusion scenarios in multi-person settings, only partial body segments might be accessible. In such cases, \emph{ThermaStrain} must demonstrate superior performance compared to the thermal and co-teaching baselines outlined in Sections \ref{EVal-comp-uni-multi} and \ref{eval-co-teachin} when dealing with partial body segment information (Discussed in Section \ref{real-deploment-partial-body}).

    \item \emph{ThermaStrain} needs to maintain its stress-assessment performance while evaluating on distances, angles, indoor settings, and scenarios that are not present during its training (Discussed in Section \ref{deployment-study-real}).

\end{enumerate}
Evaluation and discussion on these challenges are below:


\subsubsection{Segmentation and Person-Identification in Multi-Person Scenarios}\label{real-deploment-segmentation}
We developed an integrated framework combining pre-trained \textit{human body segmentation} and \textit{re-identification} models to achieve simultaneous human body segmentation and identification. Initially, thermal frames are inputted into the \textit{segmentation model} to generate pixel-wise human segmentation, producing disjoint object segments. Subsequently, these object segments are fed into the \textit{human re-identification} model to assess if it belongs to one of the target individuals whose stress is being assessed.

\par 
As discussed in Section \ref{Data-preprocessing}, for \textit{human body segmentation}, we use the pre-trained DetectorRS \cite{detectors} model on Microsoft COCO dataset \cite{lin2014microsoft}, that identifies the human body regions in the thermal frame and masks all other parts of the background.
We use pre-trained Omni-Scale Network (OSNet) \cite{ zhou2019omni} for \textit{human re-identification}. The OSNet comprises a residual block composed of multiple convolutional feature streams, each detecting features at a certain scale. This enables OSNet to learn omni-scale feature learning.  We take the pre-trained checkpoint provided by the author \cite{zhou2019torchreid}.

\par 
We conducted a \textit{small evaluation} for a five-person indoor stress assessment scenario to assess the framework's effectiveness. Initially, we gathered a few seconds of data from each of the five participants separately for fine-tuning the human re-identification model. Subsequently, we conducted three sessions where varying subsets of the five individuals appeared simultaneously in front of the camera. Two sessions involved three different participants, while the remaining session had four participants. For the evaluation of body segmentation and person identification, the data were annotated by two graduate students, achieving an inter-rater reliability rate of over $94.91$\% ($0.94+$), specifically measured using Cohen’s kappa statistic \cite{mchugh2012interrater}.

\begin{figure}[h]
\centering
\includegraphics[width=0.9\linewidth]{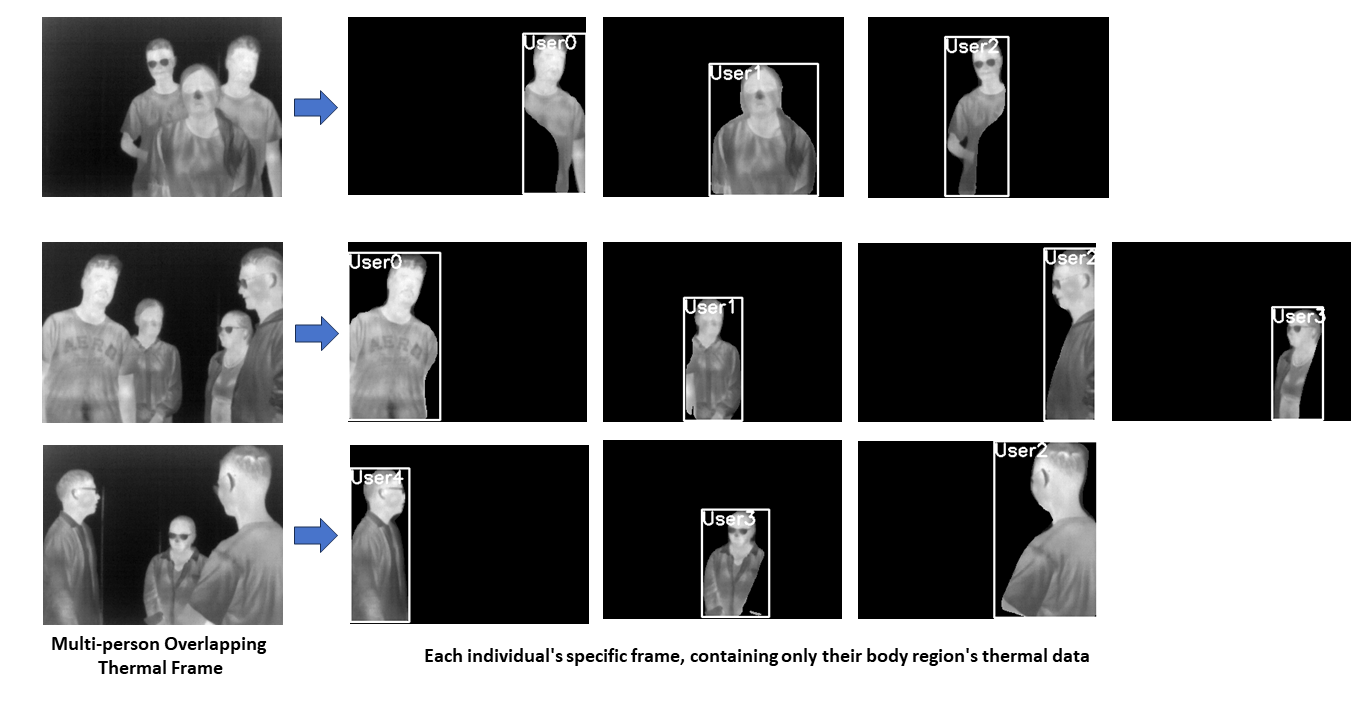}
\vskip -1ex
\caption{Integrated multi-individual body segment and identification detection}
\label{fig:multi-object-future}
\end{figure}

Frames featuring multiple individuals were processed by the DetectorRS human body segmentation model to extract object segments. We fine-tuned the Omni-Scale Network using the initial data collected individually, then evaluated the model's performance on the three multi-person sessions. The fine-tuned Omni-Scale Network achieved an impressive $97.29$\% accuracy in re-identifying participants during concurrent appearances. 

Figure \ref{fig:multi-object-future} illustrates instances of the integrated framework in action. When multiple individuals are present, the framework distinguishes and separates their respective body segments, detects corresponding identifications, and generates distinct frames for each identified body region along with the person's identification label. Each individual's specific frame, containing only their body region's thermal data, is inputted into the \emph{ThermaStrain} model for stress assessment.

Given that in multi-person scenarios, only partial body segments might be accessible, the subsequent section delves into the discussion about the effectiveness of the \emph{ThermaStrain} model, as well as different thermal and co-teaching baselines, when dealing with partially occluded body segments.

\subsubsection{Stress Assessment when Human Body Segment is Partially Masked}\label{real-deploment-partial-body}
Occlusion presents a challenge in stress detection. This section discusses a pseudo-partially-masked-body-segment data augmentation to address this challenge. In this approach, during the training process, a portion of the participant's body is randomly masked with zeros. To facilitate this, a pre-trained multi-person human body part segmentation model named CDCL (Cross-Domain Complementary Learning) \cite{lin2020cross} is adapted, utilizing which we segment the human body into eight distinct sections. The CDCL recognizes pixel-wise human body part segmentations, such as the head, torso, upper arms, and forearms, from thermal frames. Subsequently, these segmentations were further refined into left and right divisions for each body part, resulting in eight subdivisions: the left face, the right face, the left torso, the right torso, the left upper arm, the right upper arm, the left forearm, and the right forearm.
\par
\begin{figure}[h]
\centering
\includegraphics[width=1\linewidth]{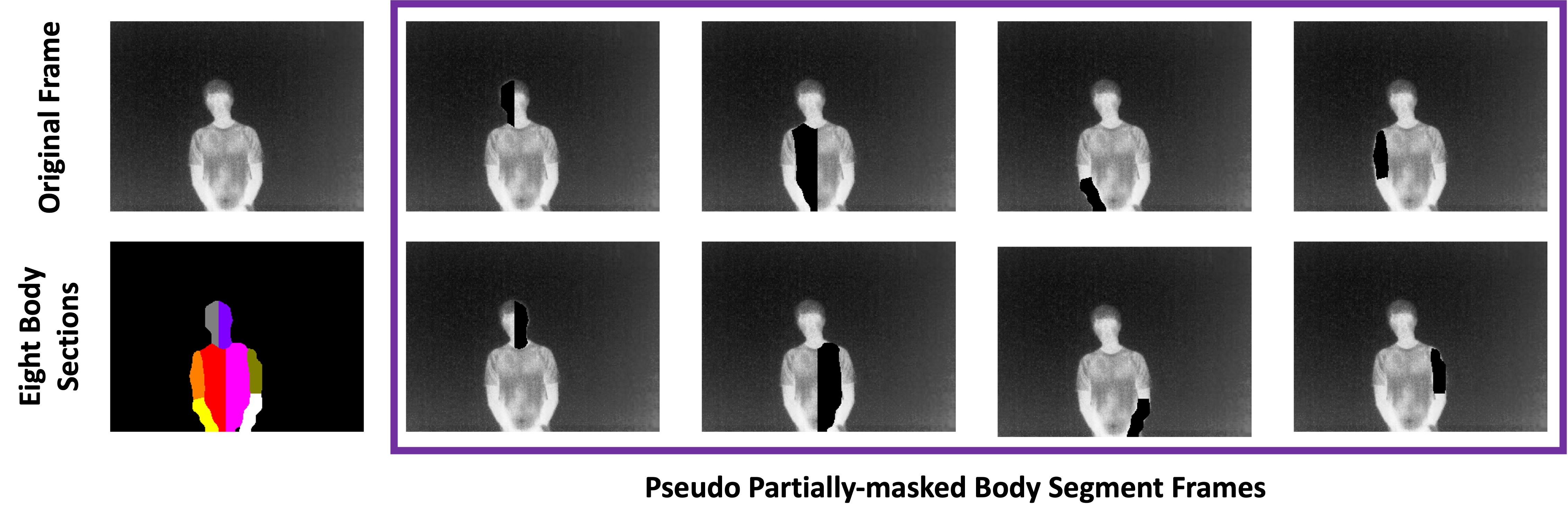}
\vskip -1ex
\caption{Pseudo one-body-part masked human body segment frames.}
\label{fig:partially-masked-future}
\end{figure}

 
We trained the \emph{ThermaStrain} model alongside the baselines outlined in Sections \ref{EVal-comp-uni-multi} and \ref{eval-co-teachin}, where there was a 23\% chance of masking one out of the eight parts, a 23\% chance of masking two parts, a 24\% chance of masking three parts, and a 30\% chance of keeping all parts unmasked. An example of pseudo one-body-part masked human-body segment frames, alongside the eight distinct body sections identified by CDCL from the original frame, is shown in Figure \ref{fig:partially-masked-future}.

The evaluation outcomes for the pseudo-augmented trained \emph{ThermaStrain} and the baselines are presented in Table \ref{table:partial-body-result}. This evaluation encompasses scenarios where different body segments are masked/occluded. Comparing these results with those in Table \ref{table:co-teaching}, it becomes apparent that the training process involving significant partial noise causes a reduction of approximately 2\% in \emph{ThermaStrain}'s F1 score for `without any masking body segments.' Nevertheless, the model sustains a satisfactory performance level across scenarios involving masking different body parts, consistently outperforming all the baselines. Notably, according to Table \ref{table:partial-body-result}, masking the head and body exerts a more pronounced impact on performance than other parts. Additionally, masking any two body sections simultaneously reduces \emph{ThermaStrain}'s accuracy and F1 score to on avg. $72$\% and $0.69$ F1 scores, outperforming the thermal baseline (on avg. $68$\% accuracy and $0.65$ F1 score) and best co-teaching baseline Hallucination (on avg. $69.7$\% accuracy and $0.66$ F1 score).
\par 
More robust strategies are available to tackle the challenge of partial body masking due to occlusion. These methods include reconstructing the missing segments \cite{wang2022feature} and incorporating pyramid perception \cite{he2019foreground}, which can potentially enhance the partial body stress sensing performance. However, considering this paper's primary focus is on co-teaching, the evaluation in this section was confined to pseudo-data augmentation. This evaluation demonstrates the viability of \emph{ThermaStrain} in scenarios involving occlusion of partial body segments and its consistent outperformance compared to the baselines.
\par
Finally, this section's evaluations and discussions demonstrate the \emph{ThermaStrain}'s robustness against diverse occlusion scenarios compared to the baselines, thereby showcasing its improved utility in real-world settings.
\begin{table}[h]
\caption{Evaluation results of stress assessment on different masked sections of the Human Body.}
\resizebox{1\textwidth}{!}{
\begin{tabular}{|c|cc|cc|cc|cc|cc|}

\hline
                        & \multicolumn{2}{c|}{ThermaStrain}        & \multicolumn{2}{c|}{\begin{tabular}[c]{@{}c@{}}Thermal\\ baseline\end{tabular}} & \multicolumn{2}{c|}{Multi-task}          & \multicolumn{2}{c|}{StressNet}           & \multicolumn{2}{c|}{Hallucination}                 \\ \hline
Metrics                 & \multicolumn{1}{c|}{Accuracy} & F1 Score & \multicolumn{1}{c|}{Accuracy}                     & F1 Score                    & \multicolumn{1}{c|}{Accuracy} & F1 Score & \multicolumn{1}{c|}{Accuracy} & F1 Score & \multicolumn{1}{c|}{Accuracy}      & F1 Score      \\ \hline
Without Masking         & \multicolumn{1}{c|}{\textbf{0.81}}     & \textbf{0.80}     & \multicolumn{1}{c|}{0.75}                         & 0.74                        & \multicolumn{1}{c|}{0.74}     & 0.73     & \multicolumn{1}{c|}{0.75}     & 0.74     & \multicolumn{1}{c|}{0.77} & 0.75 \\ \hline
Masking Left Head       & \multicolumn{1}{c|}{\textbf{0.79}}     & \textbf{0.78}     & \multicolumn{1}{c|}{0.75}                         & 0.74                        & \multicolumn{1}{c|}{0.73}     & 0.72     & \multicolumn{1}{c|}{0.72}     & 0.72     & \multicolumn{1}{c|}{0.76}          & 0.75          \\ \hline
Masking Left Body       & \multicolumn{1}{c|}{0.73}     & \textbf{0.73}     & \multicolumn{1}{c|}{0.73}                         & 0.71                        & \multicolumn{1}{c|}{0.70}     & 0.68     & \multicolumn{1}{c|}{0.71}     & 0.71     & \multicolumn{1}{c|}{\textbf{0.74}}          & \textbf{0.73}          \\ \hline
Masking Left Upper Arm  & \multicolumn{1}{c|}{\textbf{0.79}}     & \textbf{0.78}     & \multicolumn{1}{c|}{0.75}                         & 0.74                        & \multicolumn{1}{c|}{0.74}     & 0.73     & \multicolumn{1}{c|}{0.75}     & 0.74     & \multicolumn{1}{c|}{0.76}          & 0.75          \\ \hline
Masking Left Lower Arm  & \multicolumn{1}{c|}{\textbf{0.80}}     & \textbf{0.79}     & \multicolumn{1}{c|}{0.75}                         & 0.74                        & \multicolumn{1}{c|}{0.74}     & 0.73     & \multicolumn{1}{c|}{0.74}     & 0.73     & \multicolumn{1}{c|}{0.77}          & 0.75          \\ \hline
Masking Right Head      & \multicolumn{1}{c|}{0.75}     & 0.73     & \multicolumn{1}{c|}{0.73}                         & 0.72                        & \multicolumn{1}{c|}{0.73}     & 0.72     & \multicolumn{1}{c|}{0.72}     & 0.72     & \multicolumn{1}{c|}{\textbf{0.77}}          & \textbf{0.76}          \\ \hline
Masking Right Body      & \multicolumn{1}{c|}{0.72}     & 0.71     & \multicolumn{1}{c|}{0.68}                         & 0.67                        & \multicolumn{1}{c|}{0.69}     & 0.68     & \multicolumn{1}{c|}{0.70}     & 0.70     & \multicolumn{1}{c|}{\textbf{0.73}}          & \textbf{0.72}          \\ \hline
Masking Right Upper Arm & \multicolumn{1}{c|}{\textbf{0.78}}     & \textbf{0.77}     & \multicolumn{1}{c|}{0.71}                         & 0.71                        & \multicolumn{1}{c|}{0.73}     & 0.72     & \multicolumn{1}{c|}{0.73}     & 0.72     & \multicolumn{1}{c|}{0.76}          & 0.75          \\ \hline
Masking Right Lower Arm & \multicolumn{1}{c|}{\textbf{0.79}}     & \textbf{0.79}     & \multicolumn{1}{c|}{0.75}                         & 0.74                        & \multicolumn{1}{c|}{0.74}     & 0.73     & \multicolumn{1}{c|}{0.74}     & 0.74     & \multicolumn{1}{c|}{0.77}          & 0.76          \\ \hline
\end{tabular}

}
\label{table:partial-body-result}
\end{table}

\subsubsection{Real Deployment Evaluation of \emph{ThermaStrain} on Unseen Distance, Camera-Angle, Indoor-Setting and Scenario}\label{deployment-study-real}
\begin{figure}
     \centering
     \begin{subfigure}[b]{0.39\textwidth}
         \centering
         \includegraphics[width=\textwidth]{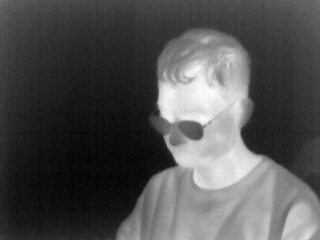}
         \caption{Thermal Frame Example}
         \label{fig:thermalworkplace}
     \end{subfigure}
     \hfill
     \begin{subfigure}[b]{0.52\textwidth}
         \centering
         \includegraphics[width=\textwidth]{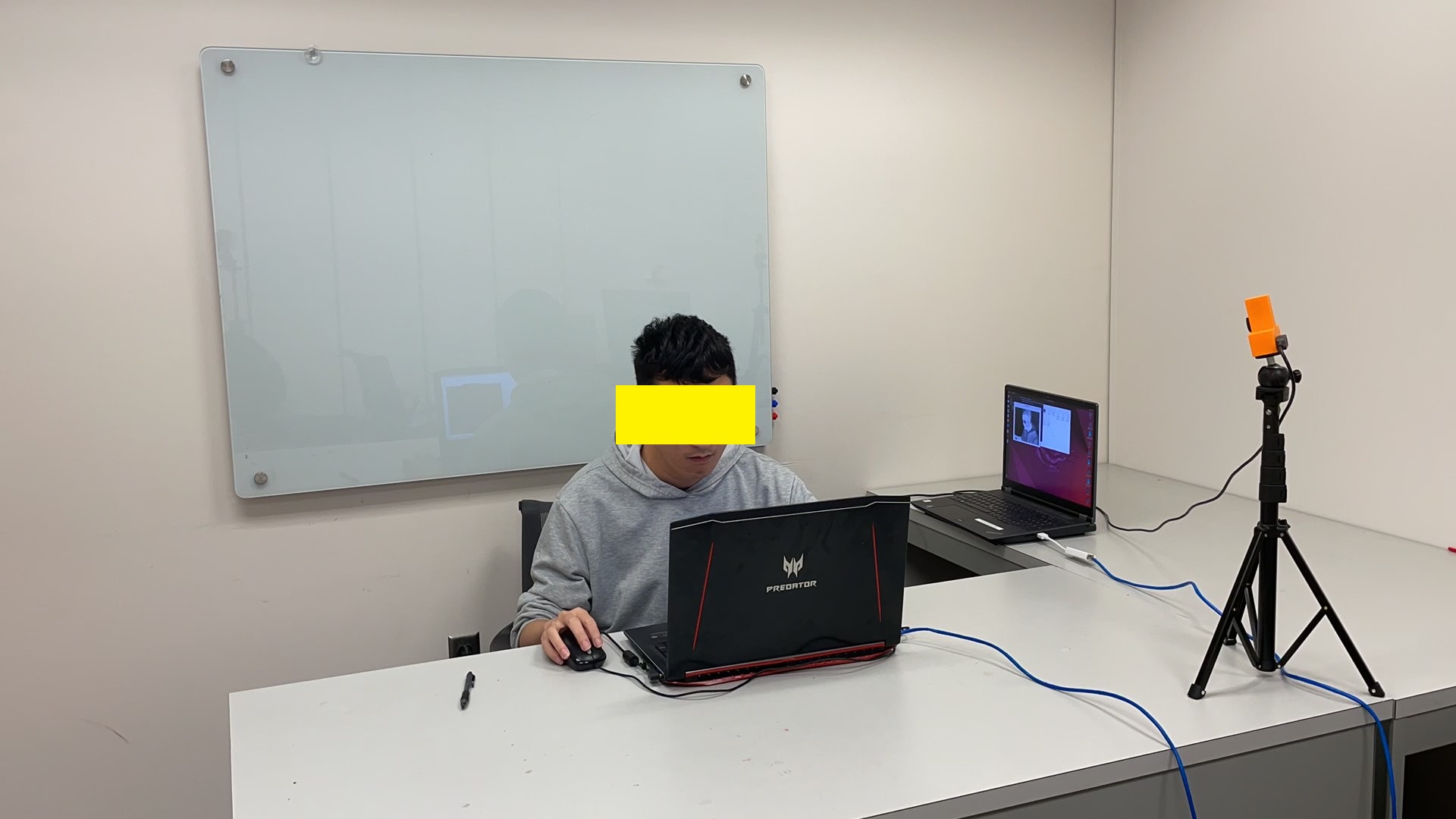}
         \caption{Work-Place setup}
         \label{fig:rgbworkplace}
     \end{subfigure}
\caption{Work-Place Application Deployment Experiment}
\label{fig:work-place}
\end{figure}

We deployed and evaluated the performance of the trained \emph{ThermaStrain} model from Section \ref{real-deploment-partial-body} on a one-person workplace scenario in an indoor lab shown in Figure \ref{fig:rgbworkplace}. The participant in this study was a male who did not appear in the training set or validation set of the \emph{ThermaStrain} model. In contrast to the data collection setup described in Section \ref{Data-collection}, the thermal camera was positioned at a forty-five-degree angle to the participant's left front and maintained a distance of about three feet from the participant. Also, the background environment setting (e.g., a whiteboard in the background) was different. The experiment spanned two consecutive days with a single individual: on the first day, the participant engaged in a LeetCode contest, simulating stress conditions, while on the following day, the participant relaxed by watching random YouTube videos of his choice. Each session lasted for approximately an hour. Data from the participant were captured using both the thermal camera and the Empatica E4 device.
\par

Following the procedure outlined in Section \ref{Stress=validation-preprocessing}, we calculated the LF/HF ratio of the participant's data collected with Empatica E4. The average LF/HF ratio for the first day was $1.71$, which exceeded the ratio of $1.57$ observed on the second day. This indicates higher stress was experienced by the participant on the first day. To statistically confirm this observed trend, we conducted a one-way ANOVA test, revealing a highly significant impact of stress (p-value=6.397e-07). No data segments were excluded from this section’s analysis. We treated all data from the first day as stress instances and all data from the second day as non-stress instances.

\par 
The data collected from this experiment represents a real-deployment scenario for \emph{ThermaStrain}, wherein the \textit{thermal camera angle, distance, participant's posture (i.e., seated behind a desk instead of standing in Section \ref{Data-collection}), stress-inducing task conditions, indoor environment, and background} were unobserved during the model's training. Without any further adjustments (i.e., re-training), we evaluated the previously trained \emph{ThermaStrain} model from Section \ref{real-deploment-partial-body} on this collected indoor one-person workplace scenario dataset. The evaluation yielded an accuracy of $84.59$\% and an F1 score of $0.8398$, aligning with \emph{ThermaStrain}'s performance in Table \ref{table:baselines}.
\par 
While a comprehensive study involving numerous participants, diverse scenarios, and varied indoor settings would be essential to establish \emph{ThermaStrain}'s robustness to unseen data scenarios, this section's evaluation showcases its potential for real-world deployments across various applications.

\subsection{Possible Ethical/Practical Considerations for Deployment}\label{Ethical-discussion}
Ethical considerations for practical deployment scenarios are critical in human sensing applications. When deploying \emph{ThermaStrain} in real-world scenarios, established strategies and protective measures from relevant literature \cite{andrews2023ethical,hanley2020ethical,maurice2018ethical,gubrium2014situated,shneiderman2020bridging} will be utilized to uphold the well-being, privacy, and rights of individuals.

\par 
Before gathering and processing thermal information, the application will secure informed consent from individuals whose data will be utilized. This ensures that participants understand data usage's purpose and potential consequences, allowing them to opt out. Clear communication regarding thermal information processing's aims, methods, and potential outcomes will foster trust and facilitate informed choices.

\par 
For those who opt out or do not pertain to the target group of individuals for stress assessment, their data will remain unused and unprocessed. As discussed in Section \ref{real-deploment-segmentation}, it is possible to identify the body segments of the target users accurately. Our proposed deployment procedure (discussed in Sections \ref{Data-preprocessing} \& \ref{discussion-multi-person}) zeros out all other content of the thermal frames except the target user’s body segment, hence other individuals' data will be masked-out (i.e., zeroed out), and no processing will be performed on their information. Similarly, no information from the indoor environment will be processed, such as furniture, personal items, books, addresses, displayed documents, content within photo frames, and similar items, safeguarding against the leakage of environmental information and preserving privacy \cite{climent2021protection,chiu2023privacy}.
\par 
Additionally, on-device computation offers superior security and privacy \cite{yang2022enabling} compared to cloud-based alternatives. Our real-time evaluation in Section \ref{eval-real-time} verifies that the \emph{ThermaStrain} approach can operate efficiently on resource-constrained edge platforms like Nvidia Jetson Nano, ensuring secure and privacy-preserving stress assessment in real-world settings.
\par 
The proliferation of thermal and image-based human-centric applications is driven by advancements in sensors and AI. Collaboration among experts spanning ethics, law, social sciences, and technology is imperative for a comprehensive ethical approach. While this paper adheres to prevailing ethical standards, future multidisciplinary collaboration endeavors will yield more well-rounded solutions for thermal human-centric sensing applications. Nevertheless, such endeavors remain beyond the scope of this paper.

\section{Discussion on Study Limitations}\label{discussion:limitation}
This section discusses the study limitations and future research scopes of \emph{ThermaStrain} approach.

\paragraph{Physiological Signals as Aiding Modality in Co-teaching:}
It is important to note that the presented solution showed that co-teaching enhances thermal stress sensing performance with the aid of EDA during training, a physiological sensing modality, during training. However, we also evaluated ECG and HR as aiding modalities. However, EDA outperformed others. Hence the presented paper includes only the EDA co-teaching solution and corresponding results. However, we cannot conclude inclusion of ECG or HR as a co-teaching aiding modality would not be beneficial. Exhaustive analysis with larger datasets and scenarios is needed to make such a conclusion, which was out of the scope of this paper.

\paragraph{In-the-wild Evaluation:} 
A limitation of this study is that we specifically analyzed indoor data derived from laboratory environments. Given the primary focus on co-teaching, a comprehensive evaluation in real-world conditions was not within the study's scope. It is worth noting that Section \ref{discussion-multi-person} offers extensive analysis and discourse regarding the deployment of \emph{ThermaStrain} in real-world multi-person scenarios, highlighting its superior viability in comparison to the baselines. However, numerous real-world factors remain unexamined. For instance, the impact of ambient temperature on thermal stress assessment was not assessed, as the data collection took place solely in temperature-controlled indoor settings with no recording of ambient temperatures for each session. Therefore, future work would benefit from sampling data from a broader range of in-the-wild situations to determine the boundaries of the \emph{ThermaStrain} model’s predictive validity.

\paragraph{Study Protocol:} As detailed in Section \ref{procedure}, this study's data collection approach adhered to established literature in behavioral science and psychology \cite{brouwer2014new,goodman2017meta,iqbal2022stress}. For instance, to avoid any bias from residual stress effects, non-stress-inducing tasks were followed by stress-inducing tasks \cite{brouwer2014new,goodman2017meta,iqbal2022stress}. Additionally, as highlighted by \cite{goodman2017meta}, no interfering activities, like questionnaires, occurred at least 15 minutes before introducing the stress-inducing tasks. Our analysis with HRV on Section \ref{Stress=validation-preprocessing} confirms heightened participant stress levels during stress-inducing tasks, indicating the protocol's effectiveness. Nonetheless, a more comprehensive assessment, such as randomizing the order of the stress-inducing tasks, would unveil the most efficacious stress-inducing protocol. This falls within the domain of behavioral science/psychology research and is beyond the scope of this paper.

\paragraph{Age, Sex, and Demography:} With respect to sex, our dataset was relatively balanced (12 male, 20 female). Our analysis showed that \emph{ThermaStrain} achieves similar F1 scores (with 1\% higher in females than males) and accuracy (almost the same). However, with respect to age and demography, the dataset was limited. Future studies involving a larger population with diverse ages, sex, and demographic distributions would be highly beneficial.
It will allow for a more comprehensive understanding of how the thermal signature of stress manifests across different populations and demographic groups, potentially uncovering any variations or patterns. However, such analysis was out of the scope of this paper.

\section{Conclusion}
Existing studies have examined uni-modal and multimodal thermal stress sensing solutions, each with its advantages and limitations. While uni-modal thermal solutions offer non-intrusive sensing, they may lack effectiveness. On the other hand, multimodal approaches can improve performance but may compromise the non-intrusive nature. \emph{ThermaStrain}  combines the benefits of both approaches, providing enhanced stress-sensing performance in a non-intrusive and passive manner. The study collected a comprehensive multimodal thermal stress sensing dataset with diverse stressors and variable distances. Extensive evaluations demonstrated \emph{ThermaStrain}'s ability to generalize and adapt to unknown scenarios, conditions, and environments. These evaluations validated \emph{ThermaStrain}'s fidelity to the co-teaching paradigm and its capacity to enhance stress sensing.
\section*{ACKNOWLEDGMENTS}
This work was partly supported by NSF IIS SCH \#2124285 and NSF CNS CPS \#2148187.

\bibliographystyle{ACM-Reference-Format}

\bibliography{sample-base}


\begin{thebibliography}{124}


\ifx \showCODEN    \undefined \def \showCODEN     #1{\unskip}     \fi
\ifx \showDOI      \undefined \def \showDOI       #1{#1}\fi
\ifx \showISBNx    \undefined \def \showISBNx     #1{\unskip}     \fi
\ifx \showISBNxiii \undefined \def \showISBNxiii  #1{\unskip}     \fi
\ifx \showISSN     \undefined \def \showISSN      #1{\unskip}     \fi
\ifx \showLCCN     \undefined \def \showLCCN      #1{\unskip}     \fi
\ifx \shownote     \undefined \def \shownote      #1{#1}          \fi
\ifx \showarticletitle \undefined \def \showarticletitle #1{#1}   \fi
\ifx \showURL      \undefined \def \showURL       {\relax}        \fi
\providecommand\bibfield[2]{#2}
\providecommand\bibinfo[2]{#2}
\providecommand\natexlab[1]{#1}
\providecommand\showeprint[2][]{arXiv:#2}

\bibitem[Abdelrahman et~al\mbox{.}(2017)]%
        {abdelrahman2017cognitive}
\bibfield{author}{\bibinfo{person}{Yomna Abdelrahman}, \bibinfo{person}{Eduardo Velloso}, \bibinfo{person}{Tilman Dingler}, \bibinfo{person}{Albrecht Schmidt}, {and} \bibinfo{person}{Frank Vetere}.} \bibinfo{year}{2017}\natexlab{}.
\newblock \showarticletitle{Cognitive heat: exploring the usage of thermal imaging to unobtrusively estimate cognitive load}.
\newblock \bibinfo{journal}{\emph{Proceedings of the ACM on Interactive, Mobile, Wearable and Ubiquitous Technologies}} \bibinfo{volume}{1}, \bibinfo{number}{3} (\bibinfo{year}{2017}), \bibinfo{pages}{1--20}.
\newblock


\bibitem[Akiba et~al\mbox{.}(2019)]%
        {optuna_2019}
\bibfield{author}{\bibinfo{person}{Takuya Akiba}, \bibinfo{person}{Shotaro Sano}, \bibinfo{person}{Toshihiko Yanase}, \bibinfo{person}{Takeru Ohta}, {and} \bibinfo{person}{Masanori Koyama}.} \bibinfo{year}{2019}\natexlab{}.
\newblock \showarticletitle{Optuna: A Next-generation Hyperparameter Optimization Framework}. In \bibinfo{booktitle}{\emph{Proceedings of the 25th {ACM} {SIGKDD} International Conference on Knowledge Discovery and Data Mining}}.
\newblock


\bibitem[Al~Qudah et~al\mbox{.}(2021)]%
        {al2021affective}
\bibfield{author}{\bibinfo{person}{Mustafa~MM Al~Qudah}, \bibinfo{person}{Ahmad~SA Mohamed}, {and} \bibinfo{person}{Syaheerah~L Lutfi}.} \bibinfo{year}{2021}\natexlab{}.
\newblock \showarticletitle{Affective State Recognition Using Thermal-Based Imaging: A Survey.}
\newblock \bibinfo{journal}{\emph{Comput. Syst. Sci. Eng.}} \bibinfo{volume}{37}, \bibinfo{number}{1} (\bibinfo{year}{2021}), \bibinfo{pages}{47--62}.
\newblock


\bibitem[Alberdi et~al\mbox{.}(2016)]%
        {alberdi2016towards}
\bibfield{author}{\bibinfo{person}{Ane Alberdi}, \bibinfo{person}{Asier Aztiria}, {and} \bibinfo{person}{Adrian Basarab}.} \bibinfo{year}{2016}\natexlab{}.
\newblock \showarticletitle{Towards an automatic early stress recognition system for office environments based on multimodal measurements: A review}.
\newblock \bibinfo{journal}{\emph{Journal of biomedical informatics}}  \bibinfo{volume}{59} (\bibinfo{year}{2016}), \bibinfo{pages}{49--75}.
\newblock


\bibitem[Alberdi et~al\mbox{.}(2018)]%
        {alberdi2018using}
\bibfield{author}{\bibinfo{person}{Ane Alberdi}, \bibinfo{person}{Asier Aztiria}, \bibinfo{person}{Adrian Basarab}, {and} \bibinfo{person}{Diane~J Cook}.} \bibinfo{year}{2018}\natexlab{}.
\newblock \showarticletitle{Using smart offices to predict occupational stress}.
\newblock \bibinfo{journal}{\emph{International Journal of Industrial Ergonomics}}  \bibinfo{volume}{67} (\bibinfo{year}{2018}), \bibinfo{pages}{13--26}.
\newblock


\bibitem[Andrews et~al\mbox{.}(2023)]%
        {andrews2023ethical}
\bibfield{author}{\bibinfo{person}{Jerone~TA Andrews}, \bibinfo{person}{Dora Zhao}, \bibinfo{person}{William Thong}, \bibinfo{person}{Apostolos Modas}, \bibinfo{person}{Orestis Papakyriakopoulos}, \bibinfo{person}{Shruti Nagpal}, {and} \bibinfo{person}{Alice Xiang}.} \bibinfo{year}{2023}\natexlab{}.
\newblock \showarticletitle{Ethical considerations for collecting human-centric image datasets}.
\newblock \bibinfo{journal}{\emph{arXiv preprint arXiv:2302.03629}} (\bibinfo{year}{2023}).
\newblock


\bibitem[Apple(2023)]%
        {apple--support}
\bibfield{author}{\bibinfo{person}{Apple}.} \bibinfo{year}{2023}\natexlab{}.
\newblock \bibinfo{title}{Apple Watch}.
\newblock \bibinfo{howpublished}{\url{https://support.apple.com/}}.
\newblock


\bibitem[Bergmann et~al\mbox{.}(2012)]%
        {bergmann2012wearable}
\bibfield{author}{\bibinfo{person}{Jeroen~HM Bergmann}, \bibinfo{person}{Vikesh Chandaria}, {and} \bibinfo{person}{Alison McGregor}.} \bibinfo{year}{2012}\natexlab{}.
\newblock \showarticletitle{Wearable and implantable sensors: The patient’s perspective}.
\newblock \bibinfo{journal}{\emph{Sensors}} \bibinfo{volume}{12}, \bibinfo{number}{12} (\bibinfo{year}{2012}), \bibinfo{pages}{16695--16709}.
\newblock


\bibitem[Boucsein(2012)]%
        {boucsein2012electrodermal}
\bibfield{author}{\bibinfo{person}{Wolfram Boucsein}.} \bibinfo{year}{2012}\natexlab{}.
\newblock \bibinfo{booktitle}{\emph{Electrodermal activity}}.
\newblock \bibinfo{publisher}{Springer Science \& Business Media}.
\newblock


\bibitem[Braithwaite et~al\mbox{.}(2013)]%
        {braithwaite2013guide}
\bibfield{author}{\bibinfo{person}{Jason~J Braithwaite}, \bibinfo{person}{Derrick~G Watson}, \bibinfo{person}{Robert Jones}, {and} \bibinfo{person}{Mickey Rowe}.} \bibinfo{year}{2013}\natexlab{}.
\newblock \showarticletitle{A guide for analysing electrodermal activity (EDA) \& skin conductance responses (SCRs) for psychological experiments}.
\newblock \bibinfo{journal}{\emph{Psychophysiology}} \bibinfo{volume}{49}, \bibinfo{number}{1} (\bibinfo{year}{2013}), \bibinfo{pages}{1017--1034}.
\newblock


\bibitem[Brouwer and Hogervorst(2014)]%
        {brouwer2014new}
\bibfield{author}{\bibinfo{person}{Anne-Marie Brouwer} {and} \bibinfo{person}{Maarten~A Hogervorst}.} \bibinfo{year}{2014}\natexlab{}.
\newblock \showarticletitle{A new paradigm to induce mental stress: the Sing-a-Song Stress Test (SSST)}.
\newblock \bibinfo{journal}{\emph{Frontiers in neuroscience}}  \bibinfo{volume}{8} (\bibinfo{year}{2014}), \bibinfo{pages}{224}.
\newblock


\bibitem[Buijs and Van~Eden(2000)]%
        {buijs2000integration}
\bibfield{author}{\bibinfo{person}{Ruud~M Buijs} {and} \bibinfo{person}{Corbert~G Van~Eden}.} \bibinfo{year}{2000}\natexlab{}.
\newblock \showarticletitle{The integration of stress by the hypothalamus, amygdala and prefrontal cortex: balance between the autonomic nervous system and the neuroendocrine system}.
\newblock In \bibinfo{booktitle}{\emph{Progress in brain research}}. Vol.~\bibinfo{volume}{126}. \bibinfo{publisher}{Elsevier}, \bibinfo{pages}{117--132}.
\newblock


\bibitem[Campanella et~al\mbox{.}(2023)]%
        {campanella2023method}
\bibfield{author}{\bibinfo{person}{Sara Campanella}, \bibinfo{person}{Ayham Altaleb}, \bibinfo{person}{Alberto Belli}, \bibinfo{person}{Paola Pierleoni}, {and} \bibinfo{person}{Lorenzo Palma}.} \bibinfo{year}{2023}\natexlab{}.
\newblock \showarticletitle{A Method for Stress Detection Using Empatica E4 Bracelet and Machine-Learning Techniques}.
\newblock \bibinfo{journal}{\emph{Sensors}} \bibinfo{volume}{23}, \bibinfo{number}{7} (\bibinfo{year}{2023}), \bibinfo{pages}{3565}.
\newblock


\bibitem[Can et~al\mbox{.}(2019)]%
        {can2019continuous}
\bibfield{author}{\bibinfo{person}{Yekta~Said Can}, \bibinfo{person}{Niaz Chalabianloo}, \bibinfo{person}{Deniz Ekiz}, {and} \bibinfo{person}{Cem Ersoy}.} \bibinfo{year}{2019}\natexlab{}.
\newblock \showarticletitle{Continuous stress detection using wearable sensors in real life: Algorithmic programming contest case study}.
\newblock \bibinfo{journal}{\emph{Sensors}} \bibinfo{volume}{19}, \bibinfo{number}{8} (\bibinfo{year}{2019}), \bibinfo{pages}{1849}.
\newblock


\bibitem[Cawley and Talbot(2010)]%
        {cawley2010over}
\bibfield{author}{\bibinfo{person}{Gavin~C Cawley} {and} \bibinfo{person}{Nicola~LC Talbot}.} \bibinfo{year}{2010}\natexlab{}.
\newblock \showarticletitle{On over-fitting in model selection and subsequent selection bias in performance evaluation}.
\newblock \bibinfo{journal}{\emph{The Journal of Machine Learning Research}}  \bibinfo{volume}{11} (\bibinfo{year}{2010}), \bibinfo{pages}{2079--2107}.
\newblock


\bibitem[CDW(2023)]%
        {CDW}
\bibfield{author}{\bibinfo{person}{CDW}.} \bibinfo{year}{2023}\natexlab{}.
\newblock \bibinfo{title}{Future Proofing \& New Work Dynamic}.
\newblock
\newblock
\urldef\tempurl%
\url{https://shorturl.at/ehAGX}
\showURL{%
Retrieved July, 2023 from \tempurl}


\bibitem[Chaudhari et~al\mbox{.}(2019)]%
        {chaudhari2019entropy}
\bibfield{author}{\bibinfo{person}{Pratik Chaudhari}, \bibinfo{person}{Anna Choromanska}, \bibinfo{person}{Stefano Soatto}, \bibinfo{person}{Yann LeCun}, \bibinfo{person}{Carlo Baldassi}, \bibinfo{person}{Christian Borgs}, \bibinfo{person}{Jennifer Chayes}, \bibinfo{person}{Levent Sagun}, {and} \bibinfo{person}{Riccardo Zecchina}.} \bibinfo{year}{2019}\natexlab{}.
\newblock \showarticletitle{Entropy-sgd: Biasing gradient descent into wide valleys}.
\newblock \bibinfo{journal}{\emph{Journal of Statistical Mechanics: Theory and Experiment}} \bibinfo{volume}{2019}, \bibinfo{number}{12} (\bibinfo{year}{2019}), \bibinfo{pages}{124018}.
\newblock


\bibitem[Chiu et~al\mbox{.}(2023)]%
        {chiu2023privacy}
\bibfield{author}{\bibinfo{person}{Sheng-Yang Chiu}, \bibinfo{person}{Yu-Ting Huang}, \bibinfo{person}{Chieh-Ting Lin}, \bibinfo{person}{Yu-Chee Tseng}, \bibinfo{person}{Jen-Jee Chen}, \bibinfo{person}{Meng-Hsuan Tu}, \bibinfo{person}{Bo-Chen Tung}, {and} \bibinfo{person}{YuJou Nieh}.} \bibinfo{year}{2023}\natexlab{}.
\newblock \showarticletitle{Privacy-preserving video conferencing via thermal-generative images}. In \bibinfo{booktitle}{\emph{2023 IEEE International Conference on Robotics and Automation (ICRA)}}. IEEE, \bibinfo{pages}{9478--9485}.
\newblock


\bibitem[Cho and Bianchi-Berthouze(2019)]%
        {cho2019physiological}
\bibfield{author}{\bibinfo{person}{Youngjun Cho} {and} \bibinfo{person}{Nadia Bianchi-Berthouze}.} \bibinfo{year}{2019}\natexlab{}.
\newblock \showarticletitle{Physiological and affective computing through thermal imaging: A survey}.
\newblock \bibinfo{journal}{\emph{arXiv preprint arXiv:1908.10307}} (\bibinfo{year}{2019}).
\newblock


\bibitem[Cho et~al\mbox{.}(2019)]%
        {cho2019instant}
\bibfield{author}{\bibinfo{person}{Youngjun Cho}, \bibinfo{person}{Simon~J Julier}, {and} \bibinfo{person}{Nadia Bianchi-Berthouze}.} \bibinfo{year}{2019}\natexlab{}.
\newblock \showarticletitle{Instant stress: detection of perceived mental stress through smartphone photoplethysmography and thermal imaging}.
\newblock \bibinfo{journal}{\emph{JMIR mental health}} \bibinfo{volume}{6}, \bibinfo{number}{4} (\bibinfo{year}{2019}), \bibinfo{pages}{e10140}.
\newblock


\bibitem[Climent-P{\'e}rez and Florez-Revuelta(2021)]%
        {climent2021protection}
\bibfield{author}{\bibinfo{person}{Pau Climent-P{\'e}rez} {and} \bibinfo{person}{Francisco Florez-Revuelta}.} \bibinfo{year}{2021}\natexlab{}.
\newblock \showarticletitle{Protection of visual privacy in videos acquired with RGB cameras for active and assisted living applications}.
\newblock \bibinfo{journal}{\emph{Multimedia Tools and Applications}} \bibinfo{volume}{80}, \bibinfo{number}{15} (\bibinfo{year}{2021}), \bibinfo{pages}{23649--23664}.
\newblock


\bibitem[Connolly and Alloy(2018)]%
        {connolly2018negative}
\bibfield{author}{\bibinfo{person}{Samantha~L Connolly} {and} \bibinfo{person}{Lauren~B Alloy}.} \bibinfo{year}{2018}\natexlab{}.
\newblock \showarticletitle{Negative event recall as a vulnerability for depression: Relationship between momentary stress-reactive rumination and memory for daily life stress}.
\newblock \bibinfo{journal}{\emph{Clinical Psychological Science}} \bibinfo{volume}{6}, \bibinfo{number}{1} (\bibinfo{year}{2018}), \bibinfo{pages}{32--47}.
\newblock


\bibitem[Cross et~al\mbox{.}(2013)]%
        {cross2013thermal}
\bibfield{author}{\bibinfo{person}{Carl~B Cross}, \bibinfo{person}{Julie~A Skipper}, {and} \bibinfo{person}{Douglas~T Petkie}.} \bibinfo{year}{2013}\natexlab{}.
\newblock \showarticletitle{Thermal imaging to detect physiological indicators of stress in humans}. In \bibinfo{booktitle}{\emph{Thermosense: thermal infrared applications XXXV}}, Vol.~\bibinfo{volume}{8705}. SPIE, \bibinfo{pages}{141--155}.
\newblock


\bibitem[Dickerson and Kemeny(2004)]%
        {dickerson2004acute}
\bibfield{author}{\bibinfo{person}{Sally~S Dickerson} {and} \bibinfo{person}{Margaret~E Kemeny}.} \bibinfo{year}{2004}\natexlab{}.
\newblock \showarticletitle{Acute stressors and cortisol responses: a theoretical integration and synthesis of laboratory research.}
\newblock \bibinfo{journal}{\emph{Psychological bulletin}} \bibinfo{volume}{130}, \bibinfo{number}{3} (\bibinfo{year}{2004}), \bibinfo{pages}{355}.
\newblock


\bibitem[Dosovitskiy et~al\mbox{.}(2020)]%
        {dosovitskiy2020image}
\bibfield{author}{\bibinfo{person}{Alexey Dosovitskiy}, \bibinfo{person}{Lucas Beyer}, \bibinfo{person}{Alexander Kolesnikov}, \bibinfo{person}{Dirk Weissenborn}, \bibinfo{person}{Xiaohua Zhai}, \bibinfo{person}{Thomas Unterthiner}, \bibinfo{person}{Mostafa Dehghani}, \bibinfo{person}{Matthias Minderer}, \bibinfo{person}{Georg Heigold}, \bibinfo{person}{Sylvain Gelly}, {et~al\mbox{.}}} \bibinfo{year}{2020}\natexlab{}.
\newblock \showarticletitle{An image is worth 16x16 words: Transformers for image recognition at scale}.
\newblock \bibinfo{journal}{\emph{arXiv preprint arXiv:2010.11929}} (\bibinfo{year}{2020}).
\newblock


\bibitem[Engert et~al\mbox{.}(2014)]%
        {engert2014exploring}
\bibfield{author}{\bibinfo{person}{Veronika Engert}, \bibinfo{person}{Arcangelo Merla}, \bibinfo{person}{Joshua~A Grant}, \bibinfo{person}{Daniela Cardone}, \bibinfo{person}{Anita Tusche}, {and} \bibinfo{person}{Tania Singer}.} \bibinfo{year}{2014}\natexlab{}.
\newblock \showarticletitle{Exploring the use of thermal infrared imaging in human stress research}.
\newblock \bibinfo{journal}{\emph{PloS one}} \bibinfo{volume}{9}, \bibinfo{number}{3} (\bibinfo{year}{2014}), \bibinfo{pages}{e90782}.
\newblock


\bibitem[Eum et~al\mbox{.}(2013)]%
        {eum2013human}
\bibfield{author}{\bibinfo{person}{Hyukmin Eum}, \bibinfo{person}{Jeisung Lee}, \bibinfo{person}{Changyong Yoon}, {and} \bibinfo{person}{Mignon Park}.} \bibinfo{year}{2013}\natexlab{}.
\newblock \showarticletitle{Human action recognition for night vision using temporal templates with infrared thermal camera}. In \bibinfo{booktitle}{\emph{2013 10th International Conference on Ubiquitous Robots and Ambient Intelligence (URAI)}}. IEEE, \bibinfo{pages}{617--621}.
\newblock


\bibitem[Fook et~al\mbox{.}(2007)]%
        {fook2007automated}
\bibfield{author}{\bibinfo{person}{Victor Foo~Siang Fook}, \bibinfo{person}{Pham~Viet Thang}, \bibinfo{person}{That~Mon Htwe}, \bibinfo{person}{Qiu Qiang}, \bibinfo{person}{Aung Aung~Phyo Wai}, \bibinfo{person}{Maniyeri Jayachandran}, \bibinfo{person}{Jit Biswas}, {and} \bibinfo{person}{Philip Yap}.} \bibinfo{year}{2007}\natexlab{}.
\newblock \showarticletitle{Automated recognition of complex agitation behavior of dementia patients using video camera}. In \bibinfo{booktitle}{\emph{2007 9th International Conference on e-Health Networking, Application and Services}}. IEEE, \bibinfo{pages}{68--73}.
\newblock


\bibitem[Fortin and Chaib-Draa(2019)]%
        {fortin2019multimodal}
\bibfield{author}{\bibinfo{person}{Mathieu~Pag{\'e} Fortin} {and} \bibinfo{person}{Brahim Chaib-Draa}.} \bibinfo{year}{2019}\natexlab{}.
\newblock \showarticletitle{Multimodal Sentiment Analysis: A Multitask Learning Approach.}. In \bibinfo{booktitle}{\emph{ICPRAM}}. \bibinfo{pages}{368--376}.
\newblock


\bibitem[Gavrilescu and Vizireanu(2019)]%
        {gavrilescu2019predicting}
\bibfield{author}{\bibinfo{person}{Mihai Gavrilescu} {and} \bibinfo{person}{Nicolae Vizireanu}.} \bibinfo{year}{2019}\natexlab{}.
\newblock \showarticletitle{Predicting Depression, Anxiety, and Stress Levels from Videos Using the Facial Action Coding System}.
\newblock \bibinfo{journal}{\emph{Sensors}} \bibinfo{volume}{19}, \bibinfo{number}{17} (\bibinfo{year}{2019}), \bibinfo{pages}{3693}.
\newblock


\bibitem[Ghosh et~al\mbox{.}(2022)]%
        {ghosh2022classification}
\bibfield{author}{\bibinfo{person}{Sayandeep Ghosh}, \bibinfo{person}{Seongki Kim}, \bibinfo{person}{Muhammad~Fazal Ijaz}, \bibinfo{person}{Pawan~Kumar Singh}, {and} \bibinfo{person}{Mufti Mahmud}.} \bibinfo{year}{2022}\natexlab{}.
\newblock \showarticletitle{Classification of mental stress from wearable physiological sensors using image-encoding-based deep neural network}.
\newblock \bibinfo{journal}{\emph{Biosensors}} \bibinfo{volume}{12}, \bibinfo{number}{12} (\bibinfo{year}{2022}), \bibinfo{pages}{1153}.
\newblock


\bibitem[Giannakakis et~al\mbox{.}(2017)]%
        {giannakakis}
\bibfield{author}{\bibinfo{person}{G. Giannakakis}, \bibinfo{person}{M. Pediaditis}, \bibinfo{person}{D. Manousos}, \bibinfo{person}{E. Kazantzaki}, \bibinfo{person}{F. Chiarugi}, \bibinfo{person}{P.G. Simos}, \bibinfo{person}{K. Marias}, {and} \bibinfo{person}{M. Tsiknakis}.} \bibinfo{year}{2017}\natexlab{}.
\newblock \showarticletitle{Stress and anxiety detection using facial cues from videos}.
\newblock \bibinfo{journal}{\emph{Biomedical Signal Processing and Control}}  \bibinfo{volume}{31} (\bibinfo{year}{2017}), \bibinfo{pages}{89 -- 101}.
\newblock
\showISSN{1746-8094}
\urldef\tempurl%
\url{https://doi.org/10.1016/j.bspc.2016.06.020}
\showDOI{\tempurl}


\bibitem[Goodfellow et~al\mbox{.}(2014)]%
        {goodfellow2014qualitatively}
\bibfield{author}{\bibinfo{person}{Ian~J Goodfellow}, \bibinfo{person}{Oriol Vinyals}, {and} \bibinfo{person}{Andrew~M Saxe}.} \bibinfo{year}{2014}\natexlab{}.
\newblock \showarticletitle{Qualitatively characterizing neural network optimization problems}.
\newblock \bibinfo{journal}{\emph{arXiv preprint arXiv:1412.6544}} (\bibinfo{year}{2014}).
\newblock


\bibitem[Goodman et~al\mbox{.}(2017)]%
        {goodman2017meta}
\bibfield{author}{\bibinfo{person}{William~K Goodman}, \bibinfo{person}{Johanna Janson}, {and} \bibinfo{person}{Jutta~M Wolf}.} \bibinfo{year}{2017}\natexlab{}.
\newblock \showarticletitle{Meta-analytical assessment of the effects of protocol variations on cortisol responses to the Trier Social Stress Test}.
\newblock \bibinfo{journal}{\emph{Psychoneuroendocrinology}}  \bibinfo{volume}{80} (\bibinfo{year}{2017}), \bibinfo{pages}{26--35}.
\newblock


\bibitem[Griffiths et~al\mbox{.}(2018)]%
        {griffiths2018privacy}
\bibfield{author}{\bibinfo{person}{Erin Griffiths}, \bibinfo{person}{Salah Assana}, {and} \bibinfo{person}{Kamin Whitehouse}.} \bibinfo{year}{2018}\natexlab{}.
\newblock \showarticletitle{Privacy-preserving image processing with binocular thermal cameras}.
\newblock \bibinfo{journal}{\emph{Proceedings of the ACM on Interactive, Mobile, Wearable and Ubiquitous Technologies}} \bibinfo{volume}{1}, \bibinfo{number}{4} (\bibinfo{year}{2018}), \bibinfo{pages}{1--25}.
\newblock


\bibitem[Gross and Jazaieri(2014)]%
        {gross2014emotion}
\bibfield{author}{\bibinfo{person}{James~J Gross} {and} \bibinfo{person}{Hooria Jazaieri}.} \bibinfo{year}{2014}\natexlab{}.
\newblock \showarticletitle{Emotion, emotion regulation, and psychopathology: An affective science perspective}.
\newblock \bibinfo{journal}{\emph{Clinical Psychological Science}} \bibinfo{volume}{2}, \bibinfo{number}{4} (\bibinfo{year}{2014}), \bibinfo{pages}{387--401}.
\newblock


\bibitem[Gubrium et~al\mbox{.}(2014)]%
        {gubrium2014situated}
\bibfield{author}{\bibinfo{person}{Aline~C Gubrium}, \bibinfo{person}{Amy~L Hill}, {and} \bibinfo{person}{Sarah Flicker}.} \bibinfo{year}{2014}\natexlab{}.
\newblock \showarticletitle{A situated practice of ethics for participatory visual and digital methods in public health research and practice: A focus on digital storytelling}.
\newblock \bibinfo{journal}{\emph{American journal of public health}} \bibinfo{volume}{104}, \bibinfo{number}{9} (\bibinfo{year}{2014}), \bibinfo{pages}{1606--1614}.
\newblock


\bibitem[Hammen(2005)]%
        {hammen2005stress}
\bibfield{author}{\bibinfo{person}{Constance Hammen}.} \bibinfo{year}{2005}\natexlab{}.
\newblock \showarticletitle{Stress and depression}.
\newblock \bibinfo{journal}{\emph{Annu. Rev. Clin. Psychol.}}  \bibinfo{volume}{1} (\bibinfo{year}{2005}), \bibinfo{pages}{293--319}.
\newblock


\bibitem[Hanley et~al\mbox{.}(2020)]%
        {hanley2020ethical}
\bibfield{author}{\bibinfo{person}{Margot Hanley}, \bibinfo{person}{Apoorv Khandelwal}, \bibinfo{person}{Hadar Averbuch-Elor}, \bibinfo{person}{Noah Snavely}, {and} \bibinfo{person}{Helen Nissenbaum}.} \bibinfo{year}{2020}\natexlab{}.
\newblock \showarticletitle{An ethical highlighter for people-centric dataset creation}.
\newblock \bibinfo{journal}{\emph{arXiv preprint arXiv:2011.13583}} (\bibinfo{year}{2020}).
\newblock


\bibitem[Helminen et~al\mbox{.}(2019)]%
        {helminen2019meta}
\bibfield{author}{\bibinfo{person}{Emily~C Helminen}, \bibinfo{person}{Melissa~L Morton}, \bibinfo{person}{Qiu Wang}, {and} \bibinfo{person}{Joshua~C Felver}.} \bibinfo{year}{2019}\natexlab{}.
\newblock \showarticletitle{A meta-analysis of cortisol reactivity to the Trier Social Stress Test in virtual environments}.
\newblock \bibinfo{journal}{\emph{Psychoneuroendocrinology}}  \bibinfo{volume}{110} (\bibinfo{year}{2019}), \bibinfo{pages}{104437}.
\newblock


\bibitem[Heo et~al\mbox{.}(2021)]%
        {heo2021stress}
\bibfield{author}{\bibinfo{person}{Seongsil Heo}, \bibinfo{person}{Sunyoung Kwon}, {and} \bibinfo{person}{Jaekoo Lee}.} \bibinfo{year}{2021}\natexlab{}.
\newblock \showarticletitle{Stress detection with single PPG sensor by orchestrating multiple denoising and peak-detecting methods}.
\newblock \bibinfo{journal}{\emph{IEEE Access}}  \bibinfo{volume}{9} (\bibinfo{year}{2021}), \bibinfo{pages}{47777--47785}.
\newblock


\bibitem[Herborn et~al\mbox{.}(2015)]%
        {herborn2015skin}
\bibfield{author}{\bibinfo{person}{Katherine~A Herborn}, \bibinfo{person}{James~L Graves}, \bibinfo{person}{Paul Jerem}, \bibinfo{person}{Neil~P Evans}, \bibinfo{person}{Ruedi Nager}, \bibinfo{person}{Dominic~J McCafferty}, {and} \bibinfo{person}{Dorothy~EF McKeegan}.} \bibinfo{year}{2015}\natexlab{}.
\newblock \showarticletitle{Skin temperature reveals the intensity of acute stress}.
\newblock \bibinfo{journal}{\emph{Physiology \& behavior}}  \bibinfo{volume}{152} (\bibinfo{year}{2015}), \bibinfo{pages}{225--230}.
\newblock


\bibitem[Hochreiter and Schmidhuber(1997)]%
        {hochreiter1997flat}
\bibfield{author}{\bibinfo{person}{Sepp Hochreiter} {and} \bibinfo{person}{J{\"u}rgen Schmidhuber}.} \bibinfo{year}{1997}\natexlab{}.
\newblock \showarticletitle{Flat minima}.
\newblock \bibinfo{journal}{\emph{Neural computation}} \bibinfo{volume}{9}, \bibinfo{number}{1} (\bibinfo{year}{1997}), \bibinfo{pages}{1--42}.
\newblock


\bibitem[Hodge et~al\mbox{.}(2018)]%
        {hodge2018anatomy}
\bibfield{author}{\bibinfo{person}{Bonnie~D Hodge}, \bibinfo{person}{Terrence Sanvictores}, {and} \bibinfo{person}{Robert~T Brodell}.} \bibinfo{year}{2018}\natexlab{}.
\newblock \showarticletitle{Anatomy, skin sweat glands}.
\newblock  (\bibinfo{year}{2018}).
\newblock


\bibitem[Im et~al\mbox{.}(2016)]%
        {im2016empirical}
\bibfield{author}{\bibinfo{person}{Daniel~Jiwoong Im}, \bibinfo{person}{Michael Tao}, {and} \bibinfo{person}{Kristin Branson}.} \bibinfo{year}{2016}\natexlab{}.
\newblock \showarticletitle{An empirical analysis of the optimization of deep network loss surfaces}.
\newblock \bibinfo{journal}{\emph{arXiv preprint arXiv:1612.04010}} (\bibinfo{year}{2016}).
\newblock


\bibitem[Iqbal et~al\mbox{.}(2022)]%
        {iqbal2022stress}
\bibfield{author}{\bibinfo{person}{Talha Iqbal}, \bibinfo{person}{Andrew~J Simpkin}, \bibinfo{person}{Davood Roshan}, \bibinfo{person}{Nicola Glynn}, \bibinfo{person}{John Killilea}, \bibinfo{person}{Jane Walsh}, \bibinfo{person}{Gerard Molloy}, \bibinfo{person}{Sandra Ganly}, \bibinfo{person}{Hannah Ryman}, \bibinfo{person}{Eileen Coen}, {et~al\mbox{.}}} \bibinfo{year}{2022}\natexlab{}.
\newblock \showarticletitle{Stress Monitoring Using Wearable Sensors: A Pilot Study and Stress-Predict Dataset}.
\newblock \bibinfo{journal}{\emph{Sensors}} \bibinfo{volume}{22}, \bibinfo{number}{21} (\bibinfo{year}{2022}), \bibinfo{pages}{8135}.
\newblock


\bibitem[James et~al\mbox{.}(2014)]%
        {james2014reliability}
\bibfield{author}{\bibinfo{person}{CA James}, \bibinfo{person}{AJ Richardson}, \bibinfo{person}{PW Watt}, {and} \bibinfo{person}{NS Maxwell}.} \bibinfo{year}{2014}\natexlab{}.
\newblock \showarticletitle{Reliability and validity of skin temperature measurement by telemetry thermistors and a thermal camera during exercise in the heat}.
\newblock \bibinfo{journal}{\emph{Journal of thermal biology}}  \bibinfo{volume}{45} (\bibinfo{year}{2014}), \bibinfo{pages}{141--149}.
\newblock


\bibitem[Jeong et~al\mbox{.}(2017)]%
        {jeong2017smartwatch}
\bibfield{author}{\bibinfo{person}{Hayeon Jeong}, \bibinfo{person}{Heepyung Kim}, \bibinfo{person}{Rihun Kim}, \bibinfo{person}{Uichin Lee}, {and} \bibinfo{person}{Yong Jeong}.} \bibinfo{year}{2017}\natexlab{}.
\newblock \showarticletitle{Smartwatch wearing behavior analysis: a longitudinal study}.
\newblock \bibinfo{journal}{\emph{Proceedings of the ACM on Interactive, Mobile, Wearable and Ubiquitous Technologies}} \bibinfo{volume}{1}, \bibinfo{number}{3} (\bibinfo{year}{2017}), \bibinfo{pages}{1--31}.
\newblock


\bibitem[Jukiewicz et~al\mbox{.}(2021)]%
        {jukiewicz2021electrodermal}
\bibfield{author}{\bibinfo{person}{Marcin Jukiewicz}, \bibinfo{person}{Pawe{\l} {\L}upkowski}, \bibinfo{person}{Radomir Majchrowski}, \bibinfo{person}{Joanna Marcinkowska}, {and} \bibinfo{person}{Dawid Ratajczyk}.} \bibinfo{year}{2021}\natexlab{}.
\newblock \showarticletitle{Electrodermal and thermal measurement of users’ emotional reaction for a visual stimuli}.
\newblock \bibinfo{journal}{\emph{Case Studies in Thermal Engineering}}  \bibinfo{volume}{27} (\bibinfo{year}{2021}), \bibinfo{pages}{101303}.
\newblock


\bibitem[Kawaguchi et~al\mbox{.}(2017)]%
        {kawaguchi2017generalization}
\bibfield{author}{\bibinfo{person}{Kenji Kawaguchi}, \bibinfo{person}{Leslie~Pack Kaelbling}, {and} \bibinfo{person}{Yoshua Bengio}.} \bibinfo{year}{2017}\natexlab{}.
\newblock \showarticletitle{Generalization in deep learning}.
\newblock \bibinfo{journal}{\emph{arXiv preprint arXiv:1710.05468}} (\bibinfo{year}{2017}).
\newblock


\bibitem[Keshan et~al\mbox{.}(2015)]%
        {keshan2015machine}
\bibfield{author}{\bibinfo{person}{N Keshan}, \bibinfo{person}{PV Parimi}, {and} \bibinfo{person}{Isabelle Bichindaritz}.} \bibinfo{year}{2015}\natexlab{}.
\newblock \showarticletitle{Machine learning for stress detection from ECG signals in automobile drivers}. In \bibinfo{booktitle}{\emph{2015 IEEE International conference on big data (Big Data)}}. IEEE, \bibinfo{pages}{2661--2669}.
\newblock


\bibitem[Keskar et~al\mbox{.}(2016)]%
        {keskar2016large}
\bibfield{author}{\bibinfo{person}{Nitish~Shirish Keskar}, \bibinfo{person}{Dheevatsa Mudigere}, \bibinfo{person}{Jorge Nocedal}, \bibinfo{person}{Mikhail Smelyanskiy}, {and} \bibinfo{person}{Ping Tak~Peter Tang}.} \bibinfo{year}{2016}\natexlab{}.
\newblock \showarticletitle{On large-batch training for deep learning: Generalization gap and sharp minima}.
\newblock \bibinfo{journal}{\emph{arXiv preprint arXiv:1609.04836}} (\bibinfo{year}{2016}).
\newblock


\bibitem[Khosrowabadi(2018)]%
        {khosrowabadi2018stress}
\bibfield{author}{\bibinfo{person}{Reza Khosrowabadi}.} \bibinfo{year}{2018}\natexlab{}.
\newblock \showarticletitle{Stress and Perception of Emotional Stimuli: Long-term Stress Rewiring the Brain}.
\newblock \bibinfo{journal}{\emph{Basic and clinical neuroscience}} \bibinfo{volume}{9}, \bibinfo{number}{2} (\bibinfo{year}{2018}), \bibinfo{pages}{107}.
\newblock


\bibitem[Kim et~al\mbox{.}(2018a)]%
        {kim2018stress}
\bibfield{author}{\bibinfo{person}{Hye-Geum Kim}, \bibinfo{person}{Eun-Jin Cheon}, \bibinfo{person}{Dai-Seg Bai}, \bibinfo{person}{Young~Hwan Lee}, {and} \bibinfo{person}{Bon-Hoon Koo}.} \bibinfo{year}{2018}\natexlab{a}.
\newblock \showarticletitle{Stress and heart rate variability: a meta-analysis and review of the literature}.
\newblock \bibinfo{journal}{\emph{Psychiatry investigation}} \bibinfo{volume}{15}, \bibinfo{number}{3} (\bibinfo{year}{2018}), \bibinfo{pages}{235}.
\newblock


\bibitem[Kim(2019)]%
        {kim2019pedestrian}
\bibfield{author}{\bibinfo{person}{JongBae Kim}.} \bibinfo{year}{2019}\natexlab{}.
\newblock \showarticletitle{Pedestrian detection and distance estimation using thermal camera in night time}. In \bibinfo{booktitle}{\emph{2019 International Conference on Artificial Intelligence in Information and Communication (ICAIIC)}}. IEEE, \bibinfo{pages}{463--466}.
\newblock


\bibitem[Kim et~al\mbox{.}(2018b)]%
        {kim2018remote}
\bibfield{author}{\bibinfo{person}{Yoonkyoung Kim}, \bibinfo{person}{Yosep Park}, \bibinfo{person}{Jinman Kim}, {and} \bibinfo{person}{Eui~Chul Lee}.} \bibinfo{year}{2018}\natexlab{b}.
\newblock \showarticletitle{Remote heart rate monitoring method using infrared thermal camera}.
\newblock \bibinfo{journal}{\emph{Int. J. Eng. Res. Technol}} \bibinfo{volume}{11}, \bibinfo{number}{3} (\bibinfo{year}{2018}), \bibinfo{pages}{493--500}.
\newblock


\bibitem[Kirimtat et~al\mbox{.}(2020)]%
        {kirimtat2020flir}
\bibfield{author}{\bibinfo{person}{Ayca Kirimtat}, \bibinfo{person}{Ondrej Krejcar}, \bibinfo{person}{Ali Selamat}, {and} \bibinfo{person}{Enrique Herrera-Viedma}.} \bibinfo{year}{2020}\natexlab{}.
\newblock \showarticletitle{FLIR vs SEEK thermal cameras in biomedicine: comparative diagnosis through infrared thermography}.
\newblock \bibinfo{journal}{\emph{BMC bioinformatics}} \bibinfo{volume}{21}, \bibinfo{number}{2} (\bibinfo{year}{2020}), \bibinfo{pages}{1--10}.
\newblock


\bibitem[Kirschbaum et~al\mbox{.}(1993)]%
        {kirschbaum1993trier}
\bibfield{author}{\bibinfo{person}{Clemens Kirschbaum}, \bibinfo{person}{Karl-Martin Pirke}, {and} \bibinfo{person}{Dirk~H Hellhammer}.} \bibinfo{year}{1993}\natexlab{}.
\newblock \showarticletitle{The ‘Trier Social Stress Test’--a tool for investigating psychobiological stress responses in a laboratory setting}.
\newblock \bibinfo{journal}{\emph{Neuropsychobiology}} \bibinfo{volume}{28}, \bibinfo{number}{1-2} (\bibinfo{year}{1993}), \bibinfo{pages}{76--81}.
\newblock


\bibitem[K{\"o}nig et~al\mbox{.}(2015)]%
        {konig2015validation}
\bibfield{author}{\bibinfo{person}{Alexandra K{\"o}nig}, \bibinfo{person}{Carlos~Fernando Crispim~Junior}, \bibinfo{person}{Alexandre Derreumaux}, \bibinfo{person}{Gregory Bensadoun}, \bibinfo{person}{Pierre-David Petit}, \bibinfo{person}{Fran{\c{c}}ois Bremond}, \bibinfo{person}{Renaud David}, \bibinfo{person}{Frans Verhey}, \bibinfo{person}{Pauline Aalten}, {and} \bibinfo{person}{Philippe Robert}.} \bibinfo{year}{2015}\natexlab{}.
\newblock \showarticletitle{Validation of an automatic video monitoring system for the detection of instrumental activities of daily living in dementia patients}.
\newblock \bibinfo{journal}{\emph{Journal of Alzheimer's Disease}} \bibinfo{volume}{44}, \bibinfo{number}{2} (\bibinfo{year}{2015}), \bibinfo{pages}{675--685}.
\newblock


\bibitem[Kross et~al\mbox{.}(2012)]%
        {kross2012asking}
\bibfield{author}{\bibinfo{person}{Ethan Kross}, \bibinfo{person}{David Gard}, \bibinfo{person}{Patricia Deldin}, \bibinfo{person}{Jessica Clifton}, {and} \bibinfo{person}{Ozlem Ayduk}.} \bibinfo{year}{2012}\natexlab{}.
\newblock \showarticletitle{“Asking why” from a distance: Its cognitive and emotional consequences for people with major depressive disorder.}
\newblock \bibinfo{journal}{\emph{Journal of abnormal psychology}} \bibinfo{volume}{121}, \bibinfo{number}{3} (\bibinfo{year}{2012}), \bibinfo{pages}{559}.
\newblock


\bibitem[Krzywicki et~al\mbox{.}(2014)]%
        {krzywicki2014non}
\bibfield{author}{\bibinfo{person}{Alan~T Krzywicki}, \bibinfo{person}{Gary~G Berntson}, {and} \bibinfo{person}{Barbara~L O'Kane}.} \bibinfo{year}{2014}\natexlab{}.
\newblock \showarticletitle{A non-contact technique for measuring eccrine sweat gland activity using passive thermal imaging}.
\newblock \bibinfo{journal}{\emph{International journal of psychophysiology}} \bibinfo{volume}{94}, \bibinfo{number}{1} (\bibinfo{year}{2014}), \bibinfo{pages}{25--34}.
\newblock


\bibitem[Kumar et~al\mbox{.}(2021)]%
        {kumar2021stressnet}
\bibfield{author}{\bibinfo{person}{Satish Kumar}, \bibinfo{person}{ASM Iftekhar}, \bibinfo{person}{Michael Goebel}, \bibinfo{person}{Tom Bullock}, \bibinfo{person}{Mary~H MacLean}, \bibinfo{person}{Michael~B Miller}, \bibinfo{person}{Tyler Santander}, \bibinfo{person}{Barry Giesbrecht}, \bibinfo{person}{Scott~T Grafton}, {and} \bibinfo{person}{BS Manjunath}.} \bibinfo{year}{2021}\natexlab{}.
\newblock \showarticletitle{StressNet: detecting stress in thermal videos}. In \bibinfo{booktitle}{\emph{Proceedings of the IEEE/CVF Winter Conference on Applications of Computer Vision}}. \bibinfo{pages}{999--1009}.
\newblock


\bibitem[Kylili et~al\mbox{.}(2014)]%
        {kylili2014infrared}
\bibfield{author}{\bibinfo{person}{Angeliki Kylili}, \bibinfo{person}{Paris~A Fokaides}, \bibinfo{person}{Petros Christou}, {and} \bibinfo{person}{Soteris~A Kalogirou}.} \bibinfo{year}{2014}\natexlab{}.
\newblock \showarticletitle{Infrared thermography (IRT) applications for building diagnostics: A review}.
\newblock \bibinfo{journal}{\emph{Applied Energy}}  \bibinfo{volume}{134} (\bibinfo{year}{2014}), \bibinfo{pages}{531--549}.
\newblock


\bibitem[Lee et~al\mbox{.}(2019)]%
        {lee2019clara}
\bibfield{author}{\bibinfo{person}{Juwon Lee}, \bibinfo{person}{Megan Lam}, {and} \bibinfo{person}{Caleb Chiu}.} \bibinfo{year}{2019}\natexlab{}.
\newblock \showarticletitle{Clara: design of a new system for passive sensing of depression, stress and anxiety in the workplace}. In \bibinfo{booktitle}{\emph{Pervasive Computing Paradigms for Mental Health: 9th International Conference, MindCare 2019, Buenos Aires, Argentina, April 23--24, 2019, Proceedings 9}}. Springer, \bibinfo{pages}{12--28}.
\newblock


\bibitem[Lee et~al\mbox{.}(2020)]%
        {lee2020toward}
\bibfield{author}{\bibinfo{person}{Kwangyoung Lee}, \bibinfo{person}{Hyewon Cho}, \bibinfo{person}{Kobiljon Toshnazarov}, \bibinfo{person}{Nematjon Narziev}, \bibinfo{person}{So~Young Rhim}, \bibinfo{person}{Kyungsik Han}, \bibinfo{person}{YoungTae Noh}, {and} \bibinfo{person}{Hwajung Hong}.} \bibinfo{year}{2020}\natexlab{}.
\newblock \showarticletitle{Toward future-centric personal informatics: Expecting stressful events and preparing personalized interventions in stress management}. In \bibinfo{booktitle}{\emph{Proceedings of the 2020 CHI Conference on Human Factors in Computing Systems}}. \bibinfo{pages}{1--13}.
\newblock


\bibitem[Leone et~al\mbox{.}(2020)]%
        {leone2020multi}
\bibfield{author}{\bibinfo{person}{Alessandro Leone}, \bibinfo{person}{Gabriele Rescio}, \bibinfo{person}{Pietro Siciliano}, \bibinfo{person}{Alessandra Papetti}, \bibinfo{person}{Agnese Brunzini}, {and} \bibinfo{person}{Michele Germani}.} \bibinfo{year}{2020}\natexlab{}.
\newblock \showarticletitle{Multi sensors platform for stress monitoring of workers in smart manufacturing context}. In \bibinfo{booktitle}{\emph{2020 IEEE International Instrumentation and Measurement Technology Conference (I2MTC)}}. IEEE, \bibinfo{pages}{1--5}.
\newblock


\bibitem[Li et~al\mbox{.}(2018)]%
        {li2018visualizing}
\bibfield{author}{\bibinfo{person}{Hao Li}, \bibinfo{person}{Zheng Xu}, \bibinfo{person}{Gavin Taylor}, \bibinfo{person}{Christoph Studer}, {and} \bibinfo{person}{Tom Goldstein}.} \bibinfo{year}{2018}\natexlab{}.
\newblock \showarticletitle{Visualizing the loss landscape of neural nets}.
\newblock \bibinfo{journal}{\emph{Advances in neural information processing systems}}  \bibinfo{volume}{31} (\bibinfo{year}{2018}).
\newblock


\bibitem[Li et~al\mbox{.}(2022)]%
        {li2022valhalla}
\bibfield{author}{\bibinfo{person}{Yi Li}, \bibinfo{person}{Rameswar Panda}, \bibinfo{person}{Yoon Kim}, \bibinfo{person}{Chun-Fu~Richard Chen}, \bibinfo{person}{Rogerio~S Feris}, \bibinfo{person}{David Cox}, {and} \bibinfo{person}{Nuno Vasconcelos}.} \bibinfo{year}{2022}\natexlab{}.
\newblock \showarticletitle{VALHALLA: Visual Hallucination for Machine Translation}. In \bibinfo{booktitle}{\emph{Proceedings of the IEEE/CVF Conference on Computer Vision and Pattern Recognition}}. \bibinfo{pages}{5216--5226}.
\newblock


\bibitem[Lin et~al\mbox{.}(2020)]%
        {lin2020cross}
\bibfield{author}{\bibinfo{person}{Kevin Lin}, \bibinfo{person}{Lijuan Wang}, \bibinfo{person}{Kun Luo}, \bibinfo{person}{Yinpeng Chen}, \bibinfo{person}{Zicheng Liu}, {and} \bibinfo{person}{Ming-Ting Sun}.} \bibinfo{year}{2020}\natexlab{}.
\newblock \showarticletitle{Cross-domain complementary learning using pose for multi-person part segmentation}.
\newblock \bibinfo{journal}{\emph{IEEE Transactions on Circuits and Systems for Video Technology}} \bibinfo{volume}{31}, \bibinfo{number}{3} (\bibinfo{year}{2020}), \bibinfo{pages}{1066--1078}.
\newblock


\bibitem[Lin et~al\mbox{.}(2014)]%
        {lin2014microsoft}
\bibfield{author}{\bibinfo{person}{Tsung-Yi Lin}, \bibinfo{person}{Michael Maire}, \bibinfo{person}{Serge Belongie}, \bibinfo{person}{James Hays}, \bibinfo{person}{Pietro Perona}, \bibinfo{person}{Deva Ramanan}, \bibinfo{person}{Piotr Doll{\'a}r}, {and} \bibinfo{person}{C~Lawrence Zitnick}.} \bibinfo{year}{2014}\natexlab{}.
\newblock \showarticletitle{Microsoft coco: Common objects in context}. In \bibinfo{booktitle}{\emph{European conference on computer vision}}. Springer, \bibinfo{pages}{740--755}.
\newblock


\bibitem[Liu(2017)]%
        {liu2017many}
\bibfield{author}{\bibinfo{person}{Bing Liu}.} \bibinfo{year}{2017}\natexlab{}.
\newblock \showarticletitle{Many facets of sentiment analysis}.
\newblock In \bibinfo{booktitle}{\emph{A practical guide to sentiment analysis}}. \bibinfo{publisher}{Springer}, \bibinfo{pages}{11--39}.
\newblock


\bibitem[Lundberg and Lee(2017)]%
        {lundberg2017unified}
\bibfield{author}{\bibinfo{person}{Scott~M Lundberg} {and} \bibinfo{person}{Su-In Lee}.} \bibinfo{year}{2017}\natexlab{}.
\newblock \showarticletitle{A unified approach to interpreting model predictions}. In \bibinfo{booktitle}{\emph{Advances in neural information processing systems}}. \bibinfo{pages}{4765--4774}.
\newblock


\bibitem[Makowski et~al\mbox{.}(2021)]%
        {makowski2021neurokit2}
\bibfield{author}{\bibinfo{person}{Dominique Makowski}, \bibinfo{person}{Tam Pham}, \bibinfo{person}{Zen~J Lau}, \bibinfo{person}{Jan~C Brammer}, \bibinfo{person}{Fran{\c{c}}ois Lespinasse}, \bibinfo{person}{Hung Pham}, \bibinfo{person}{Christopher Sch{\"o}lzel}, {and} \bibinfo{person}{SH~Annabel Chen}.} \bibinfo{year}{2021}\natexlab{}.
\newblock \showarticletitle{NeuroKit2: A Python toolbox for neurophysiological signal processing}.
\newblock \bibinfo{journal}{\emph{Behavior research methods}} (\bibinfo{year}{2021}), \bibinfo{pages}{1--8}.
\newblock


\bibitem[Mantello and Ho(2023)]%
        {mantello2023emotional}
\bibfield{author}{\bibinfo{person}{Peter Mantello} {and} \bibinfo{person}{Manh-Tung Ho}.} \bibinfo{year}{2023}\natexlab{}.
\newblock \showarticletitle{Emotional AI and the future of wellbeing in the post-pandemic workplace}.
\newblock \bibinfo{journal}{\emph{AI \& society}} (\bibinfo{year}{2023}), \bibinfo{pages}{1--7}.
\newblock


\bibitem[marcellodebernardi(2019)]%
        {marcellodebernardi}
\bibfield{author}{\bibinfo{person}{marcellodebernardi}.} \bibinfo{year}{2019}\natexlab{}.
\newblock \bibinfo{title}{loss-landscapes}.
\newblock \bibinfo{howpublished}{\url{https://github.com/marcellodebernardi/loss-landscapes}}.
\newblock


\bibitem[Mauri et~al\mbox{.}(2010)]%
        {mauri2010psychophysiological}
\bibfield{author}{\bibinfo{person}{Maurizio Mauri}, \bibinfo{person}{Valentina Magagnin}, \bibinfo{person}{Pietro Cipresso}, \bibinfo{person}{Luca Mainardi}, \bibinfo{person}{Emery~N Brown}, \bibinfo{person}{Sergio Cerutti}, \bibinfo{person}{Marco Villamira}, {and} \bibinfo{person}{Riccardo Barbieri}.} \bibinfo{year}{2010}\natexlab{}.
\newblock \showarticletitle{Psychophysiological signals associated with affective states}. In \bibinfo{booktitle}{\emph{2010 Annual International Conference of the IEEE Engineering in Medicine and Biology}}. IEEE, \bibinfo{pages}{3563--3566}.
\newblock


\bibitem[Maurice et~al\mbox{.}(2018)]%
        {maurice2018ethical}
\bibfield{author}{\bibinfo{person}{Pauline Maurice}, \bibinfo{person}{Ludivine Allienne}, \bibinfo{person}{Adrien Malais{\'e}}, {and} \bibinfo{person}{Serena Ivaldi}.} \bibinfo{year}{2018}\natexlab{}.
\newblock \showarticletitle{Ethical and social considerations for the introduction of human-centered technologies at work}. In \bibinfo{booktitle}{\emph{2018 IEEE Workshop on Advanced Robotics and its Social Impacts (ARSO)}}. IEEE, \bibinfo{pages}{131--138}.
\newblock


\bibitem[McHugh(2012)]%
        {mchugh2012interrater}
\bibfield{author}{\bibinfo{person}{Mary~L McHugh}.} \bibinfo{year}{2012}\natexlab{}.
\newblock \showarticletitle{Interrater reliability: the kappa statistic}.
\newblock \bibinfo{journal}{\emph{Biochemia medica}} \bibinfo{volume}{22}, \bibinfo{number}{3} (\bibinfo{year}{2012}), \bibinfo{pages}{276--282}.
\newblock


\bibitem[Menghini et~al\mbox{.}(2019)]%
        {menghini2019stressing}
\bibfield{author}{\bibinfo{person}{Luca Menghini}, \bibinfo{person}{Evelyn Gianfranchi}, \bibinfo{person}{Nicola Cellini}, \bibinfo{person}{Elisabetta Patron}, \bibinfo{person}{Mariaelena Tagliabue}, {and} \bibinfo{person}{Michela Sarlo}.} \bibinfo{year}{2019}\natexlab{}.
\newblock \showarticletitle{Stressing the accuracy: Wrist-worn wearable sensor validation over different conditions}.
\newblock \bibinfo{journal}{\emph{Psychophysiology}} \bibinfo{volume}{56}, \bibinfo{number}{11} (\bibinfo{year}{2019}), \bibinfo{pages}{e13441}.
\newblock


\bibitem[Merla and Romani(2007)]%
        {merla2007thermal}
\bibfield{author}{\bibinfo{person}{Arcangelo Merla} {and} \bibinfo{person}{Gian~Luca Romani}.} \bibinfo{year}{2007}\natexlab{}.
\newblock \showarticletitle{Thermal signatures of emotional arousal: a functional infrared imaging study}. In \bibinfo{booktitle}{\emph{2007 29th Annual International Conference of the IEEE Engineering in Medicine and Biology Society}}. IEEE, \bibinfo{pages}{247--249}.
\newblock


\bibitem[Neema et~al\mbox{.}(2021)]%
        {neema2021infrared}
\bibfield{author}{\bibinfo{person}{Shekhar Neema}, \bibinfo{person}{DM Tripathy}, \bibinfo{person}{Sweta Mukherjee}, \bibinfo{person}{Anwita Sinha}, \bibinfo{person}{Senkadhir Vendhan}, {and} \bibinfo{person}{Biju Vasudevan}.} \bibinfo{year}{2021}\natexlab{}.
\newblock \showarticletitle{Infrared thermography in the diagnosis of palmar hyperhidrosis: A diagnostic study}.
\newblock \bibinfo{journal}{\emph{Medical Journal Armed Forces India}} (\bibinfo{year}{2021}).
\newblock


\bibitem[Nguyen et~al\mbox{.}(2018)]%
        {nguyen2018towards}
\bibfield{author}{\bibinfo{person}{Thu Nguyen}, \bibinfo{person}{Khang Tran}, {and} \bibinfo{person}{Hung Nguyen}.} \bibinfo{year}{2018}\natexlab{}.
\newblock \showarticletitle{Towards thermal region of interest for human emotion estimation}. In \bibinfo{booktitle}{\emph{2018 10th International Conference on Knowledge and Systems Engineering (KSE)}}. IEEE, \bibinfo{pages}{152--157}.
\newblock


\bibitem[Nielsen et~al\mbox{.}(2014)]%
        {nielsen2014taking}
\bibfield{author}{\bibinfo{person}{S{\o}ren~Z Nielsen}, \bibinfo{person}{Rikke Gade}, \bibinfo{person}{Thomas~B Moeslund}, {and} \bibinfo{person}{Hans Skov-Petersen}.} \bibinfo{year}{2014}\natexlab{}.
\newblock \showarticletitle{Taking the temperature of pedestrian movement in public spaces}.
\newblock \bibinfo{journal}{\emph{Transportation Research Procedia}}  \bibinfo{volume}{2} (\bibinfo{year}{2014}), \bibinfo{pages}{660--668}.
\newblock


\bibitem[Ollander et~al\mbox{.}(2016)]%
        {ollander2016comparison}
\bibfield{author}{\bibinfo{person}{Simon Ollander}, \bibinfo{person}{Christelle Godin}, \bibinfo{person}{Aur{\'e}lie Campagne}, {and} \bibinfo{person}{Sylvie Charbonnier}.} \bibinfo{year}{2016}\natexlab{}.
\newblock \showarticletitle{A comparison of wearable and stationary sensors for stress detection}. In \bibinfo{booktitle}{\emph{2016 IEEE International Conference on Systems, Man, and Cybernetics (SMC)}}. IEEE, \bibinfo{pages}{004362--004366}.
\newblock


\bibitem[Pavlidis et~al\mbox{.}(2002)]%
        {pavlidis2002seeing}
\bibfield{author}{\bibinfo{person}{Ioannis Pavlidis}, \bibinfo{person}{Norman~L Eberhardt}, {and} \bibinfo{person}{James~A Levine}.} \bibinfo{year}{2002}\natexlab{}.
\newblock \showarticletitle{Seeing through the face of deception}.
\newblock \bibinfo{journal}{\emph{Nature}} \bibinfo{volume}{415}, \bibinfo{number}{6867} (\bibinfo{year}{2002}), \bibinfo{pages}{35--35}.
\newblock


\bibitem[Pavlidis et~al\mbox{.}(2012)]%
        {pavlidis2012fast}
\bibfield{author}{\bibinfo{person}{Ioannis Pavlidis}, \bibinfo{person}{Panagiotis Tsiamyrtzis}, \bibinfo{person}{Dvijesh Shastri}, \bibinfo{person}{Avinash Wesley}, \bibinfo{person}{Yan Zhou}, \bibinfo{person}{Peggy Lindner}, \bibinfo{person}{Pradeep Buddharaju}, \bibinfo{person}{Rohan Joseph}, \bibinfo{person}{Anitha Mandapati}, \bibinfo{person}{Brian Dunkin}, {et~al\mbox{.}}} \bibinfo{year}{2012}\natexlab{}.
\newblock \showarticletitle{Fast by nature-how stress patterns define human experience and performance in dexterous tasks}.
\newblock \bibinfo{journal}{\emph{Scientific Reports}} \bibinfo{volume}{2}, \bibinfo{number}{1} (\bibinfo{year}{2012}), \bibinfo{pages}{305}.
\newblock


\bibitem[P{\'e}rez-Rosas et~al\mbox{.}(2013)]%
        {perez2013thermal}
\bibfield{author}{\bibinfo{person}{Ver{\'o}nica P{\'e}rez-Rosas}, \bibinfo{person}{Alexis Narvaez}, \bibinfo{person}{Mihai Burzo}, {and} \bibinfo{person}{Rada Mihalcea}.} \bibinfo{year}{2013}\natexlab{}.
\newblock \showarticletitle{Thermal imaging for affect detection}. In \bibinfo{booktitle}{\emph{Proceedings of the 6th International Conference on PErvasive Technologies Related to Assistive Environments}}. \bibinfo{pages}{1--4}.
\newblock


\bibitem[Perpetuini et~al\mbox{.}(2021)]%
        {perpetuini2021regions}
\bibfield{author}{\bibinfo{person}{David Perpetuini}, \bibinfo{person}{Damiano Formenti}, \bibinfo{person}{Daniela Cardone}, \bibinfo{person}{Chiara Filippini}, {and} \bibinfo{person}{Arcangelo Merla}.} \bibinfo{year}{2021}\natexlab{}.
\newblock \showarticletitle{Regions of interest selection and thermal imaging data analysis in sports and exercise science: a narrative review}.
\newblock \bibinfo{journal}{\emph{Physiological Measurement}} \bibinfo{volume}{42}, \bibinfo{number}{8} (\bibinfo{year}{2021}), \bibinfo{pages}{08TR01}.
\newblock


\bibitem[Picard(2016)]%
        {picard2016automating}
\bibfield{author}{\bibinfo{person}{Rosalind~W Picard}.} \bibinfo{year}{2016}\natexlab{}.
\newblock \showarticletitle{Automating the recognition of stress and emotion: From lab to real-world impact}.
\newblock \bibinfo{journal}{\emph{IEEE MultiMedia}} \bibinfo{volume}{23}, \bibinfo{number}{3} (\bibinfo{year}{2016}), \bibinfo{pages}{3--7}.
\newblock


\bibitem[Puri et~al\mbox{.}(2005)]%
        {puri}
\bibfield{author}{\bibinfo{person}{Colin Puri}, \bibinfo{person}{Leslie Olson}, \bibinfo{person}{Ioannis Pavlidis}, \bibinfo{person}{James Levine}, {and} \bibinfo{person}{Justin Starren}.} \bibinfo{year}{2005}\natexlab{}.
\newblock \showarticletitle{StressCam: Non-contact measurement of users' emotional states through thermal imaging}.
\newblock \bibinfo{journal}{\emph{Proceedings of the 2005 ACM Conference on Human Factors in Computing Systems}}  \bibinfo{volume}{2}, \bibinfo{pages}{1725--1728}.
\newblock
\urldef\tempurl%
\url{https://doi.org/10.1145/1056808.1057007}
\showDOI{\tempurl}


\bibitem[Qiao et~al\mbox{.}(2020)]%
        {detectors}
\bibfield{author}{\bibinfo{person}{Siyuan Qiao}, \bibinfo{person}{Liang-Chieh Chen}, {and} \bibinfo{person}{Alan Yuille}.} \bibinfo{year}{2020}\natexlab{}.
\newblock \showarticletitle{DetectoRS: Detecting Objects with Recursive Feature Pyramid and Switchable Atrous Convolution}.
\newblock \bibinfo{journal}{\emph{arXiv preprint arXiv:2006.02334}} (\bibinfo{year}{2020}).
\newblock


\bibitem[Quesada et~al\mbox{.}(2015)]%
        {quesada2015effect}
\bibfield{author}{\bibinfo{person}{Jose Ignacio~Priego Quesada}, \bibinfo{person}{Natividad~Mart{\'\i}nez Guillam{\'o}n}, \bibinfo{person}{Rosa Ma Cibri{\'a}n~Ortiz de Anda}, \bibinfo{person}{Agnes Psikuta}, \bibinfo{person}{Simon Annaheim}, \bibinfo{person}{Ren{\'e}~Michel Rossi}, \bibinfo{person}{Jos{\'e} Miguel~Corber{\'a}n Salvador}, \bibinfo{person}{Pedro P{\'e}rez-Soriano}, {and} \bibinfo{person}{Rosario~Salvador Palmer}.} \bibinfo{year}{2015}\natexlab{}.
\newblock \showarticletitle{Effect of perspiration on skin temperature measurements by infrared thermography and contact thermometry during aerobic cycling}.
\newblock \bibinfo{journal}{\emph{Infrared Physics \& Technology}}  \bibinfo{volume}{72} (\bibinfo{year}{2015}), \bibinfo{pages}{68--76}.
\newblock


\bibitem[Rajan et~al\mbox{.}(2021)]%
        {rajan2021robust}
\bibfield{author}{\bibinfo{person}{Vandana Rajan}, \bibinfo{person}{Alessio Brutti}, {and} \bibinfo{person}{Andrea Cavallaro}.} \bibinfo{year}{2021}\natexlab{}.
\newblock \showarticletitle{Robust Latent Representations Via Cross-Modal Translation and Alignment}. In \bibinfo{booktitle}{\emph{ICASSP 2021-2021 IEEE International Conference on Acoustics, Speech and Signal Processing (ICASSP)}}. IEEE, \bibinfo{pages}{4315--4319}.
\newblock


\bibitem[Reiche et~al\mbox{.}(2004)]%
        {reiche2004stress}
\bibfield{author}{\bibinfo{person}{Edna Maria~Vissoci Reiche}, \bibinfo{person}{Sandra Odebrecht~Vargas Nunes}, {and} \bibinfo{person}{Helena~Kaminami Morimoto}.} \bibinfo{year}{2004}\natexlab{}.
\newblock \showarticletitle{Stress, depression, the immune system, and cancer}.
\newblock \bibinfo{journal}{\emph{The lancet oncology}} \bibinfo{volume}{5}, \bibinfo{number}{10} (\bibinfo{year}{2004}), \bibinfo{pages}{617--625}.
\newblock


\bibitem[Rosebrock(2016)]%
        {rosebrock2016intersection}
\bibfield{author}{\bibinfo{person}{Adrian Rosebrock}.} \bibinfo{year}{2016}\natexlab{}.
\newblock \showarticletitle{Intersection over Union (IoU) for object detection}.
\newblock \bibinfo{journal}{\emph{Diambil kembali dari PYImageSearch: https://www. pyimagesearch. com/2016/11/07/intersection-over-union-iou-for-object-detection}} (\bibinfo{year}{2016}).
\newblock


\bibitem[Rusli et~al\mbox{.}(2020)]%
        {rusli2020implementation}
\bibfield{author}{\bibinfo{person}{Nazreen Rusli}, \bibinfo{person}{Shahrul~Naim Sidek}, \bibinfo{person}{Hazlina~Md Yusof}, \bibinfo{person}{Nor~Izzati Ishak}, \bibinfo{person}{Madihah Khalid}, {and} \bibinfo{person}{Ahmad Aidil~Arafat Dzulkarnain}.} \bibinfo{year}{2020}\natexlab{}.
\newblock \showarticletitle{Implementation of wavelet analysis on thermal images for affective states recognition of children with autism spectrum disorder}.
\newblock \bibinfo{journal}{\emph{IEEE Access}}  \bibinfo{volume}{8} (\bibinfo{year}{2020}), \bibinfo{pages}{120818--120834}.
\newblock


\bibitem[SafelyYou(2022)]%
        {Safelyyou}
\bibfield{author}{\bibinfo{person}{SafelyYou}.} \bibinfo{year}{2022}\natexlab{}.
\newblock \bibinfo{title}{Transform care delivery with world-leading AI + clinical expertise}.
\newblock
\newblock
\urldef\tempurl%
\url{https://www.safely-you.com/}
\showURL{%
Retrieved July, 2023 from \tempurl}


\bibitem[Salekin et~al\mbox{.}(2017)]%
        {salekin2017dave}
\bibfield{author}{\bibinfo{person}{Asif Salekin}, \bibinfo{person}{Hongning Wang}, \bibinfo{person}{Kristine Williams}, {and} \bibinfo{person}{John Stankovic}.} \bibinfo{year}{2017}\natexlab{}.
\newblock \showarticletitle{Dave: detecting agitated vocal events}. In \bibinfo{booktitle}{\emph{2017 IEEE/ACM International Conference on Connected Health: Applications, Systems and Engineering Technologies (CHASE)}}. IEEE, \bibinfo{pages}{157--166}.
\newblock


\bibitem[Sarsenbayeva et~al\mbox{.}(2019)]%
        {sarsenbayeva2019measuring}
\bibfield{author}{\bibinfo{person}{Zhanna Sarsenbayeva}, \bibinfo{person}{Niels van Berkel}, \bibinfo{person}{Danula Hettiachchi}, \bibinfo{person}{Weiwei Jiang}, \bibinfo{person}{Tilman Dingler}, \bibinfo{person}{Eduardo Velloso}, \bibinfo{person}{Vassilis Kostakos}, {and} \bibinfo{person}{Jorge Goncalves}.} \bibinfo{year}{2019}\natexlab{}.
\newblock \showarticletitle{Measuring the effects of stress on mobile interaction}.
\newblock \bibinfo{journal}{\emph{Proceedings of the ACM on Interactive, Mobile, Wearable and Ubiquitous Technologies}} \bibinfo{volume}{3}, \bibinfo{number}{1} (\bibinfo{year}{2019}), \bibinfo{pages}{1--18}.
\newblock


\bibitem[Schaefer et~al\mbox{.}(2010)]%
        {schaefer2010assessing}
\bibfield{author}{\bibinfo{person}{Alexandre Schaefer}, \bibinfo{person}{Fr{\'e}d{\'e}ric Nils}, \bibinfo{person}{Xavier Sanchez}, {and} \bibinfo{person}{Pierre Philippot}.} \bibinfo{year}{2010}\natexlab{}.
\newblock \showarticletitle{Assessing the effectiveness of a large database of emotion-eliciting films: A new tool for emotion researchers}.
\newblock \bibinfo{journal}{\emph{Cognition and emotion}} \bibinfo{volume}{24}, \bibinfo{number}{7} (\bibinfo{year}{2010}), \bibinfo{pages}{1153--1172}.
\newblock


\bibitem[Security(2023)]%
        {Protech}
\bibfield{author}{\bibinfo{person}{ProTech Security}.} \bibinfo{year}{2023}\natexlab{}.
\newblock \bibinfo{title}{How Thermal Cameras for Businesses Can Keep Employees and Customers Safe}.
\newblock
\newblock
\urldef\tempurl%
\url{https://protechsecurity.com/how-thermal-cameras-for-businesses-can-keep-employees-and-customers-safe/}
\showURL{%
Retrieved July, 2023 from \tempurl}


\bibitem[Setz et~al\mbox{.}(2009)]%
        {setz2009discriminating}
\bibfield{author}{\bibinfo{person}{Cornelia Setz}, \bibinfo{person}{Bert Arnrich}, \bibinfo{person}{Johannes Schumm}, \bibinfo{person}{Roberto La~Marca}, \bibinfo{person}{Gerhard Tr{\"o}ster}, {and} \bibinfo{person}{Ulrike Ehlert}.} \bibinfo{year}{2009}\natexlab{}.
\newblock \showarticletitle{Discriminating stress from cognitive load using a wearable EDA device}.
\newblock \bibinfo{journal}{\emph{IEEE Transactions on information technology in biomedicine}} \bibinfo{volume}{14}, \bibinfo{number}{2} (\bibinfo{year}{2009}), \bibinfo{pages}{410--417}.
\newblock


\bibitem[Shapley(1953)]%
        {shapley1953stochastic}
\bibfield{author}{\bibinfo{person}{Lloyd~S Shapley}.} \bibinfo{year}{1953}\natexlab{}.
\newblock \showarticletitle{Stochastic games}.
\newblock \bibinfo{journal}{\emph{Proceedings of the national academy of sciences}} \bibinfo{volume}{39}, \bibinfo{number}{10} (\bibinfo{year}{1953}), \bibinfo{pages}{1095--1100}.
\newblock


\bibitem[Sharma et~al\mbox{.}(2022)]%
        {sharma2022psychophysiological}
\bibfield{author}{\bibinfo{person}{Harshit Sharma}, \bibinfo{person}{Yi Xiao}, \bibinfo{person}{Victoria Tumanova}, {and} \bibinfo{person}{Asif Salekin}.} \bibinfo{year}{2022}\natexlab{}.
\newblock \showarticletitle{Psychophysiological Arousal in Young Children Who Stutter: An Interpretable AI Approach}.
\newblock \bibinfo{journal}{\emph{Proceedings of the ACM on Interactive, Mobile, Wearable and Ubiquitous Technologies}} \bibinfo{volume}{6}, \bibinfo{number}{3} (\bibinfo{year}{2022}), \bibinfo{pages}{1--32}.
\newblock


\bibitem[Shneiderman(2020)]%
        {shneiderman2020bridging}
\bibfield{author}{\bibinfo{person}{Ben Shneiderman}.} \bibinfo{year}{2020}\natexlab{}.
\newblock \showarticletitle{Bridging the gap between ethics and practice: guidelines for reliable, safe, and trustworthy human-centered AI systems}.
\newblock \bibinfo{journal}{\emph{ACM Transactions on Interactive Intelligent Systems (TiiS)}} \bibinfo{volume}{10}, \bibinfo{number}{4} (\bibinfo{year}{2020}), \bibinfo{pages}{1--31}.
\newblock


\bibitem[Sonkusare et~al\mbox{.}(2019)]%
        {sonkusare2019detecting}
\bibfield{author}{\bibinfo{person}{Saurabh Sonkusare}, \bibinfo{person}{David Ahmedt-Aristizabal}, \bibinfo{person}{Matthew~J Aburn}, \bibinfo{person}{Vinh~Thai Nguyen}, \bibinfo{person}{Tianji Pang}, \bibinfo{person}{Sascha Frydman}, \bibinfo{person}{Simon Denman}, \bibinfo{person}{Clinton Fookes}, \bibinfo{person}{Michael Breakspear}, {and} \bibinfo{person}{Christine~C Guo}.} \bibinfo{year}{2019}\natexlab{}.
\newblock \showarticletitle{Detecting changes in facial temperature induced by a sudden auditory stimulus based on deep learning-assisted face tracking}.
\newblock \bibinfo{journal}{\emph{Scientific reports}} \bibinfo{volume}{9}, \bibinfo{number}{1} (\bibinfo{year}{2019}), \bibinfo{pages}{4729}.
\newblock


\bibitem[Thammasan et~al\mbox{.}(2017)]%
        {thammasan2017familiarity}
\bibfield{author}{\bibinfo{person}{Nattapong Thammasan}, \bibinfo{person}{Koichi Moriyama}, \bibinfo{person}{Ken-ichi Fukui}, {and} \bibinfo{person}{Masayuki Numao}.} \bibinfo{year}{2017}\natexlab{}.
\newblock \showarticletitle{Familiarity effects in EEG-based emotion recognition}.
\newblock \bibinfo{journal}{\emph{Brain informatics}}  \bibinfo{volume}{4} (\bibinfo{year}{2017}), \bibinfo{pages}{39--50}.
\newblock


\bibitem[Tumanova and Backes(2019)]%
        {tumanova2019autonomic}
\bibfield{author}{\bibinfo{person}{Victoria Tumanova} {and} \bibinfo{person}{Nicole Backes}.} \bibinfo{year}{2019}\natexlab{}.
\newblock \showarticletitle{Autonomic nervous system response to speech production in stuttering and normally fluent preschool-age children}.
\newblock \bibinfo{journal}{\emph{Journal of Speech, Language, and Hearing Research}} \bibinfo{volume}{62}, \bibinfo{number}{11} (\bibinfo{year}{2019}), \bibinfo{pages}{4030--4044}.
\newblock


\bibitem[van Dooren et~al\mbox{.}(2012)]%
        {van2012emotional}
\bibfield{author}{\bibinfo{person}{Marieke van Dooren}, \bibinfo{person}{Joris~H Janssen}, {et~al\mbox{.}}} \bibinfo{year}{2012}\natexlab{}.
\newblock \showarticletitle{Emotional sweating across the body: Comparing 16 different skin conductance measurement locations}.
\newblock \bibinfo{journal}{\emph{Physiology \& behavior}} \bibinfo{volume}{106}, \bibinfo{number}{2} (\bibinfo{year}{2012}), \bibinfo{pages}{298--304}.
\newblock


\bibitem[Vandersteegen(2018)]%
        {libseek}
\bibfield{author}{\bibinfo{person}{Maarten Vandersteegen}.} \bibinfo{year}{2018}\natexlab{}.
\newblock \bibinfo{title}{SEEK thermal compact camera driver supporting the thermal Compact, thermal CompactXR and and thermal CompactPRO}.
\newblock
\newblock
\urldef\tempurl%
\url{https://github.com/maartenvds/libseek-thermal}
\showURL{%
\tempurl}


\bibitem[Von~Rosenberg et~al\mbox{.}(2017)]%
        {von2017resolving}
\bibfield{author}{\bibinfo{person}{Wilhelm Von~Rosenberg}, \bibinfo{person}{Theerasak Chanwimalueang}, \bibinfo{person}{Tricia Adjei}, \bibinfo{person}{Usman Jaffer}, \bibinfo{person}{Valentin Goverdovsky}, {and} \bibinfo{person}{Danilo~P Mandic}.} \bibinfo{year}{2017}\natexlab{}.
\newblock \showarticletitle{Resolving ambiguities in the LF/HF ratio: LF-HF scatter plots for the categorization of mental and physical stress from HRV}.
\newblock \bibinfo{journal}{\emph{Frontiers in physiology}}  \bibinfo{volume}{8} (\bibinfo{year}{2017}), \bibinfo{pages}{360}.
\newblock


\bibitem[Walambe et~al\mbox{.}(2021)]%
        {walambe2021employing}
\bibfield{author}{\bibinfo{person}{Rahee Walambe}, \bibinfo{person}{Pranav Nayak}, \bibinfo{person}{Ashmit Bhardwaj}, {and} \bibinfo{person}{Ketan Kotecha}.} \bibinfo{year}{2021}\natexlab{}.
\newblock \showarticletitle{Employing multimodal machine learning for stress detection}.
\newblock \bibinfo{journal}{\emph{Journal of Healthcare Engineering}}  \bibinfo{volume}{2021} (\bibinfo{year}{2021}), \bibinfo{pages}{1--12}.
\newblock


\bibitem[Wang et~al\mbox{.}(2022a)]%
        {wang2022adversarial}
\bibfield{author}{\bibinfo{person}{Qianqian Wang}, \bibinfo{person}{Zhiqiang Tao}, \bibinfo{person}{Wei Xia}, \bibinfo{person}{Quanxue Gao}, \bibinfo{person}{Xiaochun Cao}, {and} \bibinfo{person}{Licheng Jiao}.} \bibinfo{year}{2022}\natexlab{a}.
\newblock \showarticletitle{Adversarial multiview clustering networks with adaptive fusion}.
\newblock \bibinfo{journal}{\emph{IEEE transactions on neural networks and learning systems}} (\bibinfo{year}{2022}).
\newblock


\bibitem[Wang et~al\mbox{.}(2022b)]%
        {wang2022feature}
\bibfield{author}{\bibinfo{person}{Zhikang Wang}, \bibinfo{person}{Feng Zhu}, \bibinfo{person}{Shixiang Tang}, \bibinfo{person}{Rui Zhao}, \bibinfo{person}{Lihuo He}, {and} \bibinfo{person}{Jiangning Song}.} \bibinfo{year}{2022}\natexlab{b}.
\newblock \showarticletitle{Feature erasing and diffusion network for occluded person re-identification}. In \bibinfo{booktitle}{\emph{Proceedings of the IEEE/CVF conference on computer vision and pattern recognition}}. \bibinfo{pages}{4754--4763}.
\newblock


\bibitem[Wright and Singh(2022)]%
        {wright2022reducing}
\bibfield{author}{\bibinfo{person}{Kay Wright} {and} \bibinfo{person}{Swaran Singh}.} \bibinfo{year}{2022}\natexlab{}.
\newblock \showarticletitle{Reducing falls in dementia inpatients using vision-based technology}.
\newblock \bibinfo{journal}{\emph{Journal of Patient Safety}} \bibinfo{volume}{18}, \bibinfo{number}{3} (\bibinfo{year}{2022}), \bibinfo{pages}{177}.
\newblock


\bibitem[Yang(2022)]%
        {yang2022enabling}
\bibfield{author}{\bibinfo{person}{Jiacheng Yang}.} \bibinfo{year}{2022}\natexlab{}.
\newblock \emph{\bibinfo{title}{Enabling Privacy-Preserving Model Personalization via On-Device Incremental Training}}.
\newblock \bibinfo{thesistype}{Ph.\,D. Dissertation}. \bibinfo{school}{University of Toronto (Canada)}.
\newblock


\bibitem[Yu et~al\mbox{.}(2018)]%
        {yu2018biofeedback}
\bibfield{author}{\bibinfo{person}{Bin Yu}, \bibinfo{person}{Mathias Funk}, \bibinfo{person}{Jun Hu}, \bibinfo{person}{Qi Wang}, {and} \bibinfo{person}{Loe Feijs}.} \bibinfo{year}{2018}\natexlab{}.
\newblock \showarticletitle{Biofeedback for everyday stress management: A systematic review}.
\newblock \bibinfo{journal}{\emph{Frontiers in ICT}}  \bibinfo{volume}{5} (\bibinfo{year}{2018}), \bibinfo{pages}{23}.
\newblock


\bibitem[Zhang et~al\mbox{.}(2020)]%
        {zhang2020video}
\bibfield{author}{\bibinfo{person}{Huijun Zhang}, \bibinfo{person}{Ling Feng}, \bibinfo{person}{Ningyun Li}, \bibinfo{person}{Zhanyu Jin}, {and} \bibinfo{person}{Lei Cao}.} \bibinfo{year}{2020}\natexlab{}.
\newblock \showarticletitle{Video-Based Stress Detection through Deep Learning}.
\newblock \bibinfo{journal}{\emph{Sensors}} \bibinfo{volume}{20}, \bibinfo{number}{19} (\bibinfo{year}{2020}), \bibinfo{pages}{5552}.
\newblock


\bibitem[Zhang et~al\mbox{.}(2022)]%
        {zhang2022real}
\bibfield{author}{\bibinfo{person}{Jing Zhang}, \bibinfo{person}{Hang Yin}, \bibinfo{person}{Jiayu Zhang}, \bibinfo{person}{Gang Yang}, \bibinfo{person}{Jing Qin}, {and} \bibinfo{person}{Ling He}.} \bibinfo{year}{2022}\natexlab{}.
\newblock \showarticletitle{Real-time mental stress detection using multimodality expressions with a deep learning framework}.
\newblock \bibinfo{journal}{\emph{Frontiers in Neuroscience}}  \bibinfo{volume}{16} (\bibinfo{year}{2022}).
\newblock


\bibitem[Zheng et~al\mbox{.}(2021)]%
        {zheng2021deep}
\bibfield{author}{\bibinfo{person}{Zhuo Zheng}, \bibinfo{person}{Ailong Ma}, \bibinfo{person}{Liangpei Zhang}, {and} \bibinfo{person}{Yanfei Zhong}.} \bibinfo{year}{2021}\natexlab{}.
\newblock \showarticletitle{Deep multisensor learning for missing-modality all-weather mapping}.
\newblock \bibinfo{journal}{\emph{ISPRS Journal of Photogrammetry and Remote Sensing}}  \bibinfo{volume}{174} (\bibinfo{year}{2021}), \bibinfo{pages}{254--264}.
\newblock


\bibitem[Zhou and Xiang(2019)]%
        {zhou2019torchreid}
\bibfield{author}{\bibinfo{person}{Kaiyang Zhou} {and} \bibinfo{person}{Tao Xiang}.} \bibinfo{year}{2019}\natexlab{}.
\newblock \showarticletitle{Torchreid: A library for deep learning person re-identification in pytorch}.
\newblock \bibinfo{journal}{\emph{arXiv preprint arXiv:1910.10093}} (\bibinfo{year}{2019}).
\newblock


\bibitem[Zhou et~al\mbox{.}(2019)]%
        {zhou2019omni}
\bibfield{author}{\bibinfo{person}{Kaiyang Zhou}, \bibinfo{person}{Yongxin Yang}, \bibinfo{person}{Andrea Cavallaro}, {and} \bibinfo{person}{Tao Xiang}.} \bibinfo{year}{2019}\natexlab{}.
\newblock \showarticletitle{Omni-scale feature learning for person re-identification}. In \bibinfo{booktitle}{\emph{Proceedings of the IEEE/CVF international conference on computer vision}}. \bibinfo{pages}{3702--3712}.
\newblock


\bibitem[Zhu et~al\mbox{.}(2022)]%
        {zhu2022feasibility}
\bibfield{author}{\bibinfo{person}{Lili Zhu}, \bibinfo{person}{Pai~Chet Ng}, \bibinfo{person}{Yuanhao Yu}, \bibinfo{person}{Yang Wang}, \bibinfo{person}{Petros Spachos}, \bibinfo{person}{Dimitrios Hatzinakos}, {and} \bibinfo{person}{Konstantinos~N Plataniotis}.} \bibinfo{year}{2022}\natexlab{}.
\newblock \showarticletitle{Feasibility study of stress detection with machine learning through eda from wearable devices}. In \bibinfo{booktitle}{\emph{ICC 2022-IEEE International Conference on Communications}}. IEEE, \bibinfo{pages}{4800--4805}.
\newblock


\bibitem[Zoë~Corbyn(2021)]%
        {guardian2021}
\bibfield{author}{\bibinfo{person}{The~Guardian Zoë~Corbyn}.} \bibinfo{year}{2021}\natexlab{}.
\newblock \bibinfo{title}{The future of elder care is here – and it’s artificial intelligence}.
\newblock
\newblock
\urldef\tempurl%
\url{https://www.theguardian.com/us-news/2021/jun/03/elder-care-artificial-intelligence-software}
\showURL{%
Retrieved July, 2023 from \tempurl}


\end{thebibliography}
\appendix
\section{Appendix}
\subsection{Loss Landscape Visualization}\label{loss-landscape}
Li et al. \cite{li2018visualizing} showed that visualizing the loss landscape for a neural network provides a richer understanding of how the different model architecture and other design choices influence the optimization of the loss function. While generating the loss landscapes, we intend to visualize the impact of model parameters or $\theta$, which is a high dimensional quantity. For interpretability, it is required to reduce its dimensionality to a one or two-dimensional hyperspace. To address this challenge prior works\cite{goodfellow2014qualitatively,im2016empirical,li2018visualizing} choose a starting point in the parameter subspace $\theta$ and choose two random Gaussian directions vectors given by  $\delta$ and $\eta$ and plot the graph for:
\begin{equation}
    f(\alpha,\beta)=L(\theta+\alpha\delta+\beta\eta)
\end{equation}
This equation generates 3D visualization of the loss landscape with XY region bounded by two scalar quantities or step sizes $\alpha$ (x-axis) and $\beta$ (y-axis) and corresponding loss for the  $L(\theta+\alpha\delta+\beta\eta)$ as the z-axis. Furthermore, Li et al. \cite{li2018visualizing} suggest using filter normalized direction vectors $\delta$ and $\eta$ helps to capture the natural distance scale of the loss surfaces (details can be found in the original work\cite{li2018visualizing}). We used the \textit{loss-landscapes} library \cite{marcellodebernardi} which also uses filter-normalization appraoch \cite{li2018visualizing} to generate the 3D loss landscape plots for the best thermal baseline, and our ThermaStrain approach which is shown in the paper (Figure \ref{fig:3dlosslandscape}). We used the cross-entropy loss for the graph generation, and both the graphs were generated for a randomly selected participant from the validation set for step sizes $(\alpha=40,\beta=40)$.

\end{document}